# METRICS FOR ASSESSING INCLUSIVITY AND EMPOWERMENT OF PEOPLE FOR SUPPORTING THE DESIGN OF INCLUSIVE PRODUCT LIFECYCLES

Naz Yaldız[1][https://orcid.org/0000-0001-8738-7137], Amaresh Chakrabarti[2][https://orcid.org/0000-0002-1809-1831]
[1,2]Department of Design and Manufacturing, Indian Institute of Science, Bengaluru, India
[1]Corresponding Author: nazyaldiz@iisc.ac.in
[2]ac123@iisc.ac.in

The design of an inclusive product lifecycle is important for empowering stakeholders through their meaningful inclusion in lifecycle processes. To achieve this, the inclusion of stakeholders must be structured in a way that supports their empowerment. Inclusivity addresses the lifecycle context to improve how diverse stakeholders are included across phases, supporting their empowerment. This study aims to build on current inclusive design approaches, which often focus on empowering users through the use of products. It proposes inclusive lifecycle processes as a way to empower a wider range of stakeholders for sustainable development. We apply a novel framework to real-life case studies from the literature to identify the dimensions of empowerment and inclusivity. By analysing the relationships between these dimensions, we propose specific metrics that show strong causal connections. These metrics allow measurable outcomes to support the fair development of stakeholders. Measuring inclusivity and empowerment can serve as a foundation for supporting sustainability, dignity, well-being, and fair stakeholder development, which are explored further in Part 2 and Part 3 of this study. This study contributes to design science by expanding the understanding of inclusive design through stakeholder inclusion in lifecycle phases and by shifting the focus from product to process design.

Keywords: empowerment, inclusivity of stakeholders, inclusive product lifecycle design

## 1. INTRODUCTION

Designing inclusive lifecycles of products is essential for empowering people, through their involvement in the various phases of the lifecycle process, to support the transition towards sustainability. Sustainability is an "… approach or the guiding principles that coordinates all facets of development with the aim of achieving a sustainable level of development" (Hodge, 1997; Ozili, 2022). It is influenced by multiple factors within its three aspects: environment, society, and economy (Gräßler and Hesse, 2022). These influencing factors include resource consumption (Chen et al., 2020), equity, empowerment, inclusion, participation, access (Dempsey et al., 2011; Colantonio, 2009; Vallance et al., 2011), income generation, and consuming interest (Goodland, 1995). Therefore, these factors can vary depending on different contexts and solutions. However, to design inclusive lifecycles, it is necessary to understand how inclusive and sustainable a product lifecycle is. This requires certain metrics through which inclusivity and sustainable development can be assessed. While there is no current definition of inclusive lifecycle design, the closest related concept is 'inclusive manufacturing', as given by Roy et al. (2018): inclusive manufacturing is "… a new paradigm concept, where all parts of the lifecycle of a manufactured product is made accessible to people from all strata of the society, so as to, accelerate sustainable development and dignified well-being for all. Inclusive manufacturing aims at empowering people, especially those who are spatially, temporally, physically, economically and culturally disadvantaged, to actively participate in the conception, creation, distribution, transaction, use, and retirement of products and systems." According to this definition, inclusion and empowerment of people are key influencing factors of sustainability in the context of designing inclusive lifecycles of products. Inclusion is a process (Hansen, 2012) aimed at ensuring equal access to opportunities (Ozili, 2020), improving the well-being of people (Shore et al., 2011), protecting human dignity, and enabling the exercise of human rights (Armstrong et al., 2011). To empower people, a practice of inclusivity is to create "a community involved in coproducing processes, policies, and programs for defining and addressing public issues" (Quick and Feldman, 2011). Empowerment is "a transformation process of a community" (Ferguson, 2010; Sianipar et al., 2013) to "help people gain control over their lives" (Page and Czuba, 1999) by "gaining power" (Avelino and Wittmayer, 2016). As emphasised by Foucault (1982), power "exists only as exercised by some on others, only when it is put into action." Cattaneo and Chapman (2010) pointed out the iterative nature of empowering processes, which aim to increase the power of people based on a purpose. For instance, the purpose of customer or user inclusion in designing an inclusive lifecycle can be to empower customers through their involvement in functions such as decision-making (Jespersen, 2011), participating in various product development tasks (Füller et al., 2009), or even acting as designers or taking on the role of marketers (Sarmah and Rahman, 2017). These context-specific functions within lifecycle design can help identify who may be included or excluded. However, there is still ambiguity in understanding what kind of inclusion will lead to empowerment in an inclusive lifecycle design of a product aimed at supporting the transition to sustainability. In addition, it is critical for lifecycle designers to assess the extent to which people can be included in lifecycle processes and how much this inclusion would empower them.

The degree of inclusion of people in product design processes is often evaluated in terms of the degree of inclusion of excluded populations in society (Ning et al., 2019; Clarkson et al., 2015; Johnson et al., 2010). Ning et al. (2019) emphasise the importance of understanding the cognitive characteristics of users to prevent exclusion by aligning product demands with users' capabilities. A user who faces a mismatch between her demand and capability may not be able to use the product, which can lead to her exclusion from its use (Clarkson et al., 2015). BSI (2005) defined inclusive design as "the design of mainstream products and/or services that are accessible to, and usable by, people with the widest range of abilities within the widest range of situations, without the need for special adaptation or design." Clarkson (2016) defined exclusion in design as a situation when "…the demands of using a particular product, within a given environment, exceed the capabilities of the user." The elimination of barriers that cause exclusion in product use is considered the purpose of inclusive design (Li and Dong, 2019).

In the literature, as discussed above, the inclusion of people is currently limited to increasing the diversity of users -especially who are excluded- of a designed solution (product, process, etc.). If a more diverse group of people can use the solution, this is often interpreted as an indicator of inclusivity. As a result, current design approaches prioritise user needs and capabilities to enhance product features. In this way, the usability and accessibility of solutions can be increased, which are interpreted as key outcomes of inclusive design. As a result, the dimensions of diversity are often limited to users only, rather than considering all stakeholders. Therefore, the outcome of inclusive design is usually limited to empowering users in terms of accessing and using products, systems, or services, based on their usability and accessibility.

As Hansen (2012) pointed out, "Inclusion and exclusion are two connected and interdependent processes. Exclusion makes inclusion possible and simultaneously makes inclusion impossible as a limitless condition. As a concept, inclusion therefore presupposes exclusion, because an inclusive society needs to include a certain degree of exclusion to ensure its own existence." The exclusion perspective of the inclusion process in inclusive lifecycle design can influence the empowering outcomes of functions or activities in the lifecycle processes.

Practices for designing inclusive lifecycle processes can be structured around the product's lifecycle phases. The design of an inclusive product lifecycle can be achieved by explicitly considering and enhancing the inclusion of people in all five phases of the product lifecycle—planning, development, production, utilisation, and recirculation—which are defined by Urakami and Vajna (2018). For instance, the practice in the planning phase can involve conducting market research, while in the utilisation phase, it can involve the modernisation of the product. Therefore, it is important to identify who can be included in a function or activity in each lifecycle process and who should be excluded from that particular function.

As an empowering process, the inclusive lifecycle design of products ensures opportunities for stakeholders who can be included in all lifecycle phases: planning, development, production, utilisation, and recirculation. In this way, the argument is that stakeholders can be better empowered during the transition to sustainable development by fulfilling their own needs and sharing resources with other included stakeholders who require them. Stakeholders are internal and external resources of the lifecycle processes and are involved in each phase. Based on the lifecycle phases, stakeholders can be design partners, manufacturers, suppliers, customers, logistics providers, app developers, recycling companies, etc. (Fernandes et al., 2019).

The overall understanding in current inclusivity approaches is that they primarily focus on empowering users by improving product attributes. However, as highlighted in the definition of inclusive manufacturing, inclusivity can also empower people to contribute to sustainable development and enhance their dignity and well-being. Although inclusivity and empowerment are widely recognised as important goals, current product lifecycle processes still lack clear strategies for meaningfully including diverse stakeholders in ways that support these broader outcomes. In inclusive design, inclusivity is often addressed by increasing the number of diverse user groups; however, there remains a need to evaluate which types of inclusivity practices truly support empowerment, and what kinds of functions people can be involved in to empower themselves.

### 1.1. Aims, Objectives and Research Question

This research aims to explore the operationalisability of the inclusive manufacturing definition given in Section 1. With this aim, an approach is developed to include stakeholders in lifecycle processes in a meaningful way to support their empowerment. As conceptualised in Figure 1, the inclusion of people in empowering processes—such as product lifecycle processes—enhances their empowerment and supports their dignity and well-being through their contribution to sustainable development. This study explores the initial phase of the concept map (see Figure 1). Therefore, we argue that the inclusion of people in lifecycle processes is key to the empowerment of people. To support this argument, a framework for the assessment of empowerment and inclusivity is used in this study, with examples provided to clarify and enhance its application. The framework aims to evaluate the relationships between empowerment and the inclusivity of stakeholders within a product lifecycle. The initial version of the framework was presented in Yaldiz and Chakrabarti (2024a), further developed in this paper, and later applied in Yaldiz and Chakrabarti (2024b) to assess the influence of inclusivity on sustainability and dignified well-being, as targeted in the second phase of the study. This study applies the framework to ten real-life case studies. Based on the results of this application, the relationship between empowerment and inclusivity is identified using a variety of metrics. Among these, the specific dimensions of inclusivity and empowerment that are most strongly correlated are selected. In conclusion, this study proposes a specific dimension of inclusivity that has the most significant effect on the dimensions of empowerment in inclusive lifecycles.

Third Phase:
Pathway proposals based on
the interrelation of variables
to support equal and fair
stakeholder development

Inclusion of people → Product lifecycle processes → Empowering processes (that lead to empowerment of included people) → Well-being and dignity / Sustainable development → Fair and Equal Development of Stakeholders

First Phase:
Inclusivity of stakeholders to
enhance their empowerment

Second Phase:
Empowered stakeholders' influence
on sustainable development,
well-being and dignity

Figure 1: Conceptual pathway of inclusivity in product lifecycle processes

With insight into these aims, the objectives of this research are as follows:

1. To identify the dimensions of inclusivity and empowerment.
2. To improve a framework to measure inclusivity and empowerment in product lifecycle processes.
3. To address the relationship between inclusivity and empowerment in the context of product lifecycle processes.

Based on these aims and objectives, the research question (RQ) is as follows:

RQ: How can inclusivity and empowerment be assessed in product lifecycle processes? In the definition of inclusive manufacturing, a relationship between inclusivity and empowerment is assumed, with the emphasis that *'inclusive manufacturing aims to empower people…'* (see Section 1). The hypothesis is that an increase in the inclusivity of stakeholders in a product lifecycle process can support the enhancement of empowerment.

Section 2 presents a summary of the literature on inclusivity and empowerment. Section 3 explains the research methodology to address the research question. Section 4 describes the revised framework with its improvements. Section 5 proposes the metrics for empowerment and inclusivity, and Section 6 presents the application of the framework to case studies. Section 7 presents the results and identifies the dimensions of empowerment and inclusivity based on the results. Section 8 discusses the research, Section 9 presents the limitations, and Section 10 outlines the conclusion and future work.

## 2. BACKGROUND

### 2.1 Understanding Inclusion and Exclusion Across Lifecycle and Social Contexts

Inclusion refers to the process of involving people in society or in inclusive lifecycle contexts, ensuring that they have equal opportunities to participate in activities or functions. This is essential for supporting human dignity and well-being (Armstrong et al., 2011; Ozili, 2020; Shore et al., 2011; Yaldiz & Chakrabarti, 2024a). These functions should be goal-oriented, as achieving specific outcomes can demonstrate the value—or merit—of inclusion. Importantly, the inclusion of people in lifecycle processes is conceptually different from participation and is not simply the opposite of exclusion. As Quick and Feldman (2011) explain, "participation is oriented to increasing input for decisions," whereas inclusive practices focus on building community capacity to implement decisions and address related issues. In their view, participation is mainly about collecting input, while inclusion involves people actively contributing to problem-solving and achieving shared goals, leading to a deeper sense of satisfaction. They further define inclusion as "connections made not only among individuals' and groups' points of view but connections across issues, sectors, and engagement efforts." Similarly, Wasserman et al. (2008) describe a culture of inclusion as one where "people of all social identity groups have the opportunity to be present, to have their voices heard and appreciated, and to engage in core activities on behalf of the collective." These perspectives highlight the empowering dimension of inclusion—not just being invited, but having meaningful opportunities to engage, contribute, and shape collective action.
To further understand inclusion, it is also necessary to differentiate it from exclusion in order to evaluate who can be involved in a particular function or activity. As Hansen (2012) highlighted, "We must focus on examining the boundary between inclusion and exclusion in the specific communities by questioning how this limit is constructed. It is not possible to put meaning into the concept of inclusion without its otherness, exclusion." In Social Inclusion and Exclusion: A Review, Rawal (2008) stated that "The meaning of social exclusion depends on the nature of the society, or the dominant model of the society from which exclusion occurs and it varies in meanings according to national and ideological contexts (Silver, 1994:539)". Armstrong et al. (2011) emphasised that "Inclusion and exclusion are interrelated processes and their interplay constantly creates new inclusive/exclusive conditions and possibilities."
In different contexts such as business, education, and management, inclusion is often assessed based on the outcomes of activities. For instance, the approach to inclusive education aims to identify students "... who may be missing out, difficult to engage, or feeling in some way to be apart from what the school seeks to provide. It involves taking account of pupils' varied life experiences and needs" (Ofsted, 2000; Armstrong et al., 2011). Therefore, the outcome can be an inclusive system that provides what each student needs and reflects the diversity of students based on their achievements, attitudes, and well-being. In inclusive business contexts, the activities can involve expanding "…opportunities for the poor and disadvantaged in developing countries" (BIF, 2011; Wach, 2012), and creating shared value by addressing social challenges while achieving profitability (Prahalad, 2005; Michelini and Fiorentino, 2012). These efforts can lead to inclusive outcomes by involving people from low-resource settings in the value chain, aiming to generate both economic and social value for companies and individuals, depending on their resource conditions and needs.

### 2.2 Inclusive Design and Inclusive Manufacturing

Table 1 presents the comparison of inclusive design and inclusive manufacturing based on the recurring concepts found in definitions or approaches to inclusive design and the definition of inclusive manufacturing (see Section 1). The 80 definitions are listed in Supplementary File 1 (also used in Yaldiz and Chakrabarti, 2024b). Based on the recurring concepts of inclusive design, four concept groups are created. The groups highlight how inclusive design and inclusive manufacturing differ with respect to these concepts. These are: lifecycle, social involvement, people, and the use phase of products. Each group is named to reflect the common theme of its elements.
The first group, titled Lifecycle, covers the phases of the process such as design, conception, and creation (see Table 1 for the full list). Although design and design process are frequently used in the definitions of inclusive design (see Supplementary File 1), inclusive manufacturing emphasises the evaluation of all lifecycle phases, including conception, creation, distribution, and other stages. Design is considered part of the conception and creation phases; therefore, inclusive manufacturing evaluates the lifecycle process more comprehensively. The second group is titled Social Involvement. Inclusive design evaluates elements such as inclusive, exclusion, inclusion, excluded and included; inclusive manufacturing comprehensively evaluates community access by making the solution accessible to people. The third group is titled People. Inclusive design focuses on user(s) and designer(s) as the stakeholders; however, inclusive manufacturing extends the evaluation by considering all strata of society, including a variety of disadvantaged people. The fourth group is titled the Use Phase of Products. In inclusive design, the focus is on product attributes based on users' needs, capabilities to support accessibility and usability of products. This considers the users' inclusion in the use phase of the product.

However, inclusive manufacturing considers the active participation of people in all parts of the lifecycle process. This comparison provides the conceptual foundation for the assessment of empowerment through the inclusivity of stakeholders within lifecycle systems.

Inclusive approaches in design focus on the capabilities of users, especially excluded users, with the aim of improving product attributes to meet their needs. This is supported through the evaluation of the inclusion of a greater number of diverse users. In this approach, the inclusive aspect of the process is evaluated through the link between product attributes and their ability to meet user needs. Therefore, inclusive practices in inclusive design are more related to product use rather than expanding the process to include diverse groups of people beyond just the users. Based on this understanding, inclusive design aims to empower users by improving product attributes, thereby enhancing the usability and accessibility of products. However, as shown in Table 1, inclusive manufacturing aims to empower diverse groups of stakeholders from all strata of society through their inclusion in all phases of the lifecycle process. Hence, inclusive manufacturing provides a more comprehensive evaluation of inclusivity to empower people across different lifecycle phases. This also highlights the need for an assessment process to evaluate whether the definition of inclusive manufacturing is being operationalised. For inclusive design, this can be examined by following the guidelines or proposed approaches discussed in the following section.

Table 1: Comparison of the elements in definitions of inclusive design and inclusive manufacturing (Grouped terms include singular and plural forms (e.g., user/users), and the frequency of the elements is shown in brackets.)

| Concept Groups | Inclusive Design | Inclusive Manufacturing |
| --- | --- | --- |
| | Elements of the Groups | Elements of the Groups |
| Group 1 – Lifecycle | Design (151), (Design) Process (22) | Conception (1), Creation (1), Distribution (1), Transaction (1), Use and Retirement (1) |
| Group 2 – Social Involvement | Inclusive (78), Exclusion (24), Inclusion (5), Excluded (4), Included (0) | Accessible to people (1) |
| Group 3 – People | User(s) (53), Designer(s) (10), Stakeholder(s) (2) | All strata of society (1); Spatially (1), Temporally (1), Physically (1), Economically (1) and Culturally (1) Disadvantaged People (1) |
| Group 4 – Stakeholder Experience | Need(s) (38), Accessibility (4), Usability (3), Capability(ies) (25), Ability(ies) (12) | Actively participate in all parts of the lifecycle (1) |

### 2.2.1 Overview of Inclusive Design Methods

The outcome of an inclusive design process is to improve the attributes of an existing product to create a match between user capabilities and product features. In this way, previously excluded user populations can be included in the use phase of the product lifecycle. This objective has been supported by various inclusive design tools and frameworks, which aim to reduce exclusion by aligning product features with user capabilities.

The Exclusion Calculator was developed by the University of Cambridge to measure the capabilities of users (EC, 2018). These capabilities are grouped as vision, hearing, thinking, dominant and non-dominant hand, and mobility. Based on the indicators for each group, exclusion for a task can be calculated, and the product can be redesigned accordingly. The opposite of exclusion is defined as inclusive design by Kat Holmes (Patrick and Hollenbeck, 2021). Based on this understanding, inclusive design aims to increase the diversity of users by including excluded populations due to the mismatch between capabilities and product attributes. Waller et al. (2015) proposed fundamental concept questions to determine the needs, how to meet these needs, how well they are met, and what the next steps should be. In order to answer these questions, Waller et al. (2015) conceptualised the inclusive design process in the Design Wheel with five stages: manage, explore, create and evaluate. They suggested estimating exclusion during the evaluation stage based on user capabilities in vision, hearing, thinking, reach and dexterity, and mobility. Zallio et al. (2022) proposed an 'Inclusive Design Canvas' to address user personas, needs, and physical, sensory, and cognitive capabilities to support the design of inclusive buildings. Kirisci et al. (2011) proposed a 'Virtual User Model' to support the product development process for increasing the usability and accessibility of products by considering the aim of inclusive design: "...to successfully integrate human factors in the product development process with the intention of making products accessible for the largest possible group of users." BS 7000-6 proposed "a phased approach to Inclusive Design" for inclusive design management in industry (Clarkson and Coleman, 2015). The phases of the inclusive design approach are defined as follows: "Phase 1: explore potential, assess demands and commitment and finalise proposition; Phase 2: establish foundation and get into gear; Phase 3: Implement changes and determine impact; Phase 4: consolidate expertise and benefits and redefine approach". Each phase has various stages. For example, to progress to Phase 2, the business case for change must be finalised and communicated. Similarly, to progress to Phases 3 and 4, promoting inclusive design, nurturing a supportive culture, evaluating progress, and reviewing the contribution of the programme are required. The final stage is to review and refine the inclusive design approach.

This approach also highlights the need to expand the limits of inclusivity in processes by addressing management, culture, and business integration. This interpretation suggests that focusing solely on the design process may limit the potential of inclusive outcomes by resulting in product-level improvements. In practice, inclusive design may involve users in decision-making, but it often stays within the boundaries of participatory practices that mainly focus on design inputs rather than supporting systemic outcomes.

The contribution of the outcome of the inclusive design process may not be fully evaluated in terms of sustainability due to its limited assessment within the product lifecycle process. This limitation may not fully support the evaluation of its empowering impact, as the process mainly focuses on product use and includes only certain stakeholders, such as designers and users. Therefore, increasing the diversity of users through the inclusion of excluded people may not be sufficient to create continuity in development under dynamic conditions. Beyond practical limitations, inclusion also involves how power, justice, and exclusion are structured in group and system dynamics. In *Inclusion and Diversity in Work Groups: A Review and Model for Future Research*, Shore et al. (2011) stated that "Diversity climate is related to the inclusion or exclusion of people from diverse backgrounds (Mor Barak et al., 1998), and ... to the justice-related events pertinent to the balance of power and relations across social groups (Kossek and Zonia, 1993)." Thus, inclusion should be evaluated together with exclusion in exchange networks (e.g. a product lifecycle process, sharing resources), which can be shaped through the design of inclusive lifecycle processes to support the empowerment of stakeholders through their inclusion.

Table 2 summarises the inclusive design approaches discussed above to highlight similarities in their scope and focus on the inclusion of people. Inclusivity is often evaluated through improvements in products or services that aim to make them usable by diverse users based on their capabilities. This understanding is valuable for supporting usability and accessibility, particularly in the utilisation phase of the lifecycle. However, it also points to the need for evaluating inclusivity from a broader perspective, as the product lifecycle extends beyond just the design and use phases. A broader view is essential to identify whether inclusion supports stakeholder empowerment throughout the entire lifecycle.

Table 2: Overview of the scope and focus of selected inclusive design approaches

| Reference | Tool/Model | Scope | Inclusion Focus | Limitation |
|---|---|---|---|---|
| EC (2018) | Exclusion Calculator | Use of products | Capabilities of users | Usability of products |
| Waller et al. (2015) | Design Wheel | Manage, explore, create, and evaluate | Capabilities of users | Usability of products |
| BS 7000-6 (described by Clarkson and Coleman, 2015) | A phased approach to Inclusive Design | Design management, product and services, business processes | Users and customers | Accessibility and usability of products for diverse users |
| Zallio and Clarkson (2022) | Inclusive Design Canvas | Architectural design | Capabilities, needs, and requirements of personas | Focused on educational use; limited to early design stages and architectural domain |
| Kirisci et al. (2011) | Virtual User Model | Product development | Capabilities of users | Accessibility and usability of products for the largest groups users |

## 2.3 Empowerment

Empowerment is a context-dependent process involving different activities or functions. In manufacturing, empowering functions can include "training to create adaptability, foster innovation, improve collaboration and enhance speed" (Stough et al., 2000; Dubey and Gunasekaran, 2015). Gunasekaran et al. (2019) identified empowering functions for employees in manufacturing as realigning power relations through interdisciplinary collaboration and creating shared value by encouraging knowledge diversity. In marketing and management contexts, the focus of empowering functions extends beyond employees to include customers. These functions aim to encourage customer participation in evaluation and decision-making processes (Füller et al., 2009; Fuchs and Schreier, 2011; Ogawa and Piller, 2006) and enhance the "… ability [of customers] to access, understand and share information" (Pires et al., 2006). Sianipar et al. (2013) evaluated the empowerment process through the logic of "give-maintain-make." This model involves giving a product to people who need to be empowered, teaching them how to use it, maintain (repair) it, and make (develop or improve) it themselves. Similarly, Hernandez et al. (2020) highlighted the importance of the "right to repair" to support sustainable consumption by extending product lifecycles. This right is considered an empowering approach that aims to "... give back to the users the right to decide what to do with their products when they fail and before they have to dispose of them." According to Lyons et al. (2001), "Fully empowered people, projects and/or communities are [...] able to contribute towards the sustainability of development projects which, in turn, contribute towards the broader notion of sustainable development."

In order to evaluate the outcome of empowerment processes, dimensions can be identified based on the empowerment contexts. Kleba and Cruz (2021) proposed dimensions of empowerment in socio-technical interventions, such as "valuing cultural difference, growing environmental awareness, fostering socio-technical inclusion," etc. Dubey and Gunasekaran (2015) classified empowerment as a part of agile manufacturing, considering dimensions such as "mutual trust, cooperation, everyone's involvement," etc. However, there is still ambiguity in how to measure empowerment dimensions, both in terms of evaluating the process and identifying the outcomes. Moreover, the dimensions are not clear enough to determine whether they are intended for evaluating the process or the outcome. For instance, the outcome of an empowering process can be an increase in a person's power, depending on the context. Suppose the dimension is valuing cultural differences; this involves a different process with various influencing factors, which makes it difficult to separate the process from the outcome.

Yaldiz and Chakrabarti (2024a) interpreted empowerment evaluation by measuring the empowering impact of each function in a product lifecycle, considering the inclusion of people in those functions. These functions may involve providing information, decision-making, training, learning, etc. For example, the inclusion of marginalised people in an inclusive lifecycle process of a product can be an empowering process, with the outcome of creating income opportunities. Shore et al. (2011) also proposed the outcomes of inclusion (e.g., career opportunities for diverse individuals, well-being, etc.), which show the connection between inclusion and empowerment. As Miller and Campbell (2006) emphasised, inclusivity and empowerment are interconnected processes: "Empowering processes are inclusive processes, and inclusion is a precondition for a group or individual becoming empowered via an empowerment evaluation." While this highlights the close relationship between inclusion and empowerment, further clarity is needed on how people can be empowered through involvement in different functions of inclusive lifecycle processes, and how this can be assessed. Therefore, it is necessary to determine metrics for empowerment in inclusive processes to assess the improvement.

Table 3 presents the frequency of elements in definitions of empowerment based on 86 definitions from different domains, listed in Supplementary File 2 (also used in Yaldiz and Chakrabarti, 2024b). The main emphasis is on the terms *community, power, control, process,* and *development,* which are commonly used to define or explain empowerment. These elements are grouped into different concepts based on the similarities in themes. Community, people and individual(s) are grouped under Stakeholders as these elements present those who are involved. Process and development are in a different group titled Transformation, reflecting the dynamic and outcome-oriented aspects of empowerment as both an empowering process and its result in development. The last group, titled *Power Relations*, includes the terms power and control, since these express the relational aspects of empowerment that emerge through interactions among stakeholders.

This conceptual grouping helps highlight the link between lifecycle processes and empowerment, showing how the inclusion of people in development processes can contribute to empowerment. In inclusive lifecycle processes, functions are tasks that drive progress within each lifecycle phase and can support the empowerment of stakeholders. Including diverse stakeholders in these functions is an essential part of fostering both inclusion and empowerment. Therefore, functions can be understood as contributing to empowerment within inclusive lifecycle processes. Each function or activity—within the phases of the lifecycle—can be seen as part of an empowering process, with measurable impacts on stakeholders.

Table 3: Frequency of elements in definitions of empowerment
(See Supplementary File 2 for the full list of definitions. The frequency of the elements is shown in brackets.)

| Concept Groups | Elements of Groups |
|---|---|
| Group 1 – Stakeholders | Community (24), People (20), Individual (14) |
| Group 2 – Transformation | Process (24), Development (23) |
| Group 3 – Power Relations | Power (29), Control (24) |

Additionally, an inclusive lifecycle process can also be defined as a lifecycle process with an exchange network based on power relations among involved stakeholders (or stakeholder groups) and the distribution of valued resources among these groups. An exchange network consists of a set of actors, the distribution of resources among them, opportunities, and an exchange structure (Cook et al., 1983). Inclusion should therefore contribute to the empowerment of stakeholders. As Dahl (1957) defined: 'A has power over B to the extent that he can get B to do something that B would not do otherwise.' Identifying such power relations is necessary to assess the impact of inclusion on empowerment in inclusive lifecycle design. In the network associated with an inclusive lifecycle process, stakeholder groups exchange opportunities and network connections, which can also be interpreted as collaboration among them. The outcome of the empowering process should be to increase the power of stakeholders (Fawcett et al., 1984), considering all phases and including a diverse group of stakeholders.

## 3. METHODOLOGY

As discussed in the literature (Section 2), a conflict exists among the theory, application, and measurement of inclusivity and empowerment, as well as the influence of inclusivity on empowerment. Empowerment is an inclusive process with empowering outcomes for people to reduce societal inequality. However, the dimensions of empowerment are ambiguous in terms of the contexts in which these dimensions should be used to measure improvement. In addition, the connection between the dimensions and empowering processes is debatable, which can cause biased measurement results.

The inclusion of people in lifecycle processes is currently assessed through the increased diversity of users, aiming to make products more accessible and useful. However, this user-focused approach limits the empowering impact of lifecycle processes and can cause imbalanced power relations, creating challenges in empowering people during the transition to sustainability.

Based on the goals of inclusive manufacturing, where inclusion has the specific aim of empowering people, it is necessary to identify operational definitions of inclusivity and empowerment that are causally connected. This means that the specific notion of inclusivity captured in the proposed definition should be the most appropriate for leading to empowerment, as defined by its own operational meaning.

In order to respond to the research question (see Section 1), the methodology follows the process presented in Figure 2, which visualises the overall approach.

The first three steps (Figure 2) identify key metrics for assessing empowerment and inclusivity in the context of the product lifecycle process. The definitions of inclusivity and empowerment in the context of the product lifecycle are the input to the research methodology.

The 80 definitions of inclusivity are identified from a comprehensive review of inclusive design literature (see Supplementary File 1). These definitions are used, not for prioritising one perspective over another, but for collectively building a comprehensive understanding of inclusivity in the context of engineering design, which is only one part of the product lifecycle. The statistics show the complete lack of several perspectives of inclusivity, including those of the product lifecycle and the variety of people.

As provided in Supplementary File 2, 86 definitions of empowerment are collected from different domains (management, psychology, design, marketing, sustainable development, etc.). The purpose of reviewing multiple domains is to develop a comprehensive understanding and definition that can be adapted to the product lifecycle context.

Word counts (see Tables 1 and 3) are used only as indicators of salience, showing which elements occur more often across definitions; they are not used to statistically generate new definitions. The recurring factors with higher frequency are therefore treated only as more influential for empowerment outcomes and grouped into dimensions to guide the analysis, while the original definitions remained the conceptual input.

To understand the pathway to greater empowerment through greater inclusivity, four different dimension groups are created for both inclusivity and empowerment.

Each dimension group of empowerment consists of metrics derived from the elements in concept groups (see Table 3). The concept groups are identified based on the similarities in the elements. The first group, Stakeholders, highlights the consideration of people involved in the process. This is adopted by measuring the empowering impact of a function, which includes the number of included stakeholders as a multiplier (see Eq. 1). The second group, Transformation, signifies the empowering inputs and outcomes in development, measured through the empowering impact. The power relations, the third group, is adopted by measuring the dependency level and the number of means (discussed in detail in Section 4).

Following the same logic, each metric of inclusivity is derived from the concept groups, which are presented in Table 1. The first group, Lifecycle, is adopted by counting the number of lifecycle phases in which a stakeholder is involved. The second and fourth groups, Social Involvement and Stakeholder Experience, are measured by counting the number of interactions of a stakeholder with other stakeholders separately for each lifecycle phase. The third group, People, is adopted by measuring the level of diversity of stakeholders and their level of hierarchy.

A similar concept identification approach was used by Sarkar and Chakrabarti (2011), who analysed multiple definitions of creativity and operationalised common elements into measurable dimensions. Surma-aho and Holtta-Otto (2022) applied a concept identification approach by reviewing multiple definitions of empathy from different domains such as design, social psychology and neuroscience. Legood et al. (2022) likewise applied a concept identification approach by reviewing definitions of cognition-based and affect-based trust to operationalise these concepts for measurement. Following this precedent, the present study applies the same principle to inclusivity and empowerment.

The values for empowerment and inclusivity are calculated as the products of the metrics within each dimension group. Therefore, four dimension groups are identified for inclusivity (I1, I2, I3, I4) and for empowerment (E1, E2, E3, E4). Each dimension group represents different formulas for inclusivity and empowerment to address the relationship between empowerment and inclusivity. The rationale for the creation of dimension groups is explained in detail in Section 5.

The dimension groups are structured in sets of four to allow multiple operationalisations of inclusivity and empowerment. By pairing these groups in different combinations, the framework explores how alternative formulas can lead to different results. Therefore, the approach does not claim a single fixed formula but provides a structured way to test alternative operationalisations transparently. The accuracy of the operational definitions lies in their consistency with the literature and their coherence across case studies. As also noted in Section 9, other formulas may be possible, and the present study should be understood as one systematic approach rather than a final rule.

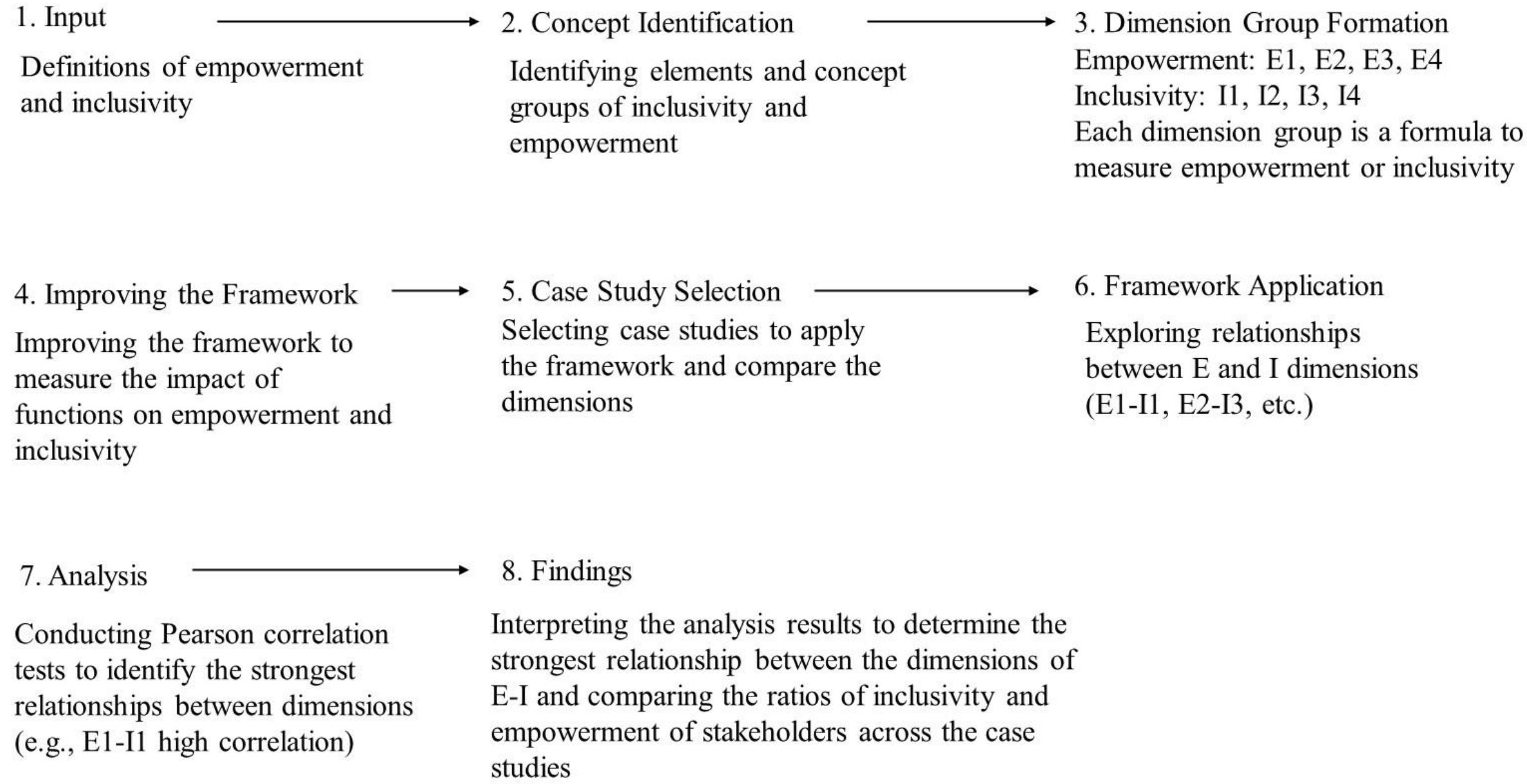

Figure 2: Conceptual map of the methodology

The framework is improved to assess empowerment and inclusivity by identifying their dimensions and applying them to case studies. The core of the framework involves transforming information from lifecycle processes (e.g., data related to product lifecycle processes in case studies) into functions, which can be grouped according to product lifecycle phases. Functions refer to the tasks or activities performed by stakeholders, where their inclusivity is evaluated. Thus, the framework assesses the inclusivity and empowerment of stakeholders based on their performance in functions within the phases of a product lifecycle process.

The framework is applied to 10 case studies involving 92 different stakeholders in total. Through this application, the metric values for empowerment and inclusivity are identified, and their dimensions are calculated. For each stakeholder, four separate values (based on the dimensions) for empowerment and inclusivity are determined. The next step involves pairing these dimensions to prepare the data for Pearson Correlation Coefficient tests. For each case study, eight dimension pairs are tested. The results involve 10 test outcomes for each pair of empowerment and inclusivity dimensions, based on data from 92 stakeholders across 10 case studies.

The strongest pair of empowerment and inclusivity dimensions is identified based on the highest observed correlation. Comparing the correlations across eight-dimension pairs supports the selection of the most relevant dimensions. This outcome enables a data-driven understanding of how stakeholder inclusivity contributes to empowerment within product lifecycle processes. In addition, comparing the ratios of empowerment and inclusivity across different stakeholders (e.g. users and other stakeholders) highlights areas for improvement, which provides a foundation for supporting more equal stakeholder development in the next phase of the study.

## 4. REVISED FRAMEWORK OF EMPOWERMENT AND INCLUSIVITY ASSESSMENT

The framework proposed in our earlier work (Yaldiz and Chakrabarti, 2024a) addressed the reasons for imbalanced development by evaluating power relations. This section presents its revised version, focusing on the standardisation of function identification. Detailed application examples are provided in Supplementary File 3.

The framework proposed for assessing inclusivity and empowerment aims to provide a method for addressing imbalanced situations in product lifecycle processes. These imbalances can be the different ratios of inclusivity and empowerment of stakeholders throughout the lifecycle process. This can have various causes; thus, the approach begins by determining the boundaries of the product lifecycle for the scope of the assessment. The product lifecycle phases (planning, development, production, utilisation, and recirculation) are integrated into the framework to help define the attributes of each phase, enabling a structured assessment of empowerment through stakeholder inclusion. The attribute of a process is linked to the tasks involved, meaning the activities that influence the product. For example, a task could be preparing the product portfolio in the planning phase or creating a conceptual design during the development phase of the lifecycle. This task identification is essential to understand which stakeholder groups can be involved in each lifecycle phase and in which tasks. Once the relevant phases are selected for evaluation (e.g., the planning phase), the associated functions within that phase can be identified, and their impact on empowerment can be assessed. This function identification—based on lifecycle phases and stakeholder roles—is the first step in the three-step framework. It provides a structured way to analyse how tasks and responsibilities are distributed to support stakeholder inclusivity.

In this study, functions are identified from case study content to assess empowerment and inclusivity. This approach allows the consistent application of empowerment and inclusivity assessment across different cases and stakeholder groups. The function identification creates an interaction-based structure by evaluating human interactions during the product lifecycle and identifying stakeholder roles and actions across lifecycle phases. This approach is related to the functional architecture ideas described in ISO/IEC/IEEE 15288:2023. However, because functions can be diverse and numerous, activity theory is used within the framework to guide the interpretation of key tasks and human activities in each phase. Since the product lifecycle is a long-term process with many varied functions, standardising function identification is necessary. In this way, function identification supports the assessment of inclusivity (through stakeholder inclusion in the functions) and empowerment (through the influence that inclusion enables) in the product lifecycle. Beyond their role in structuring the framework, functions also serve as opportunities for stakeholders to take part in activities that contribute to their empowerment. They are not always exercised individually in the lifecycle process, but act as sources of interaction that enable the inclusion of a larger number of stakeholders. Each function carries an empowering impact within the lifecycle process. Since the lifecycle involves developing a product and later recirculating it into another form (for example, through recycling), it provides empowering potential for stakeholders at different stages. The similarities between tasks in the lifecycle process and empowerment are explained in Section 2.3.

As cited in Moser (2013), Engeström (1987) emphasised that "activity theory demands analysing at least two interacting activity systems." The lifecycle processes of a product, service, or system are activity systems with various functions carried out by stakeholders to ensure progress. The interaction between two stakeholder groups creates function groups. For example, Stakeholder A may request a function to be completed by Stakeholder B. This request might require B to carry it out individually or to involve Stakeholder C, in which case B would request a function from C (e.g., providing information). In the first case, the activity is counted as one function; in the second, it is counted as two. Furthermore, Stakeholder C (which could be a group, such as a design team) may have internal interactions to respond to the request from B, which would then increase the number of functions (as per Activity Theory). The standardised model of function identification is presented in Supplementary File 3. The function models (based on the interactions between stakeholders) presented in the file are also used in a related study (Yaldiz and Chakrabarti, 2024b). In this study, functions are accepted as information sharing, task requesting, and completing tasks for the continuity of the lifecycle process. Collaboration (e.g., information sharing, contributing to the process) can be the empowering aspect of inclusivity (see discussion in Section 2.3). Therefore, functions are identified based on information sharing, requesting, and completing tasks. This limits the identification for developing an understanding of the relationship between inclusivity and empowerment.

Functions that are completed by one stakeholder group are considered individual functions in the framework due to their self-empowering impact. In a situation where the task is completed by one stakeholder group, there can be more than two individual functions. For example, suppose that the stakeholder group is a design team; the functions can be detailed to complete the development process. These individual functions show how stakeholder groups can empower themselves by progressing through the lifecycle process, which can later lead to requesting tasks or sharing information to involve other stakeholders in the process. This contributes to the understanding of the influence of inclusivity on empowerment within lifecycle phases. In such cases, the tasks of the related phase (e.g., development phase) and the information provided in the case study are the limitations in function analysis.
Stakeholders are accepted as groups to generalise the empowering effect of the functions and are categorised as providers or receivers. Requesting a task from a stakeholder group is considered providing an opportunity. Therefore, the function of requesting a task is completed by the stakeholder in the role of the provider. In addition, a provider may give information to another stakeholder. The role of the receiver is to gain the opportunity to participate in a task or to receive information needed to complete it. The roles of providers and receivers demonstrate how stakeholder interactions can create opportunities for stakeholder inclusivity, which in turn supports empowerment as stakeholders gain influence through involvement in tasks.
The second step of the framework measures the empowering impact of each function. Cattaneo and Chapman (2010) emphasised that an individual takes action toward the goal of empowerment, which reflects on the impact of their actions. Therefore, evaluating the empowering impact of each function on involved stakeholder groups is important. Additionally, its relation to power also needs to be assessed, as it is a key concept of empowerment, discussed in Section 2.3.
As shown in Figure 3, stakeholder A requests a task (Function 1) from stakeholder B. Function 1 impacts B. The response to Function 1 is Function 2, which is initiated by B and impacts A. These function exchanges reflect how empowerment can be influenced through interactions.

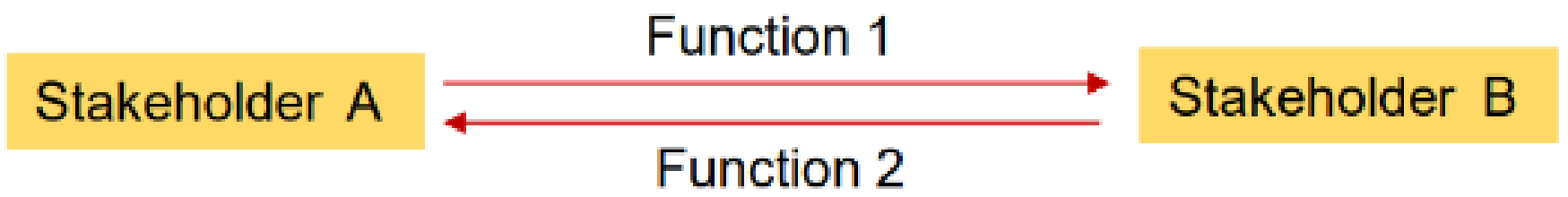

Figure 3: Function-Based Stakeholder Interaction
(reproduced from Yaldiz and Chakrabarti (2024a), under CC BY-NC-ND 4.0 licence)

The empowering impact of a function (Fj), written as EI(Fj), is influenced by two factors: the number of stakeholder groups involved (#stakeholders) and the number of tasks or functions completed by each person (#functions, #F). The method for calculating this empowering impact was introduced by Yaldiz and Chakrabarti (2024a), and is presented with the following equation:

$$EI(Fj) = \#stakeholders \times \#functions \qquad \text{(Eq. 1)}$$

This formulation (Eq. 1) is grounded in the concept that the impact of empowerment can be assessed through both the involvement of stakeholders and the number of functions in which this involvement occurs. Similarly, Pardo del Val and Lloyd (2003) highlight that empowerment measurement should consider both the number of participants and the areas or levels where participation takes place. This is also in line with Jupp and Ali (2010), who emphasise the influence of the number of people and their involvement in various roles on their empowerment. In this study, this idea is applied in a simplified way, where the empowering impact of a function is evaluated as the product of the number of stakeholders and the number of functions (Eq. 1).
For example, in Figure 3, suppose Stakeholder A is the manager and requests a task from Stakeholder B, the customer service team. The task could be gathering feedback from the users of their newly launched product. This request from Stakeholder A can be considered *Function 1*. The empowering impact of Function 1 is calculated as 2 × 1 = 2, since two stakeholders are involved in one function. The response of Stakeholder B, which completes the task by providing feedback from users, can be considered *Function 2*. Again, the empowering impact is 2 × 1 = 2, because two stakeholders (A and B) are involved in one function. However, it should be noted that empowerment is a multi-dimensional concept which can be influenced by different factors. Therefore, the detailed analysis of the empowerment of stakeholders can be assessed by a product of different concepts in a specific context, discussed in Section 5.2.

Function identification helps structure stakeholder interactions to address the influence of inclusivity on empowerment. It is also important to understand the relationship between empowerment and power, as this influences stakeholder roles in the process. The ratio of empowerment alone cannot sufficiently explain the amount of power, since power is one of the key concepts of empowerment (see Section 2.3). However, measuring power is important for supporting equal stakeholder development.

The third step of the framework is to measure the amount of power for each stakeholder. As discussed in Section 2.3, the power of a person can be evaluated based on their dependencies on others in a societal context, considering specific conditions (Dahl, 1957; Emerson, 1962). Dahl (1957) defined these conditions as "sources (domain, base), means (instruments), amount or extent, and range or scope," and highlighted their importance for the accuracy of power relations. Emerson (1962) explained that power relations are connected to the dependencies between A and B, along with the factors of availability (supply) and motivational interest (demand). Evaluating the resources is significant for understanding people's dependencies on each other and for addressing power relations as balanced or imbalanced (Emerson, 1962). Therefore, the approach is to adopt the concepts of (re)sources, means, and extent to product lifecycle processes based on the evaluation of the dependencies in interactions among the stakeholders. In short, in the lifecycle context, the influencing factors of the power of stakeholders are extension and dependency level.

As proposed by Yaldiz and Chakrabarti (2024a), the extension is adopted in lifecycle processes as the potential of a stakeholder group to influence other stakeholder groups across different phases of the lifecycle. As shown in Eq. 2, the extension is associated with the number of means used to complete a function and the frequency of the stakeholder group. The means represent the number of tools used by stakeholders to perform a task. The frequency refers to the number of functions in which the stakeholder group is involved within a specific phase of the lifecycle process. In this way, by using a greater number of means (tools) and increasing the frequency, a stakeholder group can access people to influence them for their inclusion in the process.

$$\text{Extension} = \#\text{means} \times \text{frequency} \qquad \text{(Eq. 2)}$$

In addition to extension, the dependency level of stakeholders also influences power relations. In the lifecycle context, the dependency level is evaluated by the number of needs (representing demand) and resources (representing supply). The needs and resources of stakeholders can vary depending on the functions. The matrix of needs and satisfiers proposed by Max-Neef (1992) is adopted in the framework to standardise the types of needs and resources to be used. Resources represent the differentiation of stakeholders from one another by enabling them to share their assets through fulfilling each other's needs. For instance, in Figure 3, the benefit of Stakeholder A for Stakeholder B through involvement in function pairs 1 and 2 is based on their mutual needs and available resources. Stakeholder A has a specific need that can be addressed by Stakeholder B, who possesses sufficient resources to involve in the function. A detailed example is provided in Supplementary File 3 to illustrate the identification of needs and resources.

The measurement of needs and resources is important for evaluating the fulfilment of stakeholders' needs. This supports stakeholders in becoming more independent from the system, enabling them to contribute to and improve other systems or lifecycle processes. As proposed in Yaldiz and Chakrabarti (2024a), the dependency level of stakeholders (Eq. 3) is calculated as the number of resources per need.

Dependency level = #Resources / #Needs (Eq. 3)

The connectivity among stakeholders, shaped by their resources and needs, forms the power-exchange network within the lifecycle process. The amount of power held by Stakeholder A is linked to the extension and dependency level of Stakeholder B. This is because "the power of actor A over actor B is the amount of resistance on the part of B which can potentially be overcome by A" (Emerson, 1964). Therefore, as proposed in Yaldiz and Chakrabarti (2024a), the equation for determining the amount of power held by Stakeholder A is given as follows.

Power (A) = Extension (A) × Dependency level (B) (Eq. 4)

In summary, the framework is structured around four concepts based on the function analysis. These are empowering impact, power with extension, and dependency level. Together, these form the structure of the framework. The aim is to examine how stakeholder inclusivity in lifecycle processes translates into empowerment by evaluating the potential of each function.

Furthermore, the formulas for these concepts (empowering impact, extension, dependency level, and power) are adapted from the literature (Pardo del Val and Lloyd, 2003; Jupp and Ali, 2010; Dahl, 1957; Emerson, 1962 and 1964; Yaldiz and Chakrabarti, 2024a) to operationalise them within lifecycle processes.

## 5. METRICS FOR INCLUSIVITY AND EMPOWERMENT

Building on the framework established in the previous section, this section introduces the metrics developed to evaluate stakeholder inclusivity and empowerment across lifecycle phases. These metrics are based on key concepts of inclusivity and empowerment. The framework is used to calculate the values of these metrics and to explore the relationship between inclusivity and empowerment in different lifecycle contexts.

### 5.1 Inclusivity Dimensions

Based on the discussion in Sections 1 and 2, it is argued that the level of inclusivity results from inclusion and exclusion processes designed to ensure inclusive outcomes. Figure 4 illustrates the conceptualisation of inclusion, exclusion, inclusivity, and inclusive outcomes. In an inclusive lifecycle process, the inclusion of stakeholder groups can occur based on the purpose of mobilising resources and fulfilling needs. However, this approach has limitations in understanding the extent to which stakeholders can be included in a particular function. Thus, inclusivity metrics must be determined to ensure the merit of inclusive outcomes, which aim to provide opportunities and contribute to sustainability, well-being, and dignity.

The exclusion of stakeholders, based on a mismatch among purpose, resources, needs, and inclusivity metrics, is necessary to prevent undesired situations—such as the misinterpretation of participatory practices as inclusive practices. Inclusive outcomes can be created through the inclusion of people based on inclusivity metrics. This empowers individuals through the design of an inclusive product lifecycle that supports deeper assessments of sustainability, well-being, and dignity.

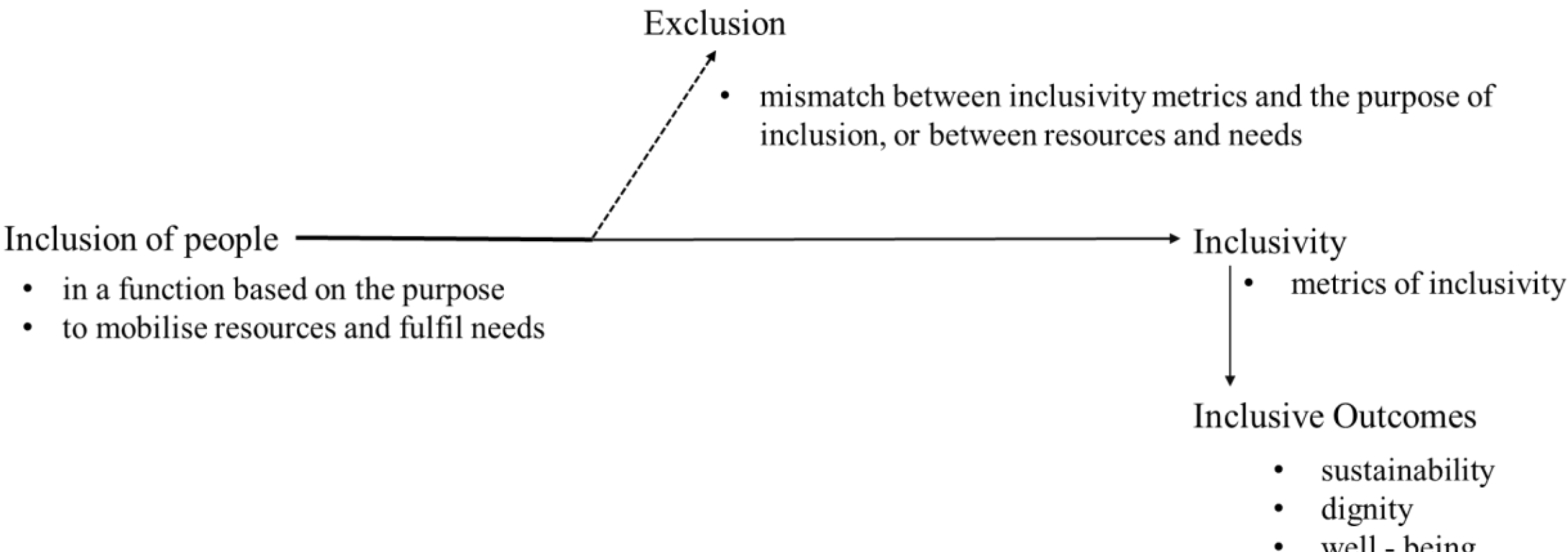

Figure 4: Conceptualisation of Inclusion, Exclusion, Inclusivity and Inclusive Outcomes
(The differences in line thickness illustrate convergence towards the outcome. The angled dotted line represents the uncertainty regarding the onset of exclusion, which depends on the process and the stakeholders included. The dots indicate that an excluded stakeholder can potentially be re-included in subsequent phases.)

As stated in the inclusive manufacturing definition (Section 1), inclusivity is understood as the involvement of spatially, temporally, physically, economically, and culturally disadvantaged people from all strata of society in all parts of the lifecycle, which aims to empower stakeholders. Based on this understanding and the conceptual comparison of inclusive design and manufacturing (see Table 1), the dimensions that influence the level of inclusivity in lifecycle processes are identified: the degree of diversity of stakeholders, the number of lifecycle phases in which they are involved, the degree of hierarchy of stakeholders, and the number of interactions between the stakeholders of one group and those of others. The approaches to evaluating inclusivity metrics are explained below.

### 5.1.1 The Degree of Diversity of Stakeholders

As discussed in Section 2.2, inclusive design approaches the evaluation of diversity mainly through the capabilities of user groups, which supports the value of the product. However, with the aim of generating inclusive lifecycle processes, it is necessary to evaluate diversity by building further on what already exists in product lifecycle processes. Therefore, in this study, diversity is evaluated through the shifts in contexts that can occur based on information sharing among stakeholders. As highlighted by Vandekerckhove et al. (2023), diversity can also be evaluated in terms of differences in *knowledge, perspectives, and inferential approaches* among stakeholders, rather than only demographic or capability-based categories. This supports the approach that stakeholder diversity can be understood through the variety of contexts and the flow of information between groups. Maj (2015) discusses diversity across multiple layers, including *personality, internal, external, and organisational dimensions,* showing that diversity is not a single fixed notion but must be interpreted according to context. This supports the rationale for adapting the evaluation of diversity to the varied contexts of lifecycle processes in this study.
Evaluating the diversity of stakeholders requires a comprehensive understanding of various contexts and dimensions. Especially in a complex system like the product lifecycle process, the diversity of stakeholders plays a critical role in the equal distribution of opportunities for the enhancement of empowerment. There are various approaches to measuring diversity, focusing on *cultural diversity* (Nijkamp and Poot, 2015), or on *variety (number of categories), balance (evenness of distribution), and disparity (degree of difference between categories)* (Van Dam, 2019). However, the evaluation should also relate to the purpose of consideration or the need for stakeholder diversity. Therefore, the diversity of people can be accepted as a dimension, as highlighted in the inclusive manufacturing definition:

"...spatially, temporally, physically, economically, and culturally disadvantaged people...". These dimensions highlight the need for context evaluation to determine the diversity of stakeholders in the process (Pringle and Ryan, 2015). One way to identify the level of diversity within a stakeholder group—accepted in this study—is to count the contexts (time, location, culture, economic) or conditions (temporal, physical) in which stakeholders participate in group functions to shift disadvantaged situations towards advantage by sharing resources and supporting their dependencies through empowerment.

Zimmermann et al. (2007) categorised the contexts as individuality, activity, location, time, and relations. In the scope of the lifecycle process, when evaluating the diversity of stakeholders based on their involvement in functions, the individuality and activity contexts are grouped together due to the similarity in their purposes and the nature of the approach. The individuality context is used to evaluate each stakeholder as a separate system with its own needs, resources, and role, with the purpose of enabling their involvement in group functions within the product lifecycle processes. The activity context refers to the relationship between stakeholders and the specific functions they perform within product lifecycle processes. Given the similarities between the explanations of individuality and activity contexts, an overlap can occur when evaluating the diversity of stakeholders in these contexts. Therefore, individuality and activity contexts are combined to evaluate the diversity of stakeholders. The location context refers to the spatial factors that influence stakeholder involvement in product lifecycle processes, considering where stakeholders are situated geographically, organisationally, or virtually. The time context refers to the temporal factors that influence stakeholder involvement in product lifecycle processes, considering when stakeholders are engaged or involved in functions. The relations context refers to the interactions between stakeholders, focusing on resource sharing, the fulfilment of needs, and the dynamics of power relations within product lifecycle processes. The detailed features of the categories of contexts are accepted as provided in Yaldiz and Chakrabarti (2023). For example, individuality and activity contexts can be identified through characteristics such as income, disability, or cognition. The location context may refer to a shift between a city and a district. The time context can be related to the duration of an activity, such as years, hours, or months. The relations context can be social or functional, depending on the purpose of the activity.

In short, the diversity of stakeholders is measured by identifying the contexts in which a stakeholder group can support the shift of another stakeholder group from a disadvantaged to an advantaged situation. This interpretation is consistent with Vandekerckhove et al. (2023), who show that diversity can be reflected through differences in knowledge and information sharing, meaning that access to knowledge can reduce disadvantage and create advantage. Similarly, Maj (2015) emphasises that stakeholders may shift from disadvantaged to advantaged positions as contexts and relationships change. For example, such a shift can occur through information sharing between stakeholders. Suppose the stakeholders are a design team and a customer user. To continue the design process, the design team needs feedback or information about customer needs from the customer user. Thus, the design team can request information from the customer user, and in return, the information can be provided for the common goal, which is the continuation of the lifecycle process. From this perspective, the design team is in a disadvantaged situation due to the lack of information. By requesting information and receiving it in return, the team shifts its situation to an advantaged one by valuing diversity in the relationship, individuality, and activity contexts, mutually with customer users. This is because receiving information fulfils the need for continuity of the lifecycle process. It can be noted that it is not just requesting information but obtaining information that results in the shift from disadvantaged to advantaged. The degree of diversity is 2 for the design team and customer user, based on this example. Therefore, the inclusivity of disadvantaged stakeholders involves shifting their situation to an advantaged one through the empowering potential of diverse stakeholder groups. Additionally, the meanings of disadvantaged and advantaged are accepted as defined in the Cambridge Dictionary (2025), where disadvantaged is described as "(of people) not having the benefits, such as enough money and a healthy social situation, that others have, and therefore having less opportunity to be successful, "and advantaged is defined as "people who have a better standard of living, more opportunities to succeed, etc. than others."

### 5.1.2 The Number of Lifecycle Phases

The other metric of inclusivity from the inclusive manufacturing definition (see Section 1) is the number of lifecycle phases in which a stakeholder group is involved, as expressed in the phrases: "...all parts of the lifecycle..." and "...to actively participate in the conception, creation, distribution, transaction, use, and retirement of products and systems." The lifecycle phases are accepted as proposed by Urakami et al. (2018), where the product lifecycle has five phases: planning, development, production, utilisation, and recirculation. The approach to measure this metric is to count the number of lifecycle phases in which a particular stakeholder group is involved. For example, if a design team is involved in the planning and development phases, the metric (#LC, the number of lifecycle phases) is two.

### 5.1.3 The Hierarchy of Stakeholders

The hierarchy of stakeholder groups is interpreted as the initiators of the function or the decision-makers in the function groups. Hierarchy is needed as a dimension for evaluating the "...all strata of society..." and "… to actively participate in…" as expressed in the definition. Therefore, hierarchy can also be considered as an aspect of the diversity of people based on their responsibilities in a lifecycle process. The approach to measuring the hierarchy is done by numbering the roles of stakeholders based on their influence on the continuity of the lifecycle process. Responsibilities of stakeholders can be various in the system; therefore, based on their potential impacts, the hierarchy value is numbered from 1 to 5. For example, suppose the stakeholders are the manager and the supplier. The attempt to perform a function can be done by the manager, who is also the decision-maker. Thus, the manager's hierarchy is 5. The supplier’s hierarchy is 1, which means having less influence in attempting a new function or being a decision-maker. In different cases, a supplier’s hierarchy can be more than 1 or even 5. However, the limitation is based on the decision-making role and the attempt to perform functions for the continuation of the lifecycle process.

### 5.1.4 The Number of Interactions

The other metric to measure the inclusivity of stakeholders is the number of unique stakeholders a stakeholder interacts with. The inclusive manufacturing definition (see Section 1) emphasises that "...a manufactured product is made accessible to people...". A manufactured product follows a lifecycle process in which people can include themselves. This inclusion can be supported by increasing the number of interactions among stakeholder groups. Therefore, the number of interactions is also an essential dimension for promoting inclusivity in lifecycle processes. Since progress in lifecycle processes depends on interactions among stakeholders (e.g., information sharing, requests for tasks, or exchange of resources), the number of unique interactions is taken as a metric of inclusivity. Moreover, this approach extends the limits of the process by fostering greater interaction among people. For example, suppose the stakeholders are the design team, customer user, and coordinator. If interactions are created indirectly through the coordinator between the design team and customer user, this limits the enhancement of interactions because the design team and customer user only interact with the coordinator to be involved in the process. In this case, the number of interactions is two for the coordinator, and one for the design team and customer user. However, if all three groups can interact directly with one another, it can support the enhancement of inclusivity by increasing the number of interactions. In this case, the design team and customer user can also have direct interactions. This can increase their number of interactions to two. Additionally, the quality of interactions is not evaluated in this phase of the research; the approach of stakeholders during interactions may influence inclusivity. As noted in Section 9, the negative impacts of such factors are not reflected in the current framework; this could be explored in future.
These metrics are grouped into four sets, representing alternative but comprehensive operational definitions of inclusivity (see below), as derived from the definition of inclusive manufacturing. Each set includes a combination of dimensions—diversity, hierarchy, number of interactions, and number of lifecycle phases—reflecting different aspects of stakeholder inclusion in product lifecycle processes. In each case, the overall inclusivity score is calculated by multiplying the values of the selected dimensions:
Inclusivity can be operationally defined (see $I_1$) as a product of the degree of diversity, the degree of hierarchy and the number of interactions. This can be expressed as follows:

- $I_1$ = degree of diversity × degree of hierarchy × number of interactions

$I_1$ highlights the role of stakeholder interactions in fostering inclusivity through diverse involvement, supported by the degree of hierarchy.
Inclusivity can alternatively be operationally defined (see $I_2$) as a product of the degree of diversity, the degree of hierarchy and the number of lifecycle phases. This can be expressed as follows:

- $I_2$ = degree of diversity × degree of hierarchy × number of lifecycle phases

$I_2$ evaluates stakeholder inclusivity based on their involvement across lifecycle phases, supported by diversity within the degree of hierarchy.
Inclusivity can alternatively be operationally defined (see $I_3$) as a product of the degree of diversity, the degree of hierarchy, the number of lifecycle phases, and the number of interactions. This can be expressed as follows:

- $I_3$ = degree of diversity × degree of hierarchy × number of lifecycle phases × number of interactions

$I_3$ evaluates inclusivity by also counting the number of interactions among stakeholders in addition to the metrics of $I_2$.
Inclusivity can alternatively be operationally defined (see $I_4$) as a product of the degree of diversity, the number of lifecycle phases and the number of interactions. This can be expressed as follows:

- $I_4$ = degree of diversity × number of lifecycle phases × number of interactions

$I_4$ evaluates inclusivity by the representation of diverse stakeholder groups in different lifecycle phases, based on the number of interactions they have through the process.
Since the aim is to assess inclusivity in a product lifecycle context, single metrics are not evaluated as an operational definition. For example, the degree of diversity is an influencing factor of inclusivity; however, alone, it is not sufficient to create inclusivity in the process. Because diversity without interactions and/or involvement in different lifecycle phases cannot be sufficient to support other stakeholders' inclusion in the process. Therefore, the operational definitions are based on the dimensions of inclusivity generated by considering the functional (i.e. operational) potential of the metric combinations. For example, diversity and hierarchy are not considered as another definition of inclusivity, since both of them, as reflections of inclusivity rather than drivers, cannot be sufficient for starting or continuing an operation.

### 5.2 Empowerment Dimensions

As mentioned in Table 3, the concept groups of empowerment are Stakeholders, Transformation and Power Relations, which can be supported by Cattaneo and Chapman (2010). They defined empowerment "...as an iterative process in which a person who lacks power sets a personally meaningful goal oriented toward increasing power, takes action toward that goal, and observes and reflects on the impact of this action, drawing on his or her evolving self-efficacy, knowledge, and competence related to the goal." Based on this definition, it can be interpreted that the elements to adopt and assess empowerment of people through product lifecycle processes can be power, impact and process.

As proposed in Section 4, impact can be measured using Eq. (1). The amount of power is formulated in Eq. (4), where means, dependency level, and frequency are considered as the multipliers of power. The means refer to the number of tools used to complete a particular function. Therefore, determining the number and quality of means depends on the context, as well as the diversity of stakeholders and functions. For example, suppose that the design team, customer user, and production team are the stakeholders. The design team interacts with both the customer user and production team to check the producibility of customer demands. As a result, the tools used for product development and communication may vary between these functional groups due to contextual differences, even though they work toward the common goal of continuing the product lifecycle process. As shown in Eq. (3), the dependency level evaluates how much stakeholder dependencies are reduced through the sharing of resources to fulfil needs. Therefore, a higher dependency level can support a greater enhancement of power.

The frequency of a stakeholder group is defined as the number of functions completed by that group. Empowering impact depends on the number of functions (see Eq. (1)). This causes a repetition of the number of functions in both the empowering impact and power equations (see Eq. (2) and Eq. (4)). To avoid double-counting the same factor, four alternative operational definitions are proposed for the empowerment metric.

Empowerment can be operationally defined (see $E_1$) as a product of empowering impact and the number of means. This can be expressed as follows:

- $E_1$ = empowering impact × number of means

$E_1$ evaluates the empowerment of stakeholders based on the influence of the number of means on the empowering impact of functions. The number of means can affect changes in the number of functions (a factor of empowering impact). For example, an increase in the number of means can either increase or decrease the number of functions, which in turn affects the empowering impact.

Empowerment can alternatively be operationally defined (see $E_2$) as a product of the number of means and the dependency level. This can be expressed as follows:

- $E_2$ = number of means × dependency level

$E_2$ evaluates empowerment based on the influence of the number of means on the dependency level of stakeholders. Since the dependency level is the ratio of needs to available resources, the number of means can influence both the fulfilment of needs and the distribution of resources.

Empowerment can alternatively be operationally defined (see $E_3$) as a product of the empowering impact and the dependency level. This can be expressed as follows:

- $E_3$ = empowering impact × dependency level

The third definition, E3, evaluates empowerment by considering the influence of dependency level on empowering impact. The dependency level of a stakeholder also highlights their involvement in various group functions. This indicates an increase in the number of included stakeholders (a factor of empowering impact; see Eq. 1) within a particular function, thereby influencing the empowering impact of that function.

Empowerment can alternatively be operationally defined (see $E_4$) as a product of the empowering impact, the number of means and the dependency level. This can be expressed as follows:

- $E_4$ = empowering impact × dependency level × number of means

The fourth definition, E4, evaluates empowerment by considering the influence of both dependency level and the number of means. The definitions are formed to support the operational structure of the empowering process; therefore, metrics are not considered alone as separate operational definitions.

Overall, different formulas are generated to measure inclusivity and empowerment using various metric pairs, depending on the focus of the relationship — for example, on how interactions influence the strength of impact, on stakeholder dependency levels, or on which metrics contribute more strongly to empowerment. This approach can help identify the critical metrics of empowerment and inclusivity of stakeholders in lifecycle processes. The purpose of quantifying these dimensions is to track improvements and changes in stakeholder inclusivity and empowerment. By measuring different metric pairs, this can support the identification of which combinations provide better assessment of inclusivity and empowerment in various contexts across lifecycle processes. Comparing the quantified results using Pearson correlation analysis (discussed in Section 7) helps to reveal the strongest relationships between empowerment and inclusivity. This suggests that the proposed metrics can effectively support the assessment of stakeholder development in different contexts.

## 6. CASE STUDY

This section applies the framework to a case study to provide a better understanding of its application. Additionally, the framework is applied to 10 real-life case studies. One case study is presented in detail to illustrate the framework in depth, while the results from all ten case studies are summarised in Supplementary File 4. Based on the application results, the inclusivity and empowerment dimensions are measured, and the dimensions are selected according to the strength of the correlation between them. A brief summary of the ten case studies is provided in Supplementary File 4. The summary highlights the purpose of each project, the involved stakeholders, the total number of identified functions, and the evaluated lifecycle phases.

### 6.1 Application of the Framework to the Case Study

This section briefly explains the application of the framework to a case study (Singh and Gosain, 2019) on the development of a low-cost groundwater-level measuring device (summarised in CS3; see Supplementary File 4). The functions are analysed across four phases—planning, development, production, and utilisation—of the lifecycle process. The recirculation phase is not discussed in the case study and is therefore not included. Eight stakeholder groups are identified from the case study. Table 4 shows a part of the analysis; the roles of the stakeholders are identified from the function explanation column. A change in role from provider to receiver, or vice versa, is considered an individual function. F1&F2, F3&F4&F5, and F9&F10 are function groups with multiple stakeholders, which create power relations among them.

Table 4: Illustrative Section of Framework Application – Step 1

| Function Group | Function | Explanation | Stakeholder | Role | Phase |
|---|---|---|---|---|---|
| F1&F2 | F1 | Provide information (during workshop) | D | Provider | Planning |
| | F2 | Receive information | C | Receiver | |
| F3&F4&F5 | F3 | Request to share problems about conditions and product | D | Provider | |
| | F4 | Receive the task | C | Receiver | |
| | F5 | | A | Receiver | |
| X | F6 | Change of role from Provider to Receiver | D | Receiver | |
| X | F7 | Change of role from Receiver to Provider | C | Provider | |
| X | F8 | Change of role from Receiver to Provider | A | Provider | |
| F9&F10 | F9 | Request to provide problems and user complaints | C | Provider | |
| | F10 | Receive the task | B | Receiver | |

In the second step, shown in Table 5, the empowering impact (Eq. (1(Eq.)) of the stakeholders is calculated for each function, considering the self-empowering impact of individual functions.

Table 5: Illustrative Section of Framework Application – Step 2

| Function Group | Function | Stakeholder | #Stakeholders | #Task | Impact |
|---|---|---|---|---|---|
| F1&F2 | F1 | D | 2 | 1 | 2 |
| | F2 | C | 2 | 1 | 2 |
| F3&F4&F5 | F3 | D | 3 | 1 | 3 |
| | F4 | C | 3 | 1 | 3 |
| | F5 | A | 3 | 1 | 3 |
| X | F6 | D | 1 | 1 | 1 |
| X | F7 | C | 1 | 1 | 1 |
| X | F8 | A | 1 | 1 | 1 |
| F9&F10 | F9 | C | 2 | 1 | 2 |
| | F10 | B | 2 | 1 | 2 |

The last step of the framework application, presented in Section 4, calculates the dependency level and the amount of power of the stakeholders. As shown in Table 6, due to the available information in the case study, the means are assumed to be one for each function. The frequency of stakeholders is listed for the planning phase of the lifecycle in Table 6; the frequencies in other phases are different (see detailed tables in Supplementary File 5). The groups of needs are identified by considering the function explanations and stakeholders. For example, Function 1 is "Providing information during the workshop." The need to provide information can be "cooperate and interact" and is grouped under "Participation and Doing" (Max-Neef, 1992). In this example, the number of need groups is two. The resources are considered as position and capability. Based on these, the dependency level of each stakeholder on the others is calculated for each function group. The dependency level of stakeholders in the function group 'F3&F4&F5' is determined using (Eq. (3)), as shown in Table 6. The receivers (Stakeholders C and A) depend individually on Function 5 (F5). Therefore, the dependency level of Stakeholder A is calculated by dividing its resources by the needs of Stakeholder D. The same applies to Stakeholder C. The dependency level of Stakeholder D is calculated by dividing the sum of the resources of Stakeholders A and C by the sum of their needs. The amount of power of the stakeholders is computed for each function group. Individual functions do not create dependency on other stakeholders, as these functions do not require support from others to be carried out. Therefore, the dependency level of stakeholders for these functions is zero.

Table 6: Illustrative Section of Framework Application – Step 3

| Function Groups | F | Means | Frequency | Extension | Needs | #Needs | Resources | #Resource | DL | Power |
|---|---|---|---|---|---|---|---|---|---|---|
| F1&F2 | F1 | 1 | 29 | 29 | Participation and Doing | 2 | Position | 1 | 0.5 | 14.5 |
| | F2 | 1 | 22 | 22 | Understanding and Doing | 2 | Position | 1 | 0.5 | 11 |
| F3&F4 &F5 | F3 | 1 | 29 | 29 | Participation and Doing | 2 | Position | 1 | 0.5 | 14.5 |
| | F4 | 1 | 22 | 22 | Participation and Doing | 2 | Position | 1 | 0.5 | 11 |
| | F5 | 1 | 8 | 8 | Creation and Interacting | 2 | Capability | 1 | 0.5 | 4 |
| X | F6 | 1 | 29 | 29 | Participation and Having | 2 | Position | 1 | 0 | 0 |
| X | F7 | 1 | 22 | 22 | Participation and Having | 2 | Position | 1 | 0 | 0 |
| X | F8 | 1 | 8 | 8 | Creation and Interacting | 2 | Capability | 1 | 0 | 0 |
| F9& F10 | F9 | 1 | 22 | 22 | Participation and Doing | 2 | Position | 1 | 0.5 | 11 |
| | F10 | 1 | 9 | 9 | Creation and Interacting | 2 | Capability | 1 | 0.5 | 4.5 |

Using the above, the dimensions of empowerment and inclusivity are calculated (see Table 7). Stakeholders E, F, and H do not participate in any function in the planning phase; therefore, their dimensions are taken as zero. As explained in Section 5.1.1, the diversity of people is identified in terms of the context and conditions that cause disadvantaged situations. The range of diversity is taken from 1 to 4, according to the number of the following contexts: individuality and activity, relations, location, and time. For instance, Stakeholder A is disadvantaged in the contexts of location and time due to the lack of resources in the location at different periods; hence, the diversity level can be considered as 2 (location and time contexts). The hierarchy of stakeholders is scaled from 1 to 5 based on their influence on decision-making and progress in the lifecycle process. For example, Stakeholders C and D have more influence on decision-making than Stakeholders B and A. The number of lifecycle phases stakeholders participate in, and the number of interacted stakeholder groups, are counted from the results of the framework application. For instance, Stakeholder A participated in one lifecycle phase, while Stakeholder B participated in two phases. Stakeholder C interacts with three stakeholders (A, B, and D). Inclusivity metrics are calculated by multiplying these dimensions. Empowerment dimensions are then computed by applying the framework in all cases.

Table 7: Calculation of the dimensions of I and E in the planning phase

| Planning | | | | | | | | |
|---|---|---|---|---|---|---|---|---|
| | A | B | C | D | E | F | G | H |
| Diversity | 2 | 2 | 1 | 1 | 0 | 0 | 1 | 0 |
| Hierarchy | 2 | 3 | 4 | 4 | 0 | 0 | 5 | 0 |
| LC | 1 | 2 | 3 | 4 | 0 | 0 | 1 | 0 |
| #interactions | 2 | 2 | 3 | 3 | 0 | 0 | 1 | 0 |
| $I_1$ | 8 | 12 | 12 | 12 | 0 | 0 | 5 | 0 |
| $I_2$ | 4 | 12 | 12 | 16 | 0 | 0 | 5 | 0 |
| $I_3$ | 8 | 24 | 36 | 48 | 0 | 0 | 5 | 0 |
| $I_4$ | 4 | 8 | 9 | 12 | 0 | 0 | 1 | 0 |
| Means | 1 | 1 | 1 | 1 | 0 | 0 | 1 | 0 |
| Impact | 14 | 15 | 36 | 46 | 0 | 0 | 15 | 0 |
| DL | 1.5 | 4 | 5.5 | 9 | 0 | 0 | 2.5 | 0 |
| $E_1$ | 14 | 15 | 36 | 46 | 0 | 0 | 15 | 0 |
| $E_2$ | 1.5 | 4 | 5.5 | 9 | 0 | 0 | 2.5 | 0 |
| $E_3$ | 21 | 60 | 198 | 414 | 0 | 0 | 37.5 | 0 |
| $E_4$ | 21 | 60 | 198 | 414 | 0 | 0 | 37.5 | 0 |

Table 8 presents the results for all dimensions of the empowerment and inclusivity metrics for each stakeholder across each phase of the lifecycle process in the case study above. The dimensions are aggregated separately for each stakeholder. For example, the E1 score of Stakeholder A is calculated by summing the E1 values obtained from each phase of the lifecycle process.

Table 8: Metrics of empowerment and inclusivity for the lifecycle process

| Stakeholders | E1 | E2 | E3 | E4 | I1 | I2 | I3 | I4 |
|---|---|---|---|---|---|---|---|---|
| A | 14 | 1.5 | 21 | 21 | 8 | 4 | 8 | 4 |
| B | 41 | 9 | 190 | 190 | 24 | 24 | 48 | 16 |
| C | 84 | 13 | 400.5 | 400.5 | 24 | 32 | 64 | 16 |
| D | 118 | 21 | 726 | 726 | 28 | 64 | 112 | 28 |
| E | 8 | 1.5 | 12 | 12 | 1 | 1 | 1 | 1 |
| F | 19 | 3.5 | 66.5 | 66.5 | 4 | 2 | 4 | 2 |
| G | 15 | 2.5 | 37.5 | 37.5 | 5 | 5 | 5 | 1 |
| H | 33 | 5 | 165 | 165 | 4 | 2 | 4 | 2 |

Supplementary File 5 provides the detailed function analysis of CS3 with the link of all case studies, and it includes the results of the dimension calculations for all phases of the lifecycle process, as a continuation of Table 7. The function analysis conducted in this research covers all stakeholders throughout the entire lifecycle. An access link to all case study data is also provided in this file. Additionally, Supplementary File 4 presents the empowerment and inclusivity results for all stakeholders across all case studies, extending the data shown in Table 8.

## 7. RESULTS

Based on the analysis results (see Supplementary File 4 for the number of functions, evaluated lifecycle phases, and the level of empowerment and inclusivity of each stakeholder across the ten real-life case studies), not all phases of the product lifecycle are covered in any of the case studies. In particular, the recirculation phase is absent from all of them. The total number of functions and stakeholders varies across the case studies.

Empowerment and inclusivity metrics have been calculated for each stakeholder group according to the lifecycle phases included in each case study. To understand the correlation between inclusivity and empowerment metrics, the dimensions are first paired, as shown in Table 9. Then, Pearson correlation coefficient tests are conducted to compare the paired dimensions of inclusivity and empowerment. A value of $r = 0.5$ is considered a moderately positive correlation, while a greater r indicates the strongest correlation. A minimum acceptable p-value of $< 0.1$ is used as the threshold for marginal statistical significance ($0.05 < p < 0.1$), which can be acceptable with careful interpretation in exploratory studies and small-sized datasets, consistent with the interpretation used in previous research (Zhang et al., 2022). Additionally, further discussion on the correlations between the dimensions is aimed for future work by supporting centrality analysis and suggesting ways to improve the correlations between empowerment and inclusivity. In this analysis, no correction for multiple comparisons was applied, as the number of correlations tested was small (small number of stakeholders; see Supplementary File 4), and the study is exploratory.

Based on the r and p values from Table 9, the metric combination that shows a consistent positive correlation across all ten case studies is E4 (empowering impact, dependency level, number of means) and I3 (diversity, hierarchy, number of lifecycle phases, number of interactions). Accepting the significance value as 0.1, the E4-I3 pair shows statistically significant positive correlations in 90% of the case studies. These significant correlations often occur alongside strong r-values ($r > 0.7$), showing that E4 and I3 offers a stable, consistent, and statistically supported relationship across the data from all case studies. Only one case (CS4) shows a weak and insignificant correlation, which does not affect the overall conclusion. The weak relationship may be due to the unequal distribution of resources among stakeholders, which can diminish the pattern between empowerment and inclusivity. Additionally, CS4 is aimed to be examined in detail in future work to address the reason for the low correlation and to find approaches to improve the strength of the relationship between empowerment and inclusivity.

A potential alternative to E4-I3 could be the dimension pair E4-I4. Although E4–I4 shows similarly strong correlations in some cases, E4–I3 consistently shows higher or comparable r-values in 60% of the case studies (based on the cases where $r_{E4-I3} \geq r_{E4-I4}$). E4–I4 fails to reach significance in CS6 ($p = 0.20$), while E4–I3 is non-significant only in CS4 ($p = 0.65$). E4–I3 was selected to ensure greater consistency and stronger correlations. However, due to the similarities between the results of E4–I3 and E4–I4, the following section details the comparison of the dimension pairs.

Table 9: Comparison of Empowerment (E) and Inclusivity (I) Metrics

| Dimension Pairs | CS1 | | CS2 | | CS3 | | CS4 | | CS5 | | CS6 | | CS7 | | CS8 | | CS9 | | CS10 | |
|---|---|---|---|---|---|---|---|---|---|---|---|---|---|---|---|---|---|---|---|---|
| | r | p | r | p | r | p | r | p | r | p | r | p | r | p | r | p | r | p | r | p |
| E1-I1 | 0.86 | 0 | 0.8 | 0.01 | 0.87 | 0 | 0.25 | 0.55 | 0.9 | 0.01 | 0.73 | 0.27 | 0.96 | 0 | 0.66 | 0.03 | 0.98 | 0 | 0.7 | 0 |
| E1-I2 | 0.36 | 0.28 | 0.64 | 0.06 | 0.97 | 0 | 0.02 | 0.96 | 0.83 | 0.02 | 0.68 | 0.32 | 0.95 | 0 | 0.88 | 0 | 0.97 | 0 | 0.71 | 0 |
| E1-I3 | 0.86 | 0 | 0.79 | 0.01 | 0.98 | 0 | 0.32 | 0.44 | 0.87 | 0.01 | 0.86 | 0.14 | 0.97 | 0 | 0.85 | 0 | 0.97 | 0 | 0.7 | 0 |
| E1-I4 | 0.9 | 0 | 0.82 | 0.01 | 0.94 | 0 | 0.74 | 0.04 | 0.87 | 0.01 | 0.81 | 0.19 | 0.96 | 0 | 0.82 | 0 | 0.97 | 0 | 0.78 | 0 |
| E2-I1 | 0.87 | 0 | 0.8 | 0.01 | 0.9 | 0 | 0.29 | 0.49 | 0.87 | 0.01 | 0.74 | 0.26 | 0.95 | 0 | 0.68 | 0.02 | 0.98 | 0 | 0.74 | 0 |
| E2-I2 | 0.4 | 0.22 | 0.59 | 0.09 | 0.99 | 0 | 0.09 | 0.83 | 0.8 | 0.03 | 0.7 | 0.3 | 0.94 | 0 | 0.88 | 0 | 0.96 | 0 | 0.74 | 0 |
| E2-I3 | 0.87 | 0 | 0.79 | 0.01 | 0.99 | 0 | 0.38 | 0.35 | 0.84 | 0.02 | 0.86 | 0.14 | 0.97 | 0 | 0.87 | 0 | 0.97 | 0 | 0.73 | 0 |
| E2-I4 | 0.9 | 0 | 0.81 | 0.01 | 0.97 | 0 | 0.79 | 0.02 | 0.85 | 0.02 | 0.84 | 0.16 | 0.96 | 0 | 0.83 | 0 | 0.94 | 0 | 0.81 | 0 |
| E3-I1 | 0.91 | 0 | 0.69 | 0.04 | 0.84 | 0.01 | 0.12 | 0.78 | 0.85 | 0.02 | 0.79 | 0.21 | 0.9 | 0 | 0.63 | 0.04 | 0.97 | 0 | 0.59 | 0.02 |
| E3-I2 | 0.34 | 0.31 | 0.39 | 0.3 | 0.99 | 0 | -0.08 | 0.85 | 0.77 | 0.04 | 0.69 | 0.31 | 0.88 | 0 | 0.82 | 0 | 0.94 | 0 | 0.61 | 0.02 |
| E3-13 | 0.89 | 0 | 0.66 | 0.05 | 0.98 | 0 | 0.19 | 0.65 | 0.82 | 0.02 | 0.91 | 0.09 | 0.93 | 0 | 0.84 | 0 | 0.96 | 0 | 0.58 | 0.02 |
| E3-14 | 0.94 | 0 | 0.66 | 0.05 | 0.95 | 0 | 0.66 | 0.07 | 0.82 | 0.02 | 0.8 | 0.2 | 0.9 | 0 | 0.79 | 0 | 0.91 | 0 | 0.67 | 0.01 |
| E4-I1 | 0.91 | 0 | 0.69 | 0.04 | 0.83 | 0.01 | 0.12 | 0.78 | 0.85 | 0.02 | 0.79 | 0.21 | 0.9 | 0 | 0.63 | 0.04 | 0.97 | 0 | 0.59 | 0.02 |
| E4-I2 | 0.34 | 0.31 | 0.39 | 0.3 | 0.97 | 0 | -0.08 | 0.85 | 0.77 | 0.04 | 0.69 | 0.31 | 0.88 | 0 | 0.82 | 0 | 0.94 | 0 | 0.61 | 0.02 |
| E4-I3 | 0.89 | 0 | 0.66 | 0.05 | 0.96 | 0 | 0.19 | 0.65 | 0.82 | 0.02 | 0.91 | 0.09 | 0.93 | 0 | 0.84 | 0 | 0.96 | 0 | 0.58 | 0.02 |
| E4-I4 | 0.94 | 0 | 0.66 | 0.05 | 0.93 | 0 | 0.66 | 0.07 | 0.85 | 0.02 | 0.8 | 0.2 | 0.9 | 0 | 0.79 | 0 | 0.91 | 0 | 0.67 | 0.01 |

### 7.1 Comparison of E4-I3 and E4-I4

This section explains the comparisons to determine which dimension pair provides the strongest assessment of the relationship between inclusivity and empowerment, using statistical strength (r-value), reliability (p-value and sample size), and robustness (mean vs. median).

The selection of E4-I3 is based on the overall statistical strength and consistency of results across all case studies. Both E4-I3 and E4-I4 show strong correlations in most dimensions, but E4-I3 presents higher effect sizes in some critical areas. For example, in CS6, the correlation is $r = 0.91$ ($p = 0.09$) in E4-I3, which is stronger than $r = 0.80$ ($p = 0.20$) in E4-I4. Although this result is marginally significant, it is still acceptable under the threshold of $p < 0.10$, especially for exploratory research with small datasets. In CS3, the effect size is also slightly stronger in E4-I3 ($r = 0.96$) than in E4-I4 ($r = 0.93$), with the same p-value of 0.05 for both. Furthermore, the non-significant result in CS4 ($p = 0.65$) for E4-I3 is based on more data points ($n = 8$), which supports a more reliable interpretation of a weak or no relationship. On the other hand, the non-significance in CS6 for E4-I4 ($p = 0.20$) is based on fewer data points ($n = 4$), which may result from limited statistical power.

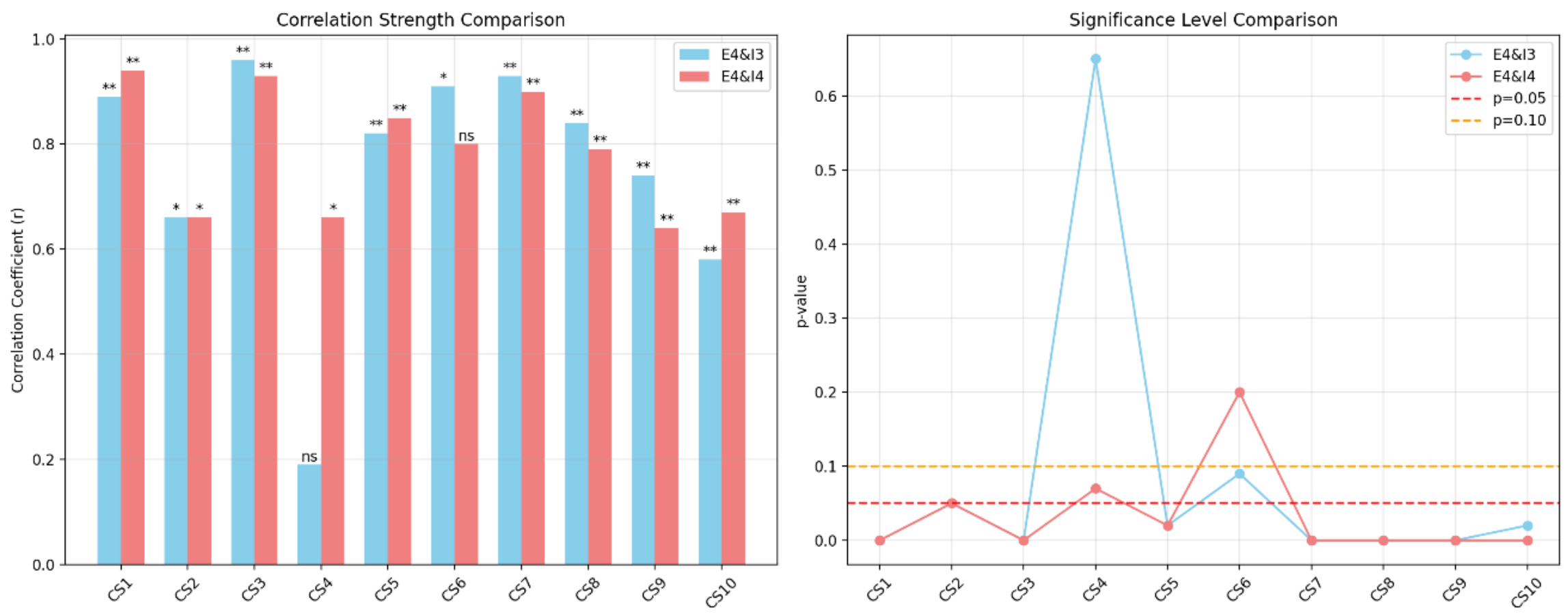

Figure 5: Comparison of E4-I3 and E4-I4

Figure 5 presents a comparison between the r and p values of E4-I3 and E4-I4 across ten case studies. The left side of the figure shows the correlation coefficients (r) for each dimension with significance levels marked as: ** ($p < 0.05$), * ($0.05 \leq p < 0.10$), and ns ($p \geq 0.10$). The right side of the figure displays the corresponding p-values, with reference lines at 0.05 and 0.10. Both E4-I3 and E4-I4 show 7 highly significant results ($p < 0.05$), 2 marginally significant ($0.05 \leq p < 0.10$), and 1 non-significant result ($p \geq 0.10$). In short, a higher r means empowerment and inclusivity tend to rise and fall together more consistently within a case, indicating the strongest association between them. A low p-value indicates greater confidence that this pattern is not due to chance, with $p < 0.05$ considered strong evidence and $p < 0.10$ considered moderate evidence appropriate for exploratory, small-sample analyses. Conversely, a high r with a high p-value is only suggestive due to limited power or variability. Since in both combinations the distribution of r and p in the ten cases are similar, both have similar statistical strength and reliability. Therefore, we use robustness as the criterion to determine which of the combinations is better. Following the suggestion of Wilcox and Rousselet (2023), who recommend combining the mean with a robust estimator such as the median when the number of cases is small and the effect-size distribution is uneven, both statistics are reported in Table 10. In this dataset, the mean and median differ, indicating skewness. Therefore, the median is prioritised for selecting the dimension pair, as it offers a more reliable summary under these conditions. The mean reflects the ordinary average of the ten r-values, but can be influenced by extreme values. In contrast, the median represents the middle value of the sorted correlations and is less affected by outliers. In this dataset (see Table 9), one unusually low correlation in CS4 (r = 0.19) reduces the E4–I3 mean to 0.774 (see Table 10), while the median remains high at 0.865, compared to 0.825 for E4–I4. A higher median correlation means that, in the study most likely to be encountered, the E4–I3 link is stronger than the E4–I4 link. Although both dimension pairs perform similarly overall, the slightly higher median correlation in E4-I3 supports its selection as a more reliable choice.

Table 10: Mean and median correlations for E4–I3 and E4–I4 dimension pairs

| Statistic | E4-I3 | E4-I4 |
|---|---|---|
| Mean | 0.774 | 0.811 |
| Median | 0.865 | 0.825 |

In other words, if E4 and I3 are adopted as operational definitions of inclusivity and empowerment, greater empowerment would be achieved by greater inclusivity in lifecycle processes. According to these definitions, the inclusivity of people for empowerment is not only about increasing the number of stakeholder groups involved. It also includes the number of people, the degree of diversity, the number of lifecycle phases in which stakeholders can be included, and the number of interactions. Inclusivity of this kind seems to have the potential to support greater empowerment by enhancing the impact of functions, the dependency level of stakeholders, and the number of means used. Thus, E4-I3 is considered the most appropriate metric combination for assessing the relationship between empowerment and inclusivity across product lifecycle processes.

In summary, this section compares the E4–I3 and E4–I4 dimension pairs to determine which provides the strongest assessment of the association between inclusivity and empowerment in lifecycle processes. Both pairs show strong correlations; however, by comparing their median values, E4–I3, which has a higher median value, is selected as the more robust combination. This means that inclusivity can be operationalised as I3 to better support the assessment of empowerment based on E4. Thus, it can guide the design of inclusive lifecycle processes for the greater inclusivity of stakeholders, which in turn enhances their greater empowerment.

It should be noted that the comparisons of dimension pairs (E4-I3 and E4-I4) show association rather than causation. The study hypothesises that greater inclusivity (as expressed in E4) can support greater empowerment (as expressed in I3), as indicated by the strength of their correlations. However, causality cannot be confirmed within the limits of this dataset.

### 7.2 Interpretation of E4–I3 Relationships Across Case Studies

Figure 6 compares the E4 and I3 values of each stakeholder in the case studies. The blue dots represent stakeholders, and the green dots represent users (in CS5, two different user groups are shown as separate green dots). The actual values are provided in Supplementary File 4. For the scatter plots in Figure 6, the $R^2$ values for each case study are provided in Supplementary File 6. For CS2, CS4, and CS10, $R^2$ is less than 0.5, while for the other cases it is above 0.67. This indicates that in most case studies, the linear model explains a large part of the variance between empowerment and inclusivity, which shows the strongest and more stable relationship in these cases. In cases with lower $R^2$, the relationship appears weaker. This can occur due to imbalanced involvement of stakeholders in the process, which causes unevenly distributed metrics of empowerment and inclusivity. The three lower-fit cases indicate the need for further evaluation of systems with imbalanced roles and small sample sizes to improve the relationship between empowerment and inclusivity. The wider confidence intervals (see Figure 6) may suggest possible non-linearity in the relationship between empowerment and inclusivity. However, this observation should not be over-interpreted, due to the small number of stakeholders in some case studies, as the limited data can cause high variability in the regression results. Therefore, the framework can be applied in more complex systems with a larger number of stakeholders, which can also support the improvement of the framework and its applications.
Based on the comparison of E4 and I3 ratios of users (green data points) with other stakeholder groups, it can be seen that increasing the inclusivity of users cannot always result in benefits for all stakeholders. Especially in CS1 and CS4, the users have a higher level of inclusivity and a low level of empowerment, indicating that for them, inclusivity could not support access to opportunities for empowerment. This also highlights the need for evaluation of the system dynamics, which can improve situations where there is high inclusivity but low empowerment. The inclusivity of users, as prioritised in inclusive design, is significant for sustainable development; however, it is not sufficient on its own. As can be seen, inclusion alone cannot support equal development of stakeholders in the context of empowerment and inclusivity.
The slopes of the trend line for CS4 (0.02), CS5 (0.04), and CS10 (0.03) are very small, indicating the need for more meaningful use of inclusivity to achieve a greater level of empowerment. This also highlights the inequity in the distribution of opportunities, since inclusivity could not be transformed into empowerment. The detailed comparison of slopes and $R^2$ values is provided in Supplementary File 6. The E4 and I3 metrics are case-specific, and the comparisons are focused on the trends within each case study, rather than comparing the absolute values across different case studies. Based on 7 out of 10 case studies, the hypothesis can be supported.
Overall, these findings indicate that if the inclusion of people in the process is supported according to inclusivity metrics, it also supports empowerment (in line with empowerment metrics). The assumed benefit of this relationship is the improvement of the well-being and dignity of the included individuals and their contribution to sustainable development, discussed in Yaldiz and Chakrabarti (2024b).
In addition, the application of the framework allows for the comparison of stakeholders' inclusivity and empowerment values. This makes the inequalities in stakeholder development visible (for example, differences in empowerment and inclusivity ratios among different stakeholders within the same case study). In this way, it becomes possible to identify the areas that need improvement and to guide targeted equity in stakeholder development. Additionally, the findings provided in Figure 6 highlight the potential of the framework to help identify what improvements are needed to achieve more inclusive stakeholder development in future lifecycle processes (e.g. like a backcasting approach). However, this requires further study to clarify which stakeholder groups should be involved in which functions and phases of the lifecycle, and who the decision-makers should be to support the inclusion of different stakeholder groups in different contexts.

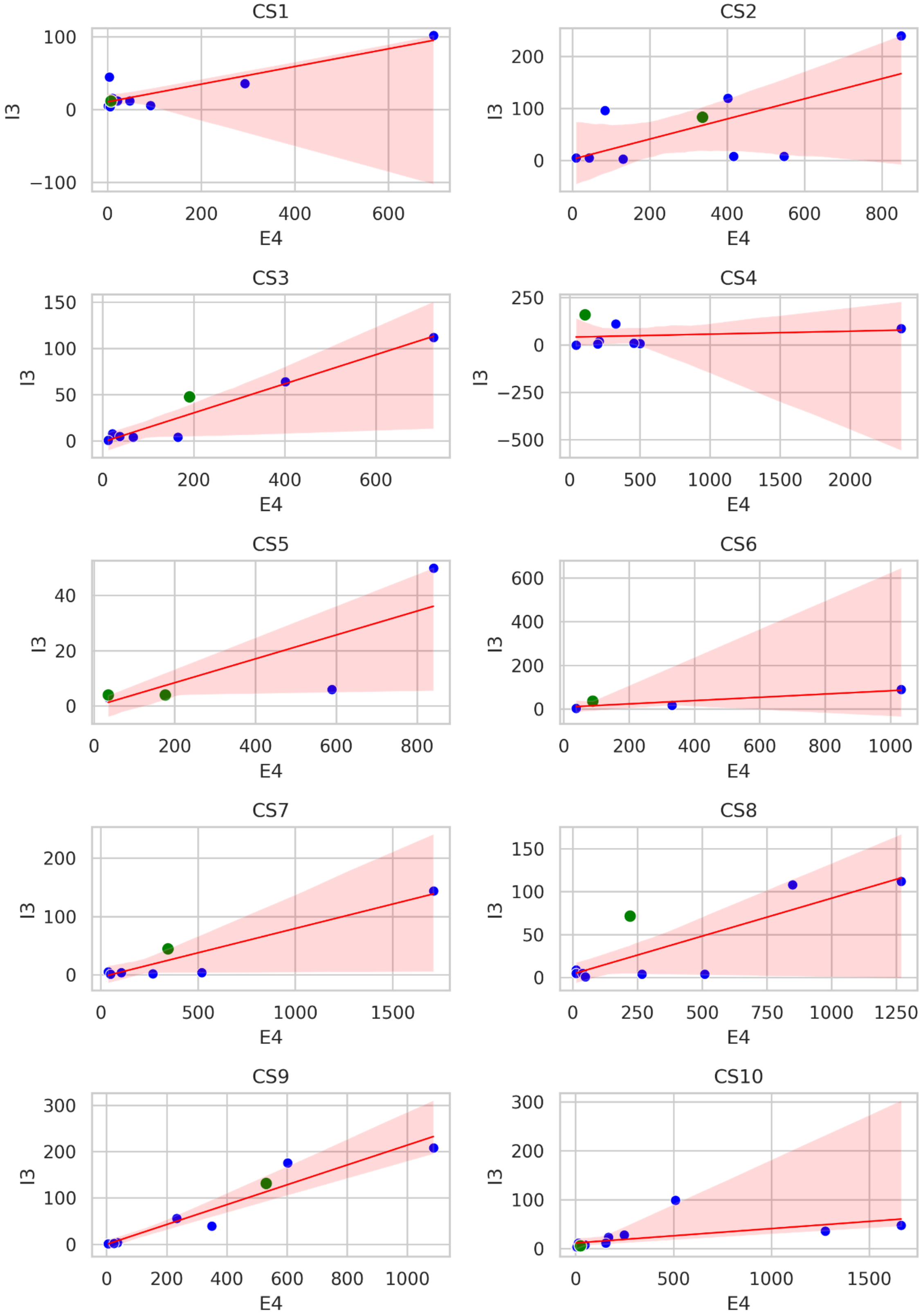

Figure 6: Comparison of E4-I3 values of stakeholders across case studies (CS1 to CS10)

## 8. DISCUSSION

The design of inclusive lifecycle processes is intended to contribute to sustainability, well-being, and the dignity of people (or stakeholder groups) by empowering them through their inclusion in different functions. Empowerment is the process of gaining power based on a goal within stakeholder inclusivity. People can become more independent and empowered through shifts in their roles within the exchange networks – in our context, it is product lifecycle processes – by sharing resources and meeting needs. Inclusion is necessary for people to become empowered in lifecycle processes, but it alone may not be sufficient unless it is supported by a linear relationship between empowerment and inclusivity based on the chosen metrics. Assessing the inclusion of people in product lifecycle design requires metrics of inclusivity that can be used to evaluate who is included in or excluded from a particular function. Understanding the exclusion of people helps identify the resources and needs required for their inclusion in the relevant phase of the lifecycle process. Therefore, the boundaries of the lifecycle context can be extended by adapting inclusion and exclusion to achieve more inclusive outcomes — in this context, sustainability, well-being, and dignity. Inclusive outcomes are usually assessed based on the increase in the number of users through the inclusion of previously excluded people. However, the merits of these concepts are not limited to simply expanding the diversity range by involving more diverse user groups. The inclusive lifecycle can be designed based on inclusivity metrics to empower people—ultimately ensuring inclusive outcomes for the transition to sustainability. Depending on the aim of the processes, inclusive outcomes can vary; however, in the context of inclusive lifecycle processes, they are intended to contribute to empowerment.

This study addresses the research question, which aims to understand the nature of empowerment and inclusivity, such that inclusivity leads to empowerment. Based on the findings, the answer to the research question (How can inclusivity and empowerment be assessed in product lifecycle processes?) is as follows:

Empowerment can be assessed as the product of impact, dependency level, and the number of means. 'Impact' is the product of the number of functions per person and the number of people participating. Impact refers to the gains that have the potential to sustain empowerment processes. These gains, achieved by people using means, can make individuals more independent by sharing resources to fulfil their own needs and those of others with whom they interact. The number of means is important for enhancing and extending empowerment processes. The argument is that empowerment is a multi-dimensional process that gives people more power through their inclusion in a product lifecycle process.

Inclusivity can be assessed as the product of diversity, hierarchy, the number of lifecycle phases, and the number of interactions. For empowerment, the inclusivity of people is not only about participation in lifecycle functions; it is about increasing the number of lifecycle phases a stakeholder group can be involved in and improving their role (i.e., hierarchy), while also considering diversity and interactions among stakeholders. A lifecycle has various phases, each with a variety of functions. Ideally, the inclusion of diverse people should be aimed for in each function in each phase of the lifecycle process.

Based on the comparison of empowerment and inclusivity dimensions (presented in Section 7.2) for each stakeholder, the proposed framework can be used to support the equal development of stakeholders by addressing inequities resulting from imbalanced situations. The proposed metrics and framework thus provide a practical basis for guiding fairer task distribution and empowerment in organisations and complex systems. The following section summarises the study's contributions to design science and practice.

### 8.1 Contributions to Design Science and Practice

This study contributes to theory by proposing an improved version of the framework and metrics to assess stakeholder inclusivity and empowerment in product lifecycle processes.

The approach addresses an important gap in the literature, where these dimensions are rarely analysed together at the system level across lifecycle phases. By systematically correlating stakeholder functions with empowerment and inclusivity, the study provides a deeper understanding of how stakeholder development can be more equitably supported in lifecycle processes. This expands the existing knowledge on inclusive practices from focusing only on product-level interventions towards including process-level dynamics.

For practice, the proposed framework and metrics provide a structured basis for identifying imbalances in stakeholder empowerment and inclusivity. The approach supports better planning of stakeholder involvement in functions, helping to improve fair task distribution and support stakeholder development. The findings can also inform future-focused strategies to promote inclusive lifecycle processes, contributing to the dignity and well-being of stakeholders and sustainable development.

## 9. LIMITATIONS

The operationalisation of inclusivity and empowerment is based solely on the theoretical and conceptual inputs discussed in Sections 1 and 2, Supplementary Files 1 and 2. Function identification is limited to the content of the case studies and is grouped based on the tasks of a product lifecycle process. The functions are detailed a maximum of two times, which limits the number of functions assigned to specific group functions. The function types are limited to the models described in Supplementary File 3. Stakeholder identification is limited to the information available in the case study content. The dependency level of stakeholders is limited to the description of the functions in which they are involved. Mainly, the number of needs is assumed to be 2, and the number of resources is assumed to be 1. This limitation is necessary to maintain focus on a specific need group, rather than expanding the scope of the study and risking inconsistency in the analysis. The diversity of stakeholders is limited to the differences in their context and the purpose of their inclusion in a function, and is assumed as 1 for many cases due to limited information in the case studies. The number of means is limited to 1 due to the lack of information in the case studies regarding tool usage.

The findings—specifically the results of inclusivity and empowerment metrics—are derived from real-life case studies based on documented project descriptions. As such, some contextual or informal aspects of stakeholder interactions may not be fully captured in the analysis. Decision-making on the dimensions of empowerment and inclusivity is limited to the results obtained by applying the framework to ten real-life case studies. The lifecycle evaluation of products is limited by the scope and detail of the case studies. In addition, the impact of time-related factors, such as process duration, timing, or delays, is not evaluated in this study and is aimed for future work. Empowerment and inclusivity are evaluated in the case studies by considering ideal conditions. This means that the negative impacts of factors that could reduce empowerment or stakeholder inclusivity are not fully reflected. The results are accepted based on the completion of the process, at least up to production or prototyping. Therefore, the framework is applied under ideal conditions, which may absorb the negative impacts of certain factors on empowerment and inclusivity. However, factors that can reduce the level of inclusivity and empowerment may cause drawbacks with real-life implications. This can be mitigated by identifying which stakeholders need to be included in which functions and combinations. For example, if Stakeholder A cannot improve their empowerment level through interaction with Stakeholder B alone, A can be involved in a function with both Stakeholder B and D. To address this, the study suggests stakeholder development strategies and applies cosine similarity analysis in Phase 3 (see Figure 1) of the study.

This study evaluates the relationships between inclusivity and empowerment using Pearson correlation, which considers only linear relationships. Other possible non-linear or complex relationships were not explored in this phase of the research. Future studies could extend this by exploring different types of relationships. The case studies are only used as sources to identify functions, and function identification is limited to the developed framework. The sample size in some case studies is small, which may affect the strength of correlation results and generalisability. Additionally, the generalisability of the framework is tested in Yaldiz and Chakrabarti (2024b).

## 10. CONCLUSIONS AND FUTURE WORK

In conclusion, this study proposes the framework for the assessment of empowerment and inclusivity by operationalising their definitions. In order to ensure operationalisability, we identify the concepts from the definitions of inclusivity and empowerment and propose approaches to measure these concepts. The findings highlight the positive correlation between empowerment and inclusivity, which can be used to support the equal development of stakeholders by shifting the scope from inclusive design to inclusive lifecycle processes. Since the existing approaches do not systematically correlate empowerment and inclusivity across lifecycle processes, this framework addresses this gap.

While the proposed framework provides a structured way to evaluate stakeholder inclusivity and empowerment, its transformative impact depends on how these concepts are interpreted and supported in real-world contexts. A framework without a clear value orientation may risk formalising existing inequalities in a more systematic way. To be truly meaningful, the application should serve those who often face unequal outcomes in the systems it aims to improve. Therefore, the future use of the framework must consider how it can guide inclusive change, rather than only measure variable performance. Future research could also explore how the framework can be adapted to dynamic environments where stakeholder roles and interactions evolve in real time, and assess its influence on broader outcomes such as sustainability, dignity, and well-being. Additionally, an important direction is to identify the underlying patterns or ratios between empowerment and inclusivity that contribute to equal stakeholder development. In this context, it is also important to identify the success criteria of products (for example, in inclusive design, usability and accessibility), in order to assess the influence of equal stakeholder development on product success. This could be an impactful contribution to design science.

## References

Armstrong, D., Armstrong, A.C. and Spandagou, I. (2011) 'Inclusion: By choice or by chance?', *International Journal of Inclusive Education*, 15(1), pp. 29–39.

BIF (Business Innovation Facility Practitioner Hub) (2011) 'What is Inclusive Business', *Briefing Note 1*, Business Innovation Facility & Innovations Against Poverty. Available at: www.globalhand.org/system/assets/b60414e3e4b8d043b46c573849258cc53df86912/original/What_do_we_mean_by_IB_20April2011.pdf?1320206957 (Accessed: 26 March 2024).

BSI (2005) *BS 7000-6: Design management systems – Managing inclusive design*. London: BSI.

British Standards Institution (2005) *BS 7000-6:2005 – Design management systems. Guide to managing inclusive design*. London: British Standards Institution.

Cattaneo, L.B. and Chapman, A.R. (2010) 'The process of empowerment: A model for use in research and practice', *American Psychologist*, 65(7), pp. 646–659. https://doi.org/10.1037/a0018854.

Chen, X., Despeisse, M. and Johansson, B. (2020) 'Environmental sustainability of digitalization in manufacturing: A review', *Sustainability*, 12(24), p. 10298.

Colantonio, A. (2009) *Social sustainability: A review and critique of traditional versus emerging themes and assessment methods*. Oxford: Oxford Institute for Sustainable Development (OISD), Oxford Brookes University.

Clarkson, P.J. and Coleman, R. (2015) 'History of inclusive design in the UK', *Applied Ergonomics*, 46, pp. 235–247.

Clarkson, P.J., Waller, S. and Cardoso, C. (2015) 'Approaches to estimating user exclusion', *Applied Ergonomics*, 46 (Part B), pp. 304–310. http://dx.doi.org/10.1016/j.apergo.2013.03.001.

Clarkson, J. (2016) 'Countering design exclusion–theory and practice', in *Design for Inclusivity*. London: Routledge, pp. 165–180.

Cook, K.S. and Hegtvedt, K.A. (1983) 'Distributive justice, equity, and equality', *Annual Review of Sociology*, 9(1), pp. 217–241. https://doi.org/10.1146/annurev.so.09.080183.001245.

Cambridge University Press (2025) *Cambridge Dictionary*. Available at: https://dictionary.cambridge.org/dictionary/.

Dahl, R.A. (1957) 'The Concept of Power', *Behavioral Science*, 2(3), pp. 201–215.

Dubey, R. and Gunasekaran, A. (2015) 'Agile manufacturing: framework and its empirical validation', *The International Journal of Advanced Manufacturing Technology*, 76, pp. 2147–2157. https://doi.org/10.1007/s00170-014-6455-6.

Dempsey, N., Bramley, G., Power, S. and Brown, C. (2009) 'The social dimension of sustainable development: Defining urban social sustainability', *Sustainable Development*, 19(5), pp. 289–300. https://doi.org/10.1002/sd.417.

EC (2018) *Exclusion Calculator Inclusive Design Toolkit*. Available at: http://calc.inclusivedesigntoolkit.com/ (Accessed: 12 June 2024).

Emerson, R.M. (1962) 'Power-Dependence Relations', *American Sociological Review*, 27(1), pp. 31–41. https://doi.org/10.2307/2089716.

Fawcett, S.B., White, G.W., Balcazar, F.E., Suarez-Balcazar, Y., Mathews, R.M., Paine-Andrews, A. et al. (1994) 'A contextual-behavioral model of empowerment: Case studies involving people with physical disabilities', *American Journal of Community Psychology*, 22(4), pp. 471–496.

Fernandes, S.C., Martins, L.D. and Rozenfeld, H. (2019) 'Who are the Stakeholders Mentioned in Cases of Product-Service System (PSS) Design?', *Proceedings of the Design Society: International Conference on Engineering Design*, 1(1), pp. 3131–3140. https://doi.org/10.1017/dsi.2019.320.

Foucault, M. (1994) 'The Subject and Power', in Faubion, J.D. (ed.) *Power: Essential Works of Foucault 1954–84*. London: Penguin Books.

Ferguson, L.J. (2010) 'Transformational empowerment: Change your world from the inside out', *Interbeing*, 4(2), pp. 35–37.
Füller, J., Mühlbacher, H., Matzler, K. and Jawecki, G. (2009) 'Consumer Empowerment Through Internet-Based Co-creation', *Journal of Management Information Systems*, 26(3), pp. 71–102. https://doi.org/10.2753/MIS0742-122226030003.
Fuchs, C. and Schreier, M. (2011) 'Customer empowerment in new product development', *Journal of Product Innovation Management*, 28(1), pp. 17–32. https://doi.org/10.1111/j.1540-5885.2010.00778.x.
Goodland, R. (1995) 'The concept of environmental sustainability', *Annual Review of Ecology and Systematics*, pp. 1–24.
Gunasekaran, A., Yusuf, Y.Y., Adeleye, E.O., Papadopoulos, T., Kovvuri, D. and Geyi, D.A.G. (2019) 'Agile manufacturing: an evolutionary review of practices', *International Journal of Production Research*, 57(15–16), pp. 5154–5174. https://doi.org/10.1080/00207543.2018.1530478.
Gräßler, I. and Hesse, P. (2022) 'Approach to sustainability-based assessment of solution alternatives in early stages of product engineering', *Proceedings of the Design Society*, 2, pp. 1833–1842. https://doi.org/10.1017/pds.2022.186.
Hansen, J.H. (2012) 'Limits to inclusion', *International Journal of Inclusive Education*, 16(1), pp. 89–98.
Hernandez, R.J., Miranda, C. and Goñi, J. (2020) 'Empowering sustainable consumption by giving back to consumers the "right to repair"', *Sustainability*, 12(3). https://doi.org/10.3390/su12030850.
Hodge, T. (1997) 'Toward a conceptual framework for assessing progress toward sustainability', *Social Indicators Research*, 40(1), pp. 5–98. https://doi.org/10.1023/A:1006859511756.
ISO/IEC/IEEE (2023) ISO/IEC/IEEE 15288:2023 Systems and software engineering — System life cycle processes. Geneva: International Organization for Standardization.
Jolanta, M. A. J. (2015). Diversity management's stakeholders and stakeholders management. In *Proceedings of the 9th International Management Conference „Management and Innovation For Competitive Advantage* (pp. 780-793).
Jespersen, K.R. (2011) 'Online channels and innovation: Are users being empowered and involved?', *International Journal of Innovation Management*, 15(6), pp. 1141–1159.
Johnson, D., Clarkson, J. and Huppert, F. (2010) 'Capability measurement for inclusive design', *Journal of Engineering Design*, 21(2–3), pp. 275–288.
Jupp, D., Ali, S.I. and Barahona, C., 2010. *Measuring empowerment? Ask them*. Sida Studies in Evaluation, 1, pp.7–99.
Kleba, J.B. and Cruz, C. (2021) 'Empowerment, emancipation and engaged engineering', *International Journal of Engineering, Social Justice, and Peace*, 8(2), pp. 28–49. https://doi.org/10.24908/ijesjp.v8i2.14380.
Kirisci, P.T., Thoben, K.D., Klein, P. and Modzelewski, M. (2011) 'Supporting inclusive product design with virtual user models at the early stages of product development', *Proceedings of the 18th International Conference on Engineering Design (ICED 11)*, Copenhagen, Denmark.
Larsson, I. and Jormfeldt, H. (2017) 'Perspectives on power relations in human health and well-being', *International Journal of Qualitative Studies on Health and Well-being*, 12(sup2), p. 1358581.
Legood, A., van der Werff, L., Lee, A., den Hartog, D. & van Knippenberg, D. (2022) 'A Critical Review of the Conceptualization, Operationalization, and Empirical Literature on Cognition-Based and Affect-Based Trust', *Journal of Management Studies*. doi: 10.1111/joms.12811.
Li, F. and Dong, H. (2019) 'The economic explanation of inclusive design in different stages of product life time', *Proceedings of the Design Society: International Conference on Engineering Design*, 1(1), pp. 2377–2386.

Lyons, M., Smuts, C. and Stephens, A. (2001) ‘Participation, empowerment and sustainability: (How) do the links work?’, *Urban Studies*, 38(8), pp. 1233–1251. https://doi.org/10.1080/00420980120061007.
Maj, J.A., 2015. Diversity management’s stakeholders and stakeholders management. In: *Proceedings of the 9th International Management Conference: Management and Innovation for Competitive Advantage*. pp.780–793.
Max-Neef, M. (1992) ‘Development and Human Needs’, in Ekins, P. and Max-Neef, M. (eds.) *Real Life Economics: Understanding Wealth Creation*. London: Routledge, pp. 197–213.
Michelini, L. and Fiorentino, D. (2012) *New business models for social innovation: Creating shared value in low-income markets*. Springer.
Miller, R.L. and Campbell, R. (2006) ‘Taking stock of empowerment evaluation: An empirical review’, *American Journal of Evaluation*, 27(3), pp. 296–319. https://doi.org/10.1177/109821400602700303.
Moser, H.A. (2013) *Systems engineering, systems thinking, and learning: A case study in space industry*. Springer.
Ning, W., Goodman-Deane, J. and Clarkson, P.J. (2019) ‘Addressing cognitive challenges in design – a review on existing approaches’, *Proceedings of the Design Society: International Conference on Engineering Design*, 1(1), pp. 2775–2784.
Nijkamp, P. and Poot, J. (2015) *Cultural diversity: A matter of measurement* (IZA Discussion Paper No. 8782). Institute of Labor Economics (IZA). Available at: https://papers.ssrn.com/sol3/papers.cfm?abstract_id=2558361.
Ogawa, S. and Piller, F.T. (2006) ‘Reducing the risks of new product development’, *MIT Sloan Management Review*, 47(2).
Ozili, P.K. (2020) ‘Social inclusion and financial inclusion: International evidence’, *International Journal of Development Issues*, 19(2), pp. 169–186.
Ozili, P.K. (2022) ‘Sustainability and sustainable development research around the world’, *Managing Global Transitions*, 20, pp. 259–293. https://doi.org/10.26493/1854-6935.20.259-293.
Page, N. and Czuba, C.E. (1999) ‘Empowerment: What is it’, *Journal of Extension*, 37(5).
Patrick, V.M. and Hollenbeck, C.R. (2021) ‘Designing for all: Consumer response to inclusive design’, *Journal of Consumer Psychology*, 31(2), pp. 360–381.
Pardo del Val, M. and Lloyd, B., 2003. Measuring empowerment. *Leadership & Organization Development Journal*, 24(2), pp.102–108.
Pires, G.D., Stanton, J. and Rita, P. (2006) ‘The Internet, consumer empowerment and marketing strategies’, *European Journal of Marketing*, 40(9/10), pp. 936–949. https://doi.org/10.1108/03090560610680943.
Prahalad, C.K. (2005) *The Fortune at the Bottom of the Pyramid: Eradicating Poverty through Profits*. Upper Saddle River, NJ: Wharton School Publishing.
Pringle, J.K. and Ryan, I. (2015) ‘Understanding context in diversity management: A multi-level analysis’, *Equality, Diversity and Inclusion: An International Journal*, 34(6), pp. 470–482. https://doi.org/10.1108/EDI-05-2015-0031.
Quick, K.S. and Feldman, M.S. (2011) ‘Distinguishing participation and inclusion’, *Journal of Planning Education and Research*, 31(3), pp. 272–290.
Rawal, N. (2008) ‘Social inclusion and exclusion: A review’, *Dhaulagiri Journal of Sociology and Anthropology*, 2, pp. 161–180.
Roy, R. et al. (2018) ‘Inclusive manufacturing: What it means and how it can accelerate growth of India’, *RITES Journal*, 1(20).
Sarkar, P. and Chakrabarti, A., 2011. Assessing design creativity. *Design Studies*, 32(4), pp.348–383.
Sarmah, B. and Rahman, Z. (2017) ‘Transforming jewellery designing: Empowering customers through crowdsourcing in India’, *Global Business Review*, 18(5), pp. 1325–1344.

Shore, L.M. et al. (2011) ‘Inclusion and diversity in work groups: A review and model for future research’, *Journal of Management*, 37(4), pp. 1262–1289.
Sianipar, C.P.M. et al. (2013) ‘Community empowerment through appropriate technology: Sustaining the sustainable development’, *Procedia Environmental Sciences*, 17, pp. 1007–1016. https://doi.org/10.1016/j.proenv.2013.02.120.
Stough, S., Eom, S. and Buckenmyer, J. (2000) ‘Virtual teaming: A strategy for moving your organisation into the new millennium’, *Industrial Management & Data Systems*, 100(8), pp. 370–378.
Surma-aho, A. & Holtta-Otto, K. (2022) ‘Conceptualization and operationalization of empathy in design research’, *Design Studies*, 78, p. 101075. doi: 10.1016/j.destud.2021.101075.
Urakami, J. and Vajna, S. (2018) ‘Human centricity in integrated design engineering’, *Proceedings of the DESIGN 2018 15th International Design Conference*, Dubrovnik, Croatia. https://doi.org/10.21278/idc.2018.0154.
Vandekerckhove, P., Timmermans, J., de Bont, A., & de Mul, M. (2023). Diversity in stakeholder groups in generative co-design for digital health: assembly procedure and preliminary assessment. *JMIR Human Factors*, *10*, e38350.
Van Dam, A. (2019) ‘Diversity and its decomposition into variety, balance and disparity’, arXiv preprint. https://arxiv.org/abs/1902.09167.
Vallance, S., Perkins, H.C. and Dixon, J.E. (2011) ‘What is social sustainability? A clarification of concepts’, *Geoforum*, 42(3), pp. 342–348. https://doi.org/10.1016/j.geoforum.2011.01.002.
Wach, E. (2012) ‘Measuring the “inclusivity” of inclusive business’, *IDS Practice Papers*, 2012(9), pp. 1–30.
Waller, S. et al. (2015) ‘Making the case for inclusive design’, *Applied Ergonomics*, 46, pp. 297–303. https://doi.org/10.1016/j.apergo.2013.03.012.
Wasserman, I.C., Gallegos, P.V. and Ferdman, B.M. (2008) ‘Dancing with resistance: Leadership challenges in fostering a culture of inclusion’, in Thomas, K.M. (ed.) *Diversity Resistance in Organizations*, pp. 175–200.
Yaldiz, N. and Chakrabarti, A. (2024a) ‘Assessment of empowerment via inclusion of people in product lifecycle processes’, *Proceedings of the Design Society*, 4, pp. 1527–1536. https://doi.org/10.1017/pds.2024.155.
Yaldiz, N. and Chakrabarti, A. (2024b) ‘Assessment of sustainability value and dignified well-being in inclusive product lifecycle’, arXiv. https://arxiv.org/abs/2412.18583.
Zallio, M. and Clarkson, P.J. (2022) ‘The Inclusive Design Canvas: A strategic design template for architectural design professionals’, *Proceedings of the Design Society*, 2, pp. 81–90. https://doi.org/10.1017/pds.2022.9.
Zimmermann, A., Lorenz, A. and Oppermann, R. (2007) ‘An operational definition of context’, in Kokinov, B. et al. (eds.) *Modeling and Using Context* (Lecture Notes in Computer Science, Vol. 4635). Springer, pp. 558–571. https://doi.org/10.1007/978-3-540-74255-5_42.
Zhang, X. et al. (2022) ‘Avocado consumption for 12 weeks and cardiometabolic risk factors: A randomized controlled trial’, *The Journal of Nutrition*, 152(8), pp. 1851–1861. https://doi.org/10.1093/jn/nxac126.
Wilcox, R.R. and Rousselet, G.A. (2023). *An updated guide to robust statistical methods in neuroscience*. Current Protocols, 3, e719. https://doi.org/10.1002/cpz1.719**:contentReference[oaicite:0]{index=0}.

**Supplementary File 1**

This supplementary file is also used to support Yaldiz and Chakrabarti (2024b).

**Note:** All entries in the “Definition/Quote” column are direct excerpts from the referenced sources.

Quotation marks have been omitted for visual clarity.

| No | Domain | Definition/Quote | References |
|---|---|---|---|
| 1 | Psychology | A formal definition and four evaluative constructs of psychosocial inclusivity in design are described: Cognitive, Emotional, Social and Value. | Lim, Y., Giacomin, J., & Nickpour, F. (2021). What is psychosocially inclusive design? A definition with constructs. The Design Journal, 24(1), 5-28. |
| 2 | Design/ Engineering | Inclusive design meets the needs of people of all abilities. It extends and preserves independence but it need not be expensive. For products to be inclusively designed, all stakeholders need to be informed and active: demanding consumers, influential consumer organisations, creative designers and design colleges, even reluctant manufacturers, legislators and standards-makers. | Etchell, L., & Yelding, D. (2004). Inclusive design: products for all consumers. Consumer Policy Review, 14(6), 186-193. |
| 3 | Design/ Engineering | Inclusive design at its simplest means designing for as many people as possible, taking into account the diversity of their abilities. | Etchell, L., & Yelding, D. (2004). Inclusive design: products for all consumers. Consumer Policy Review, 14(6), 186-193. |
| 4 | Design/ Engineering | Inclusive design should mean that the need for assistive products or adaptations is reduced. | Etchell, L., & Yelding, D. (2004). Inclusive design: products for all consumers. Consumer Policy Review, 14(6), 186-193. |
| 5 | Design/ Engineering | Inclusive design is a design philosophy that aims to consider the needs of people with reduced functional capability during the design process. The goal is to design consumer products that are accessible to as many people as possible, without being stigmatising or resorting to special aids and adaptations [25]. | Persad, U., Langdon, P., & Clarkson, J. (2007). Characterising user capabilities to support inclusive design evaluation. Universal Access in the Information Society, 6, 119-135. |
| 6 | Design/ Engineering | Inclusive design rejects the notion of the 'average user' as a myth, and embraces diversity in users' sensory, cognitive and motor characteristics. Though ageing, disease and trauma account for this diversity, the characterisations of users based on levels of capability to perform actions in the world appear to be the most appropriate for product design and evaluation [9, 21, 52]. | Persad, U., Langdon, P., & Clarkson, J. (2007). Characterising user capabilities to support inclusive design evaluation. Universal Access in the Information Society, 6, 119-135. |
| 7 | Design/ Engineering | In order to predict exclusion, it is necessary to determine if a task demands fall outside<br>the user's performance envelope as previously described. | Persad, U., Langdon, P., & Clarkson, J. (2007). Characterising user capabilities to support inclusive design evaluation. Universal Access in the Information Society, 6, 119-135. |
| 8 | Design/ Engineering | inclusive design considers the needs and capabilities of the whole population and is based on the assumption that considering<br>the full diversity of users will result in a better product (Johnson, Clarkson, & Huppert, 2010). | Patrick, V. M., & Hollenbeck, C. R. (2021). Designing for all: Consumer response to inclusive design. Journal of consumer psychology, 31(2), 360-381. |
| 9 | Design/ Engineering | Inclusive design does not aim to change, but rather to improve the design of products and environments by making them more usable, safe, and appealing to people with a wide range of abilities (Carroll & Kincade, 2007). | Patrick, V. M., & Hollenbeck, C. R. (2021). Designing for all: Consumer response to inclusive design. Journal of consumer psychology, 31(2), 360-381. |
| 10 | Design/ Engineering | Inclusive design, according to design expert Kat Holmes, is simply the opposite of exclusion. | Patrick, V. M., & Hollenbeck, C. R. (2021). Designing for all: Consumer response to inclusive design. Journal of consumer psychology, 31(2), 360-381. |
| 11 | Design/ Engineering | The implementation of inclusive design needs adherence to the following five basic principles (adapted from CABE 2008) in order to provide all<br>people with dignity, comfort, and convenience, as well as the ability to be independent and participate equally in activities:<br>• People—Place people at the heart of the design process.<br>• Diversity—Acknowledge diversity and individual difference.<br>• Choices—Offer choices where a single design solution cannot accommodate all users.<br>• Flexibility—Ensure flexibility in use.<br>• Convenience—Ensure that design products are convenient and enjoyable to use for everyone. | Patrick, V. M., & Hollenbeck, C. R. (2021). Designing for all: Consumer response to inclusive design. Journal of consumer psychology, 31(2), 360-381. |
| 12 | Design/ Engineering | we use the term inclusive design and conceptualize it as the actual or perceived match between the user and the design object. We propose that when the identity of the user actually does not—or feels as if it does not—map onto the form and function of the design object, the user is excluded. As such, inclusive design makes products, services, spaces, and experiences not only effective and easy to use, but also emotionally positive, accessible to, and usable by people with the widest range of abilities within the widest range of situations without the need for special adaptation or accommodation. In sum, inclusive design does not reject users because of who they are, but instead works with excluded users to create solutions that work well and benefit everyone. | Patrick, V. M., & Hollenbeck, C. R. (2021). Designing for all: Consumer response to inclusive design. Journal of consumer psychology, 31(2), 360-381. |
| 13 | Design/ Engineering | identity is multidimensional (Oyserman & Schwarz, 2020) and includes a combination of age, developmental and acquired disabilities, religion, ethnicity, socioeconomic status, sexual orientation, indigenous heritage, national origin, and gender, we use the ADDRESSING framework, first defined by Hays (2008), to capture the full range of identity considerations that can inform inclusive design (see Table 2). The ADDRESSING framework can help designers understand the scope of possible users and the complexities of human experience and identity (Hays, 1996, 2008). | Patrick, V. M., & Hollenbeck, C. R. (2021). Designing for all: Consumer response to inclusive design. Journal of consumer psychology, 31(2), 360-381. |
| 14 | Design/ Engineering | inclusive design does not focus solely on a process, the designer, the user, or adapt ability for the sake of profits alone; rather, inclusivity involves a moral responsibility and an ethical commitment to decreasing the mismatch between a user and the design object. | Patrick, V. M., & Hollenbeck, C. R. (2021). Designing for all: Consumer response to inclusive design. Journal of consumer psychology, 31(2), 360-381. |
| 15 | Design/ Engineering | a three-dimensional framework for inclusive design: (a) Recognize, respect, and design for human uniqueness and variability; (b) Use inclusive, open, and transparent processes, and codesign with people who hold a diversity of perspectives, including those who can't use or have difficulty using the current designs; and (c) Realize that you are designing in a complex adaptive system (Treviranus, 2018). | Patrick, V. M., & Hollenbeck, C. R. (2021). Designing for all: Consumer response to inclusive design. Journal of consumer psychology, 31(2), 360-381. |
| 16 | Design/ Engineering | Inclusive design demands diverse teams that are committed to empathizing with consumers and exploring all the elements and possibilities that make up an activity (Etchell & Yelding, 2004). | Patrick, V. M., & Hollenbeck, C. R. (2021). Designing for all: Consumer response to inclusive design. Journal of consumer psychology, 31(2), 360-381. |

| 17 | Design/ Engineering | Mismatches between products and their users represent the "building blocks of exclusion. They can feel like  little moments of exasperation when a technology product doesn't work the way we think it should. Or they can feel like running into a locked door marked with a big sign that says Keep Out" (Holmes, p. 2). | Patrick, V. M., & Hollenbeck, C. R. (2021). Designing for all: Consumer response to inclusive design. Journal of consumer psychology, 31(2), 360-381. |
|---|---|---|---|
| 18 | Design/ Engineering | inclusivity is about creating the best possible match between the environment and all potential users, recognizing that inclusion and beautification improve the overall experience for everyone. Indeed, numerous inventions that we all use today are the direct or indirect result of inclusive design: touchscreens, flexible straws, closed captioning, automatic doors, curb cuts, audiobooks, and speech-to-text software to name a few. | Patrick, V. M., & Hollenbeck, C. R. (2021). Designing for all: Consumer response to inclusive design. Journal of consumer psychology, 31(2), 360-381. |
| 19 | Design/ Engineering | the best inclusive designs are user-led and aim for usability by the greatest number of people, including those with disabilities. | Patrick, V. M., & Hollenbeck, C. R. (2021). Designing for all: Consumer response to inclusive design. Journal of consumer psychology, 31(2), 360-381. |
| 20 | Design/ Engineering | conceptualizing inclusivity using three levels (see Figure 2) based on the degree of mismatch between the individual and the designed object (e.g., a product, service, or space). Level 1 Inclusive Design merely addresses accessibility, typically reflecting industry regulations. Level 2 Inclusive Design incorporates greater engagement and evokes positive emotions from consumers, while Level 3 Inclusive Design represents the ideal wherein there is little or no mismatch between any consumer and the designed object. | Patrick, V. M., & Hollenbeck, C. R. (2021). Designing for all: Consumer response to inclusive design. Journal of consumer psychology, 31(2), 360-381. |
| 21 | Design/ Engineering | In particular, where some users are excluded from using a product or service, many more are likely to find it difficult or frustrating to use (Clarkson et al., 2003). Hence, inclusive design has become synonymous with good design, where accessibility, as an extension of usability, can lead to increased commercial success. | Langdon, P., Johnson, D., Huppert, F., & Clarkson, P. J. (2015). A framework for collecting inclusive design data for the UK population. Applied ergonomics, 46, 318-324. |

| 22 | Design/ Engineering | barriers remain to industry uptake of the ethos of inclusive design (Dong et al., 2004) in the form of: the lack of a perceived justifiable business case to support inclusive design;<br>the lack of knowledge and tools to practice inclusive design; a lack of understanding of the difficulties experienced by users of new technology products; poor access to appropriate user sets. | Langdon, P., Johnson, D., Huppert, F., & Clarkson, P. J. (2015). A framework for collecting inclusive design data for the UK population. Applied ergonomics, 46, 318-324. |
|---|---|---|---|
| 23 | Design/ Engineering | an inclusive design approach for bringing disabled people into mainstream life by improving product design. In order to predict product exclusion and difficulty, recent accurate data on capability in the population, from post-census, disability surveys, has been combined with models of human product interaction. The research to do this has generated new models and methods for assessing levels of design inclusion. | Langdon, P., Johnson, D., Huppert, F., & Clarkson, P. J. (2015). A framework for collecting inclusive design data for the UK population. Applied ergonomics, 46, 318-324. |
| 24 | Design/ Engineering | There is an increasingly urgent need to design products to accommodate the rising prevalence of functional impairments, both at more extreme levels,such as dementia,and at the extension to the mainstream, for example, in elderly friendly ICT. Analytical inclusive design is nowsupported bya plethora of tools, techniques and resources but there is a dearth of data sets that can support quantitative prediction of performance resulting from the way in which capability interacts with the demand made by product features. | Langdon, P., Johnson, D., Huppert, F., & Clarkson, P. J. (2015). A framework for collecting inclusive design data for the UK population. Applied ergonomics, 46, 318-324. |
| 25 | Design/ Engineering | "Inclusive Design as neither a new genre of design, nor a separate specialism, but as a general approach to designing in which designers ensure that their products and services address the needs of the widest possible audience, irrespective of age or ability. Inclusive Design (also known [in Europe] as Design for All and as Universal Design in the USA) is in essence the inverse of earlier approaches to designing for disabled and elderly people as a sub-set of the population, and an integral part of a more recent international trend towards the integration of older and disabled people in the mainstream of society. " (Design Council, 2008). | Clarkson, P. J., & Coleman, R.<br>(2015). History of inclusive design in the UK. Applied ergonomics, 46, 235-247. |
| 26 | Design/ Engineering | Inclusive Design is in essence the inverse of earlier approaches to designing for disabled and elderly people as a sub-set of the population, and an integral part of a more recent international trend towards the integration of older and disabled people in the mainstream of society. | Clarkson, P. J., & Coleman, R.<br>(2015). History of inclusive design in the UK. Applied ergonomics, 46, 235-247. |
| 27 | Design/ Engineering | By using the Inclusive Design Cube model and the user-aware design that it promotes, designers can better understand users' capabilities and create intuitive interfaces, easy-open packaging, well structured, logical and clear signage, power assisted steering and braking, and many other products that are regularly taken for granted. Modular and customisable designs can greatly enhance usability. | Clarkson, P. J., & Coleman, R. (2015). History of inclusive design in the UK. Applied ergonomics, 46, 235-247. |
| 28 | Design/ Engineering | Inclusively designed products, services and environments will likely benefit the whole population and can lead to an increase of business for companies. | Clarkson, P. J., & Coleman, R. (2015). History of inclusive design in the UK. Applied ergonomics, 46, 235-247. |
| 29 | Design/ Engineering | The economic case for Inclusive Design is built on two key factors. First, the Potential Support Ratio (PSR) e the numberof people aged 15-64 who could support one person over 65 e is declining rapidly, particularly in the developed world, while care costs as aproportion of GDP are escalating. Second, without an effective consumeroffer addressing people's lifestyle, needs and aspirations, older people in particular will have little incentive to spend what disposable income they have, removing what could be a significant economic driver. Inclusive Design, especially in the workplace, offers the possibility for older and disabled people to enter or continue work in gainful employment and, therefore, extend in dependent living, which in turn can contribute to lowering care costs and help stimulate the economy. | Clarkson, P. J., & Coleman, R. (2015). History of inclusive design in the UK. Applied ergonomics, 46, 235-247. |
| 30 | Design/ Engineering | In addition, there is a strong social case for Inclusive Design based on: (1) the desirability of social cohesion and inclusivity and (2) the accessibility of public buildings, spaces and services, which can promote social inclusion. | Clarkson, P. J., & Coleman, R. (2015). History of inclusive design in the UK. Applied ergonomics, 46, 235-247. |
| 31 | Design/ Engineering | "fresh approaches of the kind referred to here are needed to bridge the present gulf between main stream design and design for the elderly, especially with regard to the scale of demographic change. The concept of Inclusive Design coupled with story telling and scenario building techniques can turn what is often considered as a branch of design for disability into an exciting gateway to product innovation and a more user-friendly future for all. As attitudes towards ageing and disability change an important role is emerging for ergonomics in the design and assessment of everyday products and environments, to ensure that they allow for the broadest possible range of abilities in their user profiles" (Coleman, 1994). | Clarkson, P. J., & Coleman, R. (2015). History of inclusive design in the UK. Applied ergonomics, 46, 235-247. |
| 32 | Design/ Engineering | 'a simple concept to help them [manufacturers and retailers] see potential commercial benefits for their businesses', along with 'examples of how that concept could be applied in practice'. | Clarkson, P. J., & Coleman, R. (2015). History of inclusive design in the UK. Applied ergonomics, 46, 235-247. |
| 33 | Design/ Engineering | to measure design exclusion, the project realised the multi-faceted nature of exclusion: the combination of capability levels demanded by products and services across a variety of environmental contexts and the diversity of capability across the whole population and particularly across the lifespan. This led to a summative approach tothe measureof design exclusion, to account for the multiple, minor capacity deficiencies typical of ageing populations, and revealed important differences between the 'disabled' population and older adults. | Clarkson, P. J., & Coleman, R. (2015). History of inclusive design in the UK. Applied ergonomics, 46, 235-247. |

| 34 | Design/ Engineering | The concept of design exclusion as a quantifiable aspect of products and services is unique to Inclusive Design and differentiates it from the more aspirational Universal Design and Design for All approaches by assuming as a start point that no design will work perfectly for everyone. | Clarkson, P. J., & Coleman, R. (2015). History of inclusive design in the UK. Applied ergonomics, 46, 235-247. |
|---|---|---|---|
| 35 | Design/ Engineering | Inclusive Design based on countering design exclusion. By identifying users vulnerable to design exclusion and designing to accommodate their capability levels, products can often be improved for all users and thus compete more effectively and appeal to a wider range of consumers. | Clarkson, P. J., & Coleman, R. (2015). History of inclusive design in the UK. Applied ergonomics, 46, 235-247. |
| 36 | Design/ Engineering | Inclusive Design can therefore be seen as an iterative process of knowledge acquisition leading tocontinuous designimprovement, increased customer satisfaction and hence brand loyalty. | Clarkson, P. J., & Coleman, R. (2015). History of inclusive design in the UK. Applied ergonomics, 46, 235-247. |
| 37 | Design/ Engineering | Inclusive Design approaches for new product and service development, incorporating better data for designers that will result in products and services that facilitate independence at home, at work, and in other environments and contexts of use (Persadetal.,2007;EltonandNicolle, 2010). To achieve this there has been a need to: (1) understand the capability demand made by a product within its operating environment; (2) define a specification for and collect new population based capability data; (3) calculate levels of product exclusion and difficulty; and (4) present such data in anaccessible anduseful way. | Clarkson, P. J., & Coleman, R. (2015). History of inclusive design in the UK. Applied ergonomics, 46, 235-247. |
| 38 | Design/ Engineering | The conceptof Inclusive Design, as described by the matching of product demandstousercapabilities in a given environment is also transferable to other domains. | Clarkson, P. J., & Coleman, R. (2015). History of inclusive design in the UK. Applied ergonomics, 46, 235-247. |
| 39 | Design/ Engineering | Good design is about making conscious and well-informed decisions throughout the designprocess. A great product or service is typically built on a foundation of understanding the real needs of the userandother stakeholders. In short, good design and Inclusive Design should be seen as inseparable and essential. | Clarkson, P. J., & Coleman, R. (2015). History of inclusive design in the UK. Applied ergonomics, 46, 235-247. |

| 40 | Design/ Engineering | inclusive design applies an understanding of customer diversity to inform decisions throughout the development process, in order to better satisfy the needs of more people. Products that are more inclusive can reach a wider market, improve customer satisfaction and drive business success. | Waller, S., Bradley, M., Hosking, I., & Clarkson, P. J. (2015). Making the case for inclusive design. Applied ergonomics, 46, 297-303. |
|---|---|---|---|
| 41 | Design/ Engineering | Every design decision has the potential to include or exclude customers. Inclusive design emphasises the contribution that understanding user diversity makes to informing these decisions. | Waller, S., Bradley, M., Hosking, I., & Clarkson, P. J. (2015). Making the case for inclusive design. Applied ergonomics, 46, 297-303. |
| 42 | Design/ Engineering | Understanding diversity within the population. Responding to this diversity with informed design decisions | Waller, S., Bradley, M., Hosking, I., & Clarkson, P. J. (2015). Making the case for inclusive design. Applied ergonomics, 46, 297-303. |
| 43 | Design/ Engineering | Inclusive design does not suggest that it is always possible (or appropriate) to design one product to address the needs of the entire population. Instead, inclusive design guides an appropriate design response to diversity in the population through:<br>Developing a portfolio of products and derivatives to provide the best possible coverage of diversity within the 'Population Pyramid'.<br>Ensuring that each individual product has clear and distinct target users.<br>Making informed decisions to improve the success criteria for each product. | Waller, S., Bradley, M., Hosking, I., & Clarkson, P. J. (2015). Making the case for inclusive design. Applied ergonomics, 46, 297-303. |
| 44 | Design/ Engineering | inclusive design applies an understanding of customer diversity to the design of mainstream products, in order to better satisfy the needs of more people and deliver commercial success. | Waller, S., Bradley, M., Hosking, I., & Clarkson, P. J. (2015). Making the case for inclusive design. Applied ergonomics, 46, 297-303. |
| 45 | Design/ Engineering | Inclusive design is about choosing an appropriate target population for a particular design, and making informed decisions to maximise the success criteria for the target market. | Waller, S., Bradley, M., Hosking, I., & Clarkson, P. J. (2015). Making the case for inclusive design. Applied ergonomics, 46, 297-303. |

| 46 | Design/ Engineering | Why design inclusively? Having introduced and defined inclusive design, this section<br> contains the business case. The key messages are:- Ageingpopulations lead to growingopportunities for inclusive products.- Inclusive design mitigates business risk.- Simplicity can offer competitive advantage | Waller, S., Bradley, M., Hosking, I., &<br>Clarkson, P. J. (2015). Making the case for inclusive design. Applied ergonomics, 46, 297-303. |
|---|---|---|---|
| 47 | Design/ Engineering | products that are more inclusive can reach a wider market, improve customer satisfaction and drive business success, especially given the ageing population. | Waller, S., Bradley, M., Hosking, I., &<br>Clarkson, P. J. (2015). Making the case for inclusive design. Applied ergonomics, 46, 297-303. |
| 48 | Design/ Engineering | successful in clusive design requires informed decision-making at the concept stage, because it can become prohibitively expensive to make changes later on. The process of concept generation is summarised here according to the core activities of exploration, creativity, evaluation and project management.<br>- Whatare the needs?- Howcan the needs be met?- Howwell are the needs met?- Whatshould we do next? | Waller, S., Bradley, M., Hosking, I., &<br>Clarkson, P. J. (2015). Making the case for inclusive design. Applied ergonomics, 46, 297-303. |
| 49 | Design/ Engineering | the specific emphases for inclusive design are:<br>Understanding the true diversity of user and business needs.<br>Applying this knowledge to better inform the design decisions taken throughout the development process.<br>Evaluating rough prototypes with real users, before all the important decisions are finalised. | Waller, S., Bradley, M., Hosking, I., &<br>Clarkson, P. J. (2015). Making the case for inclusive design. Applied ergonomics, 46, 297-303. |
| 50 | Design/ Engineering | The ‘Design Wheel’ shows the specific activities that help to answer the four fundamental questions of concept design.  - Whatare the needs?- Howcan the needs be met?- Howwell are the needs met?- Whatshould we do next? the particularly critical activities, namely ‘observe users’, ‘generate personas’, ‘test with experts’, ‘test with users’ and ‘estimate exclusion’. | Waller, S., Bradley, M., Hosking, I., &<br>Clarkson, P. J. (2015). Making the case for inclusive design. Applied ergonomics, 46, 297-303. |
| 51 | Design/ Engineering | Estimating exclusion identifies the task steps where a product or prototype places the highest demands on the following user capabilities:<br>Vision<br>Hearing<br>Thinking<br>Reach and Dexterity<br>Mobility.<br>The process of estimating exclusion highlights the causes of frustration and exclusion, and prioritises these on a population basis. An exclusion calculator is freely available on the Inclusive Design Toolkit website (Clarkson et al., 2011). | Waller, S., Bradley, M., Hosking, I., &<br>Clarkson, P. J. (2015). Making the case for inclusive design. Applied ergonomics, 46, 297-303. |
| 52 | Design/ Engineering | At the heart of inclusive design is a better understanding of diversity in the population and its relevance for design. | Hosking, I., Waller, S., & Clarkson, P.<br>J. (2010). It is normal to be different: Applying inclusive design in industry. Interacting with computers, 22(6), 496-501. |
| 53 | Design/ Engineering | The aim of inclusive design is to extend the reach of mainstream products as far up the pyramid as possible, while maintaining commercial viability, and without compromising the design for those at the bottom of the pyramid | Hosking, I., Waller, S., & Clarkson, P.<br>J. (2010). It is normal to be different: Applying inclusive design in industry. Interacting with computers, 22(6), 496-501. |

| 54 | Design/ Engineering | The aim of inclusive product design is to successfully integrate human factors in the product development process with the intention of making products accessible for the largest possible group of users. | Kirisci, P. T., Thoben, K. D., Klein, P., & Modzelewski, M. (2011). Supporting inclusive product design with virtual user models at the early stages of product development. In DS 68-9: Proceedings of the 18th International Conference on Engineering Design (ICED 11), Impacting Society through Engineering Design, Vol. 9: Design Methods and Tools pt. 1, Lyngby/Copenhagen, Denmark, 15.-19.08. 2011 (pp. 80-90). |
|---|---|---|---|
| 55 | Design/ Engineering | Inclusive design is a process that results in inclusive products or environments which can be used by everyone regardless of age, gender or disability [1]. The main barriers for adopting inclusive product design include technical complexity, lack of time, lack of knowledge and techniques, and lack of guidelines [2]. | Kirisci, P. T., Thoben, K. D., Klein, P., & Modzelewski, M. (2011). Supporting inclusive product design with virtual user models at the early stages of product development. In DS 68-9: Proceedings of the 18th International Conference on Engineering Design (ICED 11), Impacting Society through Engineering Design, Vol. 9: Design Methods and Tools pt. 1, Lyngby/Copenhagen, Denmark, 15.-19.08. 2011 (pp. 80-90). |

| 56 | Design/ Engineering | According to Clarkson and Coleman (2015), inclusive design emerged "(…) not as a new approach to design, but rather as a synthesis of initiatives, experiments and insights dating back to the 1960s and beyond". | Carli Lorenzini, G., & Olsson, A. (2015). Design towards better life experience: closing the gap between pharmaceutical packaging design and elderly people. In DS 80-9 Proceedings of the 20th International Conference on Engineering Design (ICED 15) Vol 9: User-Centred Design, Design of Socio-Technical systems, Milan, Italy, 27-30.07. 15 (pp. 065-076). |
|---|---|---|---|
| 57 | Design/ Engineering | As stated by the European Design for All e-Accessibility Network (EDeAN, 2007), inclusive design is a "process-driven approach whereby designers and industry ensure that products and services address the needs of the widest possible consumer base, regardless of age or ability. Emphasis is placed on working with critical users' to stretch design brief". | Carli Lorenzini, G., & Olsson, A. (2015). Design towards better life experience: closing the gap between pharmaceutical packaging design and elderly people. In DS 80-9 Proceedings of the 20th International Conference on Engineering Design (ICED 15) Vol 9: User-Centred Design, Design of Socio-Technical systems, Milan, Italy, 27-30.07. 15 (pp. 065-076). |
| 58 | Design/ Engineering | Inclusive design works in respect to the full range of human diversity in terms of ability, language, culture, gender, age and other forms of human difference (Inclusive Design Institute, 2014). | Carli Lorenzini, G., & Olsson, A. (2015). Design towards better life experience: closing the gap between pharmaceutical packaging design and elderly people. In DS 80-9 Proceedings of the 20th International Conference on Engineering Design (ICED 15) Vol 9: User-Centred Design, Design of Socio-Technical systems, Milan, Italy, 27-30.07. 15 (pp. 065-076). |
| 59 | Design/ Engineering | What universal design, inclusive design and design for all have in common is the principle of integration of elderly and disabled people into mainstream society, instead of viewing them as sub-sets of the population (Clarkson and Coleman, 2015). These three design approaches share similar issues and interests, and have clarified that there are possibilities to achieve market goals through a more social orientation | Carli Lorenzini, G., & Olsson, A. (2015). Design towards better life experience: closing the gap between pharmaceutical packaging design and elderly people. In DS 80-9 Proceedings of the 20th International Conference on Engineering Design (ICED 15) Vol 9: User-Centred Design, Design of Socio-Technical systems, Milan, Italy, 27-30.07. 15 (pp. 065-076). |
| 60 | Design/ Engineering | the implementation of inclusivity struggles with lack of time, budget limitations, lack of knowledge and tools for practicing, and lack of perception of inclusive design as a need for the end user (Goodman et al., 2006). | Carli Lorenzini, G., & Olsson, A. (2015). Design towards better life experience: closing the gap between pharmaceutical packaging design and elderly people. In DS 80-9 Proceedings of the 20th International Conference on Engineering Design (ICED 15) Vol 9: User-Centred Design, Design of Socio-Technical systems, Milan, Italy, 27-30.07. 15 (pp. 065-076). |
| 61 | Design/ Engineering | The capability-demand approach provides a more pragmatic way of understanding the interplay between users' cognition and product features, because it directly relates the two within a product interaction context. According to this approach, products place demands on their users' capabilities. A user whose capability does not meet a demand will not be able to use the product effectively and can be considered to be 'excluded' from its use (Clarkson et al., 2015). | Ning, W., Goodman-Deane, J., & Clarkson, P. J. (2019, July). Addressing cognitive challenges in design–a review on existing approaches. In Proceedings of the design society: International conference on engineering design (Vol. 1, No. 1, pp. 2775-2784). Cambridge University Press. |
| 62 | Design/ Engineering | This approach (The capability-demand approach ) originates from the field of inclusive design. Inclusive design is ‘the design of mainstream products and/or services that are accessible to, and usable by, people with the widest range of abilities within the widest range of situations without the need for special adaptation or design’ (BSI, 2005). | Ning, W., Goodman-Deane, J., & Clarkson, P. J. (2019, July). Addressing cognitive challenges in design–a review on existing approaches. In Proceedings of the design society: International conference on engineering design (Vol. 1, No. 1, pp. 2775-2784). Cambridge University Press. |
| 63 | Design/ Engineering | Inclusive design employs commonly-used HCD methods, but has also developed additional design supports particularly for tackling the incompatibility between products’ demands and users’ capabilities. The capability-demand approach is one example of this. In this approach, population-based capability data is used to assess design exclusion. | Ning, W., Goodman-Deane, J., & Clarkson, P. J. (2019, July). Addressing cognitive challenges in design–a review on existing approaches. In Proceedings of the design society: International conference on engineering design (Vol. 1, No. 1, pp. 2775-2784). Cambridge University Press. |
| 64 | Design/ Engineering | ‘The design of mainstream products and/or services that are accessible to, and usable by, as many people as reasonably possible ... without the need for special adaptation or specialised design (BSI, 2005)’. | Li, F., & Dong, H. (2019, July). The economic explanation of inclusive design in different stages of product life time. In Proceedings of the Design Society: International Conference on Engineering Design (Vol. 1, No. 1, pp. 2377-2386). Cambridge University Press. |
| 65 | Design/ Engineering | Design exclusion occurs when the capability required for using a product (in a broad sense) exceeds the user’s actual capability (Clarkson and Keates, 2002). The purpose of inclusive design is often to eliminate the barriers and exclusion, so that products, services and built environment could meet as diverse needs as possible. | Li, F., & Dong, H. (2019, July). The economic explanation of inclusive design in different stages of product life time. In Proceedings of the Design Society: International Conference on Engineering Design (Vol. 1, No. 1, pp. 2377-2386). Cambridge University Press. |
| 66 | Design/ Engineering | barriers, such as disability; and older people often have one or more types of physical capability loss or decline, which leads to difficulties when using products designed only for ‘ordinary people’ or ‘mainstream people’ (Hitchcock et al., 2001). | Li, F., & Dong, H. (2019, July). The economic explanation of inclusive design in different stages of product life time. In Proceedings of the Design Society: International Conference on Engineering Design (Vol. 1, No. 1, pp. 2377-2386). Cambridge University Press. |

| 67 | Design/ Engineering | The BSI definition of inclusive design is given at the beginning of this paper, from which we can  derive design exclusion as: some people’s needs are not considered from the existing mainstream products, services and environment, and they are not satisfied. | Li, F., & Dong, H. (2019, July). The economic explanation of inclusive design in different stages of product life time. In Proceedings of the Design Society: International Conference on Engineering Design (Vol. 1, No. 1, pp. 2377-2386). Cambridge University Press. |
|---|---|---|---|
| 68 | Design/ Engineering | inclusive design hopes to bridge the inequality gap caused by diversity and enable as many users as possible to gain the value and experience provided by the design; on the other hand, the universal design makes diversified users reduce diversity to universality and reduce the richness of choice through a materialized uniqueness (Winance, 2014). | Li, F., & Dong, H. (2019, July). The economic explanation of inclusive design in different stages of product life time. In Proceedings of the Design Society: International Conference on Engineering Design (Vol. 1, No. 1, pp. 2377-2386). Cambridge University Press. |
| 69 | Design/ Engineering | study the inclusion and exclusion from the law of human behaviour, and focus on analysing the mechanism that affects inclusion/exclusion and comparing different situations. | Li, F., & Dong, H. (2019, July). The economic explanation of inclusive design in different stages of product life time. In Proceedings of the Design Society: International Conference on Engineering Design (Vol. 1, No. 1, pp. 2377-2386). Cambridge University Press. |

| 70 | Design/ Engineering | There is a need for responsible engineering design to accommodate the diverse user requirements that come with the global phenomenon of population ageing. Inclusive design can address these diverse requirements through the consideration of a wide diversity of user needs within the design process. | Wilson, N., Thomson, A., Thomson, A., & Holliman, A. F. (2019, July). Understanding inclusive design education. In Proceedings of the Design Society: International Conference on Engineering Design (Vol. 1, No. 1, pp. 619-628). Cambridge University Press. |
|---|---|---|---|
| 71 | Design/ Engineering | The Design Council defines inclusive design (ID) as “a general approach to designing in which designers ensure their products and services address the needs of the widest population possible, irrespective of age and ability” (Design Council, 2008). This philosophy includes all users within the engineering design process to ensure products and services are usable and accessible to as wide a range of the population as possible. With the diverse capabilities resulting from disability and population ageing, there is an immediate opportunity ID can address. | Wilson, N., Thomson, A., Thomson, A., & Holliman, A. F. (2019, July). Understanding inclusive design education. In Proceedings of the Design Society: International Conference on Engineering Design (Vol. 1, No. 1, pp. 619-628). Cambridge University Press. |
| 72 | Design/ Engineering | Inclusive design (ID) looks to provide designers with a more accurate understanding of the requirements of different user groups, with the aim of driving informed decisions throughout the engineering design process (Waller et al., 2015). The primary objective of ID is to create equality within society through the eradication of social exclusion (Clarkson & Coleman, 2015). | Wilson, N., Thomson, A., Thomson, A., & Holliman, A. F. (2019, July). Understanding inclusive design education. In Proceedings of the Design Society: International Conference on Engineering Design (Vol. 1, No. 1, pp. 619-628). Cambridge University Press. |

| 73 | Design/ Engineering | ID is an approach that can tackle the diverse population needs that come with disability and ageing, such as declines in physical and cognitive capabilities (Bouma, 2013) that make it difficult for older users to carry out simultaneous tasks. Heterogeneity also increases with age - as the population gets older, the needs and capabilities of users becomes more diverse (Johnson et al., 2010). ID aims to make products more usable by the wider population (Dong et al., 2004a), helping individuals become less reliant on healthcare and social services (Cremers et al., 2014) and maintaining independence for longer. | Wilson, N., Thomson, A., Thomson, A., & Holliman, A. F. (2019, July). Understanding inclusive design education. In Proceedings of the Design Society: International Conference on Engineering Design (Vol. 1, No. 1, pp. 619-628). Cambridge University Press. |
|---|---|---|---|
| 74 | Design/ Engineering | ID relies heavily on the collection of user information to justify design decisions (Keates et al., 2000). Involvement of the end user throughout the design process drives informed decision making (Vavik & Keitsch, 2010) and ensures their needs are incorporated within the design process, making the designer less reliant on their own perception of the problem. | Wilson, N., Thomson, A., Thomson, A., & Holliman, A. F. (2019, July). Understanding inclusive design education. In Proceedings of the Design Society: International Conference on Engineering Design (Vol. 1, No. 1, pp. 619-628). Cambridge University Press. |
| 75 | Design/ Engineering | An ID approach benefits the design process as a whole, with additional time and resources in the early stages of the process ensuring user requirements are satisfied and minimising the need for costly design alterations later in the process (Waller et al., 2015). | Wilson, N., Thomson, A., Thomson, A., & Holliman, A. F. (2019, July). Understanding inclusive design education. In Proceedings of the Design Society: International Conference on Engineering Design (Vol. 1, No. 1, pp. 619-628). Cambridge University Press. |
| 76 | Design/<br>Engineering | An ID approach also ensures user needs are satisfied early in the design process, minimising the need for costly alterations and rework later in the process (Waller et al., 2015). | Wilson, N., Thomson, A., Thomson, A., & Holliman, A. F. (2019, July). Understanding inclusive design education. In Proceedings of the Design Society: International Conference on Engineering Design (Vol. 1, No. 1, pp. 619-628). Cambridge University Press. |
| 77 | Design/<br>Engineering | Inclusive design is for those who want to make great products for the greatest number of people. | Microsoft. Inclusive Design. Available at: https://inclusive.microsoft.design/ |
| 78 | Design/<br>Engineering | Inclusive design: A design methodology that enables and draws on the full range of<br>human diversity. | Microsoft. Inclusive Design. Available at: https://inclusive.microsoft.design/ |
| 79 | Design/<br>Engineering | Designing inclusively doesn't mean you're making one thing for all people. You're designing a diversity of ways for everyone to participate in an experience with a sense of belonging. | Microsoft. Inclusive Design. Available at: https://inclusive.microsoft.design/ |
| 80 | Design/<br>Engineering | Recognize exclusion<br>Learn from diversity<br>Solve for one, extend to many | Microsoft. Inclusive Design. Available at: https://inclusive.microsoft.design/ |

**Supplementary File 2**

This supplementary file is also used to support Yaldiz and Chakrabarti (2024b).

**Note:** All entries in the “Definition/Quote” column are direct excerpts from the referenced sources.

Quotation marks have been omitted for visual clarity.

| No | Domain | Definition/Quote | References |
|---|---|---|---|
| 1 | Management | Defined as the individual experience of increased self-determination and efficacy, empowerment generally leads to increased trust in the empowering the person or organization and an enhanced tendency to repeat the empowered behavior [12,33,39] | Füller, J., Mühlbacher, H., Matzler, K., & Jawecki, G. (2009). Consumer empowerment through internet-based co-creation. Journal of management information systems, 26(3), 71-102. |
| 2 | Management | empowerment is often equated with the sharing of power with subordinates and with participative management. In this sense, empowerment describes the perceived power or control that an individual actor or organizational subunit has over others [12]. | Füller, J., Mühlbacher, H., Matzler, K., & Jawecki, G. (2009). Consumer empowerment through internet-based co-creation. Journal of management information systems, 26(3), 71-102. |
| 3 | Management | In this context, empowerment refers to how the new technologies enable people to interact with the world on different levels (personal, dyad, group, or community) and to do or to achieve things that they found difficult to do or to achieve before. | Füller, J., Mühlbacher, H., Matzler, K., & Jawecki, G. (2009). Consumer empowerment through internet-based co-creation. Journal of management information systems, 26(3), 71-102. |
| 4 | Management | Perceived empowerment is conceptualized as consumers' perceived influence on the product design and decision-making [112]. When consumers feel enabled and competent to solve the product development task assigned to them, when they feel they have some impact on the NPD decisions, they may feel empowered [33]. | Füller, J., Mühlbacher, H., Matzler, K., & Jawecki, G. (2009). Consumer empowerment through internet-based co-creation. Journal of management information systems, 26(3), 71-102. |
| 5 | Management | Empowerment becomes the process by which a leader or manager shares his or her power with subordinates. | Conger, J. A., & Kanungo, R. N. (1988). The empowerment process: Integrating theory and practice. Academy of management review, 13(3), 471-482. |
| 6 | Management | Power, in this context, is interpreted as the possession of formal authority or control over organizational resources. | Conger, J. A., & Kanungo, R. N. (1988). The empowerment process: Integrating theory and practice. Academy of management review, 13(3), 471-482. |
| 7 | Management | The emphasis is primarily on the notion of sharing authority. | Conger, J. A., & Kanungo, R. N. (1988). The empowerment process: Integrating theory and practice. Academy of management review, 13(3), 471-482. |
| 8 | Management | To empower, implies the granting of power-delegation of authority | Burke, W. (1986). Leadership as empowering others. Executive power/Jossey-Bass. |
| 9 | Management | empower as "to enable." (the Oxford English dictionary) | Conger, J. A., & Kanungo, R. N. (1988). The empowerment process: Integrating theory and practice. Academy of management review, 13(3), 471-482. |
| 10 | Management | a process of enhancing feelings of self-efficacy among organizational members through the identification of conditions that foster powerlessness and through their removal by both formal organizational practices and informal techniques of providing efficacy information. | Conger, J. A., & Kanungo, R. N. (1988). The empowerment process: Integrating theory and practice. Academy of management review, 13(3), 471-482. |
| 11 | Management | Fuchs and Schreier (2011) introduced the concept 'consumer empowerment' that signifies customers' involvement into a firm's internal NPD processes, as well as the impact of customer perception of a particular firm in the market. | Sarmah, B., & Rahman, Z. (2017). Transforming jewellery designing: Empowering customers through crowdsourcing in India. Global Business Review, 18(5), 1325-1344.<br><br>Fuchs, C., & Schreier, M. (2011). Customer empowerment in new product development. Journal of Product Innovation Management, 28(1), 17–32. |
| 12 | Management | Consumers are turned designers, and through positive word of mouth can also play the role of a marketer. Thus, it is an empowering process of the customers and elevates their positions from a mere purchaser to an internal employee and a marketer. | Sarmah, B., & Rahman, Z. (2017). Transforming jewellery designing: Empowering customers through crowdsourcing in India. Global Business Review, 18(5), 1325-1344. |
| 13 | Psychology | An iterative process in which a person who lacks power sets a personally meaningful goal-oriented toward increasing power, takes action toward that goal, and observes and reflects on the impact of this action, drawing on his or her evolving self-efficacy, knowledge, and competence related to the goal. | Cattaneo, L. B., & Chapman, A. R. (2010). The process of empowerment: a model for use in research and practice. American psychologist, 65(7), 646. |
| 14 | Psychology | empowerment; the motivational concept of self-efficacy (Conger and Kanungo (1988)) | Spreitzer, G. M. (1995). Psychological empowerment in the workplace: Dimensions, measurement, and validation. Academy of management Journal, 38(5), 1442-1465. |
| 15 | Psychology | Thomas and Velthouse (1990) defined empowerment as increased intrinsic task motivation manifested in a set of four cognitions reflecting an individual's orientation to his or her work role: meaning, competence (which is synonymous with Conger and Kanungo's self-efficacy), self-determination, and impact. | Spreitzer, G. M. (1995). Psychological empowerment in the workplace: Dimensions, measurement, and validation. Academy of management Journal, 38(5), 1442-1465. |
| 16 | Marketing | "customer empowerment" reflects consumers' enhanced ability to access, understand and share information. | Pires, G. D., Stanton, J., & Rita, P. (2006).<br>The internet, consumer empowerment and marketing strategies. European journal of marketing, 40(9/10), 936-949. |

| 17 | Marketing | As a process, empowerment requires mechanisms for individuals to gain control over issues that concern them, including opportunities to develop and practice skills necessary to exert control over their decision-making. | Pires, G. D., Stanton, J., & Rita, P. (2006).<br>The internet, consumer empowerment and marketing strategies. European journal of marketing, 40(9/10), 936-949. |
|---|---|---|---|
| 18 | Marketing | Empowerment as an outcome is subjective. Empowered individuals "would be expected to feel a sense of control, understand their sociopolitical environment, and become active in efforts to exert control" (Zimmerman and Warschausky, 1998, p. 6). | Pires, G. D., Stanton, J., & Rita, P. (2006).<br>The internet, consumer empowerment and marketing strategies. European journal of marketing, 40(9/10), 936-949. |
| 19 | Design & Psychology | Empowerment is a process by which people, organizations, and communities gain mastery over issues they perceive as concerning them (Rappaport 1987). | Hussain, S. (2010). Empowering marginalised<br>children in developing countries through participatory design processes. *CoDesign* , *6* (2), 99-117. |
| 20 | Design & Psychology | Empowerment is a multilevel construct with all levels of analysis being interdependent with each other. | Hussain, S. (2010). Empowering marginalised<br>children in developing countries through participatory design processes. *CoDesign* , *6* (2), 99-117. |
| 21 | Design & Psychology | As pointed out by Zimmerman (1995), psychological empowerment and power are not the same: 'Power suggests authority, whereas psychological empowerment is a feeling of control, a critical awareness of one's environment, and an active engagement in it. [. . .] Actual power or control is not necessary for empowerment because in some context and for some populations real control or power may not be the desired goal. Rather, goals such as being more informed, more skilled, more healthy [sic.], or more involved in decision making may be the desired outcome' (pp. 592–593). | Hussain, S. (2010). Empowering marginalised<br>children in developing countries through participatory design processes. *CoDesign* , *6* (2), 99-117. |
| 22 | Design & Psychology | Physiological empowerment is context-dependent. Therefore, the development of<br>a universal and global measure of it is not an appropriate goal; empowerment does<br>not mean the same thing for all people, everywhere, at any time (Zimmerman 1995).<br>This means that in each project, designers have to decide how changes in psychological empowerment should be evaluated. | Hussain, S. (2010). Empowering marginalised<br>children in developing countries through participatory design processes. *CoDesign* , *6* (2), 99-117. |
| 23 | Design & Psychology | 'Psychological empowerment [. . .] includes beliefs that goals can be achieved, awareness about recourses and factors that hinder or enhance one's efforts to achieve those goals, and efforts to fulfil the goals' (Zimmerman 1995, p. 582). | Hussain, S. (2010). Empowering marginalised<br>children in developing countries through participatory design processes. *CoDesign* , *6* (2), 99-117. |
| 24 | Design & Psychology | Psychological empowerment is not a static trait, but an open-ended dynamic construct. Every individual has the potential of experiencing both empowering and<br>disempowering processes. The perceptions, skills, or actions for increasing someone's sense of empowerment depend on a person's age, socioeconomic status, gender, etc. | Hussain, S. (2010). Empowering marginalised<br>children in developing countries through participatory design processes. *CoDesign* , *6* (2), 99-117. |

| 25 | Design | “the process of increasing the capacity of individuals or groups to make choices and to transform those choices into desired actions and outcomes.” *the World Bank* | Fladvad Nielsen, B. (2012). Participate! A critical investigation into the relationship between participation and empowerment in design for development. In DS 71: Proceedings of NordDesign 2012, the 9th NordDesign conference, Aarlborg University, Denmark. 22-24.08. 2012. |
|---|---|---|---|
| 26 | Design | Zimmermann [25] for example describes psychological empowerment as a potential within the individual, while Ife[26] focuses also on top-down conditions for empowerment. | Fladvad Nielsen, B. (2012). Participate! A critical investigation into the relationship between participation and empowerment in design for development. In DS 71: Proceedings of NordDesign 2012, the 9th NordDesign conference, Aarlborg University, Denmark. 22-24.08. 2012. |
| 27 | Design | “exists only as exercised by some on others, only when it is put into action” [27] | Fladvad Nielsen, B. (2012). Participate! A critical investigation into the relationship between participation and empowerment in design for development. In DS 71: Proceedings of NordDesign 2012, the 9th NordDesign conference, Aarlborg University, Denmark. 22-24.08. 2012. |
| 28 | Design | The purpose of empowerment should also include an understanding of how the individual empowerment can increase decision power and in which circumstances it can represent a real and not only potential change that they did not have before the outsider intervened; if not, one may even argue that our efforts are unethical and indeed neocolonialist in nature. | Fladvad Nielsen, B. (2012). Participate! A critical investigation into the relationship between participation and empowerment in design for development. In DS 71: Proceedings of NordDesign 2012, the 9th NordDesign conference, Aarlborg University, Denmark. 22-24.08. 2012. |
| 29 | Design | Real empowerment means that we increase someone’s power, which depends on both top-down and bottom-up approaches[26] to change while design for capabilities is focused on the individual and perhaps already assuming that the individual has the opportunity to do what they are capable of. | Fladvad Nielsen, B. (2012). Participate! A critical investigation into the relationship between participation and empowerment in design for development. In DS 71: Proceedings of NordDesign 2012, the 9th NordDesign conference, Aarlborg University, Denmark. 22-24.08. 2012. |
| 30 | Marketing | Consumer empowerment has been defined (Wathieu et al., 2002) as letting consumers take control of variables that are conventionally pre-determined by marketers; one of these variables is brand meaning. | Cova, B., & Pace, S. (2006). Brand community of convenience products: new forms of customer empowerment–the case “my Nutella The Community”. European journal of marketing, 40(9/10), 1087-1105. |
| 31 | Marketing | Ferrero enables Nutella fans to live their passion. It is in this sense that consumer<br>empowerment has occurred. Conversely, Ferrero does not encourage exchange and community-oriented creations, remaining relatively distant from any form of communal empowerment. | Cova, B., & Pace, S. (2006). Brand community of convenience products: new forms of customer empowerment–the case “my Nutella The Community”. European journal of marketing, 40(9/10), 1087-1105. |
| 32 | Management | customer empowerment – engagement strategies that give customers a sense of control over a brand’s general offerings | Acar, O. A., & Puntoni, S. (2016). Customer empowerment in the digital age. Journal of Advertising Research, 56(1), 4-8. |
| 33 | Management | Customer empowerment therefore means a deeper connection with the brand. Customers who are empowered develop positive attitudes towards a brand. | Acar, O. A., & Puntoni, S. (2016). Customer empowerment in the digital age. Journal of Advertising Research, 56(1), 4-8. |
| 34 | Marketing | consumer’s ability to specify and adjust the choice context. According to this view, the experience of empowerment derives not from more choices, but from one’s flexibility in defining one’s choices. | Wathieu, L., Brenner, L., Carmon, Z., Chattopadhyay, A., Wertenbroch, K., Drolet, A., ... & Wu, G. (2002). Consumer control and empowerment: a primer. Marketing Letters, 13, 297-305. |
| 35 | Engineering & Education | “multi-dimensional social process that helps people gain control over their own lives” by fostering power in people to operate the changes they may want “in their own lives, their communities, and in their society” (Page and Czuba, 1999) | Kleba, J.B., Cruz, C.C. (2021). From empowerment to emancipation - A framework for empowering sociotechnical interventions. International Journal of Engineering, Social Justice and Peace, 8(2),<br>pages 28-49. |
| 36 | Engineering & Education | “while we cannot give people power [or] make them “empowered,” we can provide the opportunities, resources and support that they need to become involved themselves” (Page and Czuba, 1999, p.4) | Kleba, J.B., Cruz, C.C. (2021). From empowerment to emancipation - A framework for empowering sociotechnical interventions. International Journal of Engineering, Social Justice and Peace, 8(2),<br>pages 28-49. |
| 37 | Political Science | Community empowerment implies that people will have the necessary information, as well as power and influence to exercise some control over the future of their area. | Perrons, D., & Skyers, S. (2003). Empowerment through participation? Conceptual explorations and a case study. International Journal of Urban and Regional Research, 27(2), 265-285. |
| 38 | Psychology | Empowerment - the process by which people gain some control over valued events, outcomes, and resources - is an important construct for understanding and improving the lives of people of marginal status. | Fawcett, S. B., White, G. W., Balcazar, F. E., Suarez-Balcazar, Y., Mathews, R. M., Paine-Andrews, A., ... & Smith, J. F. (1994). A contextual-behavioral model of empowerment: Case studies involving people with physical disabilities. *American Journal of Community Psychology* , *22* (4), 471-496. |
| 39 | Psychology | "...to enhance the possibilities for people to control their own lives..." (Rappaport, 1981, p. 15), whether as individuals or as groups of people sharing common experiences, turf, or concerns. | Fawcett, S. B., White, G. W., Balcazar, F. E., Suarez-Balcazar, Y., Mathews, R. M., Paine-Andrews, A., ... & Smith, J. F. (1994). A contextual-behavioral model of empowerment: Case studies involving people with physical disabilities. *American Journal of Community Psychology* , *22* (4), 471-496. |
| 40 | Psychology | Empowerment is viewed as both an individual or psychological process (Zimmerman, 1990; Zimmerman & Rappaport, 1988) and a group experience (Chavis & Wandersman, 1990). | Fawcett, S. B., White, G. W., Balcazar, F. E., Suarez-Balcazar, Y., Mathews, R. M., Paine-Andrews, A., ... & Smith, J. F. (1994). A contextual-behavioral model of empowerment: Case studies involving people with physical disabilities. *American Journal of Community Psychology* , *22* (4), 471-496. |
| 41 | Engineering/ Industrial management and organisation | Empowerment does not mean that the management has no role to play or no responsibility. In fact, the management has more responsibilities. They have to monitor the skills continuously required for carrying out the ever-changing complexity of jobs of the teams. The management must be willing to help the teams when they are unable to solve issues. The responsibilities of the management are to control the processes and not the individual team members. | Thamizhmanii, S., & Hasan, S. (2010). A review on an employee empowerment in TQM practice. Journal of Achievements in Materials and Manufacturing Engineering, 39(2), 204-210. |
| 42 | Engineering/ Industrial management and organisation | Empowerment is a concept that links individual strengths and competencies, natural helping systems and proactive behaviour to social policy and social change. In the other words, empowerment links individual and his or her well-being to wider social and political environment in which he or she functions [11]. | Thamizhmanii, S., & Hasan, S. (2010). A review on an employee empowerment in TQM practice. Journal of Achievements in Materials and Manufacturing Engineering, 39(2), 204-210. |
| 43 | Engineering/ Industrial management and organisation | Empowerment is, an organizational state, where people are obliged to direct business and understand their performance boundaries, thus it enables them to take responsibility and ownership while seeking improvements, identifying the best course of action and imitative steps to meet customer requirements [12]. | Thamizhmanii, S., & Hasan, S. (2010). A review on an employee empowerment in TQM practice. Journal of Achievements in Materials and Manufacturing Engineering, 39(2), 204-210. |

| 44 | Engineering/ Industrial management and organisation | Empowerment means engaging employees in the thinking processes of an organization. Involvement means having input. Empowerment means having input that is heard and seriously considered. | Thamizhmanii, S., & Hasan, S. (2010). A review on an employee empowerment in TQM practice. Journal of Achievements in Materials and Manufacturing Engineering, 39(2), 204-210. |
|---|---|---|---|
| 45 | Engineering/ Industrial management and organisation | Empowerment requires a change in an organization culture, but does not mean that top management  abdicate their responsibility or authority. An employee empowerment is necessary for the effective functioning of the skill of employee. | Thamizhmanii, S., & Hasan, S. (2010). A review on an employee empowerment in TQM practice. Journal of Achievements in Materials and Manufacturing Engineering, 39(2), 204-210. |
| 46 | Social and Behavioral Sciences | Community empowerment is one of the main purpose of development in developing countries. While development only result in a ratio between the "before" and "after" condition, empowerment can keep the continuity of the development through the<br>good relationship between related parties, in terms of knowledge production and field application. | Sianipar, C. P. M., & Widaretna, K. (2012). NGO as Triple-Helix axis: Some lessons from Nias community empowerment on cocoa production. Procedia-Social and Behavioral Sciences, 52, 197-206. |
| 47 | Environmental Sciences | Outsiders give hoe to local people, teach them how to use and maintain it with better method, and also teach them how to make it by themself, adjusting the function based on required conditions. These efforts will sustain the sustainable development, empower local people. When the outsiders leave them, local people will maintain the sustainable development by themself. | Sianipar, C. P. M., Yudoko, G., Adhiutama, A., & Dowaki, K. (2013). Community empowerment through appropriate technology: Sustaining the sustainable development. Procedia Environmental Sciences, 17, 1007-1016. |
| 48 | Environmental Sciences | appropriate technology can fill in the gap to transform community development into community empowerment. The appropriateness of appropriate technology can help community members in maintaining the ongoing development (sustaining the development) as well as in making adjustment of technology to overcome future conditions in changing environment. | Sianipar, C. P. M., Yudoko, G., Adhiutama, A., & Dowaki, K. (2013). Community empowerment through appropriate technology: Sustaining the sustainable development. Procedia Environmental Sciences, 17, 1007-1016. |
| 49 | Environmental Sciences | Ferguson [2] stated that empowerment is categorized as transformation process of a community. In order to fasten the process, technology is always needed to bring multiplier effect for the transformation process [3]. | Sianipar, C. P. M., Yudoko, G., Adhiutama, A., & Dowaki, K. (2013). Community empowerment through appropriate technology: Sustaining the sustainable development. Procedia Environmental Sciences, 17, 1007-1016. |

| 50 | Environmental Sciences | While development often interpreted as the flow of resources from outside into community, empowerment push-and-pull full participation of all community members to change their world by themself, from inside to outside [2]. | Sianipar, C. P. M., Yudoko, G., Adhiutama, A., & Dowaki, K. (2013). Community empowerment through appropriate technology: Sustaining the sustainable development. Procedia Environmental Sciences, 17, 1007-1016. |
|---|---|---|---|
| 51 | Environmental Sciences | Community with lack of development must be developed through community development. Then, the increasing of knowledges will sustain the development. After that, the transformation process should be continued to reach beyond sustainable development, that is empowerment. | Sianipar, C. P. M., Yudoko, G., Adhiutama, A., & Dowaki, K. (2013). Community empowerment through appropriate technology: Sustaining the sustainable development. Procedia Environmental Sciences, 17, 1007-1016. |
| 52 | Environmental Sciences | Empowerment as transformation process of a community should be understood completely through scientific thoughts as well as real field problems. | Sianipar, C. P. M., Yudoko, G., Adhiutama, A., & Dowaki, K. (2013). Community empowerment through appropriate technology: Sustaining the sustainable development. Procedia Environmental Sciences, 17, 1007-1016. |
| 53 | Environmental Sciences | As a transformation process [2], empowerment will ensure the sustainability of given<br>sustainable development. | Sianipar, C. P. M., Yudoko, G., Adhiutama, A., & Dowaki, K. (2013). Community empowerment through appropriate technology: Sustaining the sustainable development. Procedia Environmental Sciences, 17, 1007-1016. |
| 54 | Environmental Sciences | The development maybe still maintained someway, but community members will face many problems because outsiders do not teach them any capabilities to solve their future problems, how to adjust the ongoing development in the changes environment. Through this gap, empowerment approach provides the solution. In that approach, outsiders teach community members to 'make' the development by themselves. It is the next keyword after the 'give' and 'maintain.' By teaching such making activities related to the ongoing sustainable development, community members can always adapt their efforts in sustaining the development by themself, even when otsiders have already leave their areas and/or the project has ended. That is "empowerment" level. | Sianipar, C. P. M., Yudoko, G., Adhiutama, A., & Dowaki, K. (2013). Community empowerment through appropriate technology: Sustaining the sustainable development. Procedia Environmental Sciences, 17, 1007-1016. |
| 55 | Sustainability | empowering consumers by giving them more possibilities of repairing their products instead of discarding them. - Right to repair | Hernandez, R. J., Miranda, C., & Goñi, J. (2020). Empowering sustainable consumption by giving back to consumers the ‘right to repair’. Sustainability, 12(3), 850. |
| 56 | Sustainability | (we believe are underexplored, particularly about the impact directives like “right to repair” can have on empowering sustainable consumption from a consumer point of view and not only as a production mechanism.) - Right to repair | Hernandez, R. J., Miranda, C., & Goñi, J. (2020). Empowering sustainable consumption by giving back to consumers the ‘right to repair’. Sustainability, 12(3), 850. |
| 57 | Sustainability | “empower consumers to make informed choices and play an active role in the ecological transition” [39] | Hernandez, R. J., Miranda, C., & Goñi, J. (2020). Empowering sustainable consumption by giving back to consumers the ‘right to repair’. Sustainability, 12(3), 850. |
| 58 | Sustainability | Empowerment, we believe, will be based on fostering more community action and autonomy. | Hernandez, R. J., Miranda, C., & Goñi, J. (2020). Empowering sustainable consumption by giving back to consumers the ‘right to repair’. Sustainability, 12(3), 850. |
| 59 | Manufacturin | Empowerment is one of the important step adopted by top management and has achieved desired results in terms of productivity and waste minimization. Further, empowerment has improved information sharing and coordination among workers assigned to different work stations or within the same work station (e.g. [68, 70]). Empowerment includes different activities and training to create adaptability, foster innovation, improve collaboration and enhance speed [65]. | Dubey, R., & Gunasekaran, A. (2015). Agile manufacturing: framework and its empirical validation. The International Journal of Advanced Manufacturing Technology, 76, 2147-2157. |
| 60 | Manufacturin | Empowerment of workforce may help to achieve AM (agile manufacturing) environment, and thus, it can be said based on preceding discussions that, empowerment is an important construct of the AM framework. | Dubey, R., & Gunasekaran, A. (2015). Agile manufacturing: framework and its empirical validation. The International Journal of Advanced Manufacturing Technology, 76, 2147-2157. |
| 61 | Innovation Management | User empowerment refers to the types of new product decisions that are outsourced to users. In various ways are users invited to take more active part in corporate NPD. | Jespersen, K. R. (2011). Online channels and innovation: Are users being empowered and involved?. International Journal of Innovation Management, 15(06), 1141-1159. |
| 62 | Sustainable Transition (Environmental Policy & Planning) | In the broadest sense of the word, empowerment refers to the process of gaining power. There is a diversity of understandings of empowerment in transition studies, ranging from empowerment as a systemic transition pattern (De Haan & Rotmans, 2011) and functional property of niches (Smith & Raven, 2012) to empowerment at the level of individual or group capacities (Avelino, 2009,2011). | Avelino, F., & Wittmayer, J. M. (2016). Shifting power relations in sustainability transitions: a multi-actor perspective. Journal of environmental policy & planning, 18(5), 628-649. |
| 63 | Sustainable Transition (Environmental Policy & Planning) | Thomas and Velthouse (1990) operationalize empowerment in terms of ‘intrinsic motivation’ and argue that the extent to which individuals are intrinsically motivated to engage in an activity depends on the extent to which they have a sense of impact, competence, meaning and choice regarding that activity. | Avelino, F., & Wittmayer, J. M. (2016). Shifting power relations in sustainability transitions: a multi-actor perspective. Journal of environmental policy & planning, 18(5), 628-649. |

| 64 | Business & Industrial Marketing | Team empowerment, which refers to the authority and power the team has in order to manage and lead itself (Manz and Sims, 1991), is also imperative for team reflexivity. | Dayan, M., & Basarir, A. (2009). Antecedents and consequences of team reflexivity in new product development projects. Journal of Business & Industrial Marketing, 25(1), 18-29. |
|---|---|---|---|
| 65 | Design Education | teaching relevant design thinking methods and processes to community members, empowering them to design for and by themselves. | Hasselknippe, K. S., Flygenring, T., & Kirah, A. (2017). Empowering refugee and host-community youth with design thinking skills for community development. In DS 88: Proceedings of the 19th International Conference on Engineering and Product Design Education (E&PDE17), Building Community: Design Education for a Sustainable Future, Oslo, Norway, 7 & 8 September 2017 (pp. 092-097). |
| 66 | Purchasing and Materials Management | Empowerment is a philosophy congruent with the goals of Total Quality Management (TQM) and continuous improvement. | Giunipero, L. C., & Vogt, J. F. (1997). Empowering the purchasing function: moving to team decisions. International Journal of Purchasing and Materials Management, 33(4), 8-15. |
| 67 | Purchasing and Materials Management | Part of empowerment is maintaining one’s positive self-esteem and connectedness to others and to the organization. | Giunipero, L. C., & Vogt, J. F. (1997). Empowering the purchasing function: moving to team decisions. International Journal of Purchasing and Materials Management, 33(4), 8-15. |
| 68 | Purchasing and Materials Management | Empowerment means to enable, to allow, or to permit, and can be either self-initiated or initiated by others.3<br>It is a method for creating and redistributing power.4<br>In the traditional sense, power is one’s ability to control or change another’s behavior. Empowerment occurs when the power goes to the employees, who then experience a sense of ownership and control over their jobs.5 | Giunipero, L. C., & Vogt, J. F. (1997). Empowering the purchasing function: moving to team decisions. International Journal of Purchasing and Materials Management, 33(4), 8-15. |
| 69 | Design / Education | Empowering patients through their grip on surfaces - designing aesthetic products by using 3D printer | Lyche, W., & Berg, A. (2017). Empowerment through Product Design: Digital textile pattern design for grip development in healthcare. In DS 88: Proceedings of the 19th International Conference on Engineering and Product Design Education (E&PDE17), Building Community: Design Education for a Sustainable Future, Oslo, Norway, 7 & 8 September 2017 (pp. 568-573). |

| 70 | Manufacturin | Employee empowerment, therefore, requires a complete realignment of power relations within which managers emphasize interdisciplinary collaboration and leadership, shared values, and motivation for knowledge diversity (Fawcett and Myers 2001; Forsythe 1997). | Gunasekaran, A., Yusuf, Y. Y., Adeleye, E. O., Papadopoulos, T., Kovvuri, D., & Geyi, D. A. G. (2019). Agile manufacturing: an evolutionary review of practices. International Journal of Production Research, 57(15-16), 5154-5174. |
|---|---|---|---|
| 71 | Social Scienc | Empowerment is defined as a group's or individual's capacity to make effective choices, that is, to make choices and then to transform those choices into desired actions and outcomes. | Alsop, R., Bertelsen, M. F., & Holland, J. (2006). Empowerment in practice: From analysis to implementation. World Bank Publications. |
| 72 | Finance | women's empowerment is defined as improving the ability of women to access the constituents of development - in particular health, education, earning opportunities, rights, and political participation.<br>Empowerment can, in other words, accelerate development. | Duflo, E. (2012). Women empowerment<br>and economic development. Journal of Economic literature, 50(4), 1051-1079. |
| 73 | Management | The best way to define empowerment is to consider it as part of a process or an evolution – an evolution that goes on whenever you have two or more people in a relationship, personally or professionally. | Pastor, J. (1996). Empowerment: what it<br>is and what it is not. Empowerment in organizations, 4(2), 5-7. |
| 74 | Management | empowerment is the management style where managers share with the rest of the organisational members their influence in the decision making process – that is to say, the collaboration in the decision making process is not limited to those positions with formal power –, with certain characteristics as far as information systems, training, rewarding, power sharing, leadership style and organisational culture concerns. | del Val, M. P., & Lloyd, B. (2003).<br>Measuring empowerment. Leadership & organization development journal, 24(2), 102-108. |
| 75 | Sustainable Developmen t | Empowerment can be defined as a "multi-dimensional social process that helps people gain control over their own lives. It is a process that fosters power (that is, the capacity to implement) in people, for use in their own lives, their communities, and in their society, by acting on issues that they define as important". | Lohani, M., & Aburaida, L. (2017). Women empowerment: A key to sustainable development. The Social ION, 6(2), 26-29. |
| 76 | Sustainable Developmen t | Women's empowerment means women gaining more power and control over their own lives. This entails the idea of women's continued disadvantage compared to men which is apparent in different economic, socio-cultural and political spheres. Therefore, women's empowerment can also be seen as an important process in reaching gender equality, which is understood to mean that the "rights, responsibilities and opportunities of individuals will not depend on whether they are born male or female". According to the UN Population Fund, an empowered woman has a sense of self-worth. She can determine her own choices, and has access to opportunities and resources providing her with an array of options she can pursue. She has control over her own life, both within and outside the home and she has the ability to influence the direction of social change to create a more just social and economic order, both nationally and internationally. | Lohani, M., & Aburaida, L. (2017). Women empowerment: A key to sustainable development. The Social ION, 6(2), 26-29. |
| 77 | Social Science and Technology | Community empowerment is the process of increasing control by groups over consequences that are important to their members and to others in the broader community. For example, by forming a tenants' rights organization individual residents of a public housing project may increase their ability to redress grievances and obtain improvements in housing conditions.<br><br>Community empowerment requires; a critical balancing act between extending opportunities for control over community decisions and maximizing - efficiency in community functioning. | Fawcett, S. B., Seekins, T., Whang, P. L., Muiu, C., & Balcazar, Y. S. D. (1984). Creating and using social technologies for community empowerment. Prevention in Human Services, 3(2-3), 145-171. |
| 78 | Social Science and Technology | From a behavioral perspective, the ability to produce empowerment may be a function of such critical variables as knowledge of problems and solution alternatives and skill in presenting issues, leading groups, and implementing other empowerment tactics. Further, it may be related to the capacity to control consequences for critical actors in the system, including positive consequences (e.g., votes for elected officials) and negative ones (e.g., unfavorable newspaper reports). Finally, empowerment may be enhanced by environmental or structural variables, such as the opportunity to gain access to agendas of meetings of elected officials. An analysis of those with and without power might suggest the importance of similar variables in community empowerment. | Fawcett, S. B., Seekins, T., Whang, P. L., Muiu, C., & Balcazar, Y. S. D. (1984). Creating and using social technologies for community empowerment. Prevention in Human Services, 3(2-3), 145-171. |
| 79 | Social Science and Technology | the empowerment strategy as it relates to the choice of goals, the selection of the targets of change, the choice of means, and the analysis of consequences for the targets, change agents, and society in general. | Fawcett, S. B., Seekins, T., Whang, P. L., Muiu, C., & Balcazar, Y. S. D. (1984). Creating and using social technologies for community empowerment. Prevention in Human Services, 3(2-3), 145-171. |
| 80 | Social Science and Technology | the empowerment effort should increase the power of the target group relative to other groups in society | Fawcett, S. B., Seekins, T., Whang, P. L., Muiu, C., & Balcazar, Y. S. D. (1984). Creating and using social technologies for community empowerment. Prevention in Human Services, 3(2-3), 145-171. |
| 81 | Marketing | To ensure effective outcomes, marketers need to understand the factors that affect consumers' mental statement of being empowered and constitute customers' perception of appropriate empowerment. We label this customer psychological empowerment (CPE) in the current study. | Prentice, C., Han, X. Y., & Li, Y. Q. (2016). Customer empowerment to co-create service designs and delivery: Scale development and validation. Services Marketing Quarterly, 37(1), 36-51. |
| 82 | Management | For one thing, customers are now empowered with greater access to information — thanks, in part, to the Internet — so that many want to have a greater say about the products they purchase. | Ogawa, S., & Piller, F. T. (2006). Reducing the risks of new product development. MIT Sloan management review. |
| 83 | Health | The etymological denvation of 'empowerment' suggests 'to be able' or 'to enable' as meanings of the word Chandler (1991) defines 'to empower' as 'to enable to act' | Rodwell, C. M. (1996). An analysis of the concept of empowerment. Journal of advanced nursing, 23(2), 305-313. |
| 84 | Psychology | One’s degree of empowerment depends jointly on the affordances of one’s ecology and one’s capacities in relation to one’s goals. Hence, agency – the ability to act –is not the same as power. One can act (or have agency) without being able to fulfil one’s goals – to be empowered – because of deficiencies in one’s context, capacities, or their relation. One’s degree of empowerment is variable across agents, time, and location.<br>One’s degree of agency is universal, singular, and constant. | Pratto, F. (2016). On power and empowerment. British Journal of Social Psychology, 55(1), 1-20. |
| 85 | Psychology | The empowerment approach provides a clear general way to gauge people’s life conditions: Their level of well-being<br>and the quality of people’s choice sets. | Pratto, F. (2016). On power and empowerment. British Journal of Social Psychology, 55(1), 1-20. |
| 86 | Management | The way to truly empower a woman is to make her less poor, financially independent, and better educated;  we need social and cultural changes that eliminate the prejudices that are the cause of her deprivations. | Karnani, A. G. (2006). Fortune at the Bottom of the Pyramid: A Mirage. California Management Review, Forthcoming. |

**Supplementary File 3**

Figure 1 demonstrates the types of functions in an exchange network among stakeholders A, B, and C. F1 and F2 are group functions, or function pairs, that create a dependency among stakeholders based on their needs and resources. The other group functions are F4–5, F8–10, F13–15, F18–20, F23–25, and F28–30. Through the mobilisation of stakeholder resources by participating in these function groups, stakeholders can increase their power by empowering themselves. Individual functions are also a way to empowerment; however, they do not contribute to an increase in power because they do not create an exchange network with other stakeholders. Thus, individual functions cannot influence the inclusion of new stakeholders in a product lifecycle process. Activity theory is adopted in the context of exchange networks, which rely on both group functions and individual functions. In exchange networks, we detail the interactions a maximum of two times to complete a goal. As shown in Figure 1, F1 is the goal of the involved stakeholders and requires a coalition among A, B, and C to achieve it. F30 represents the achievement of the coalition goal (A–B–C). The links between B–C and A–C show interactions that occurred two times. The number of individual functions is connected with different group functions, and the detailing of interactions is limited to a maximum of two functions.

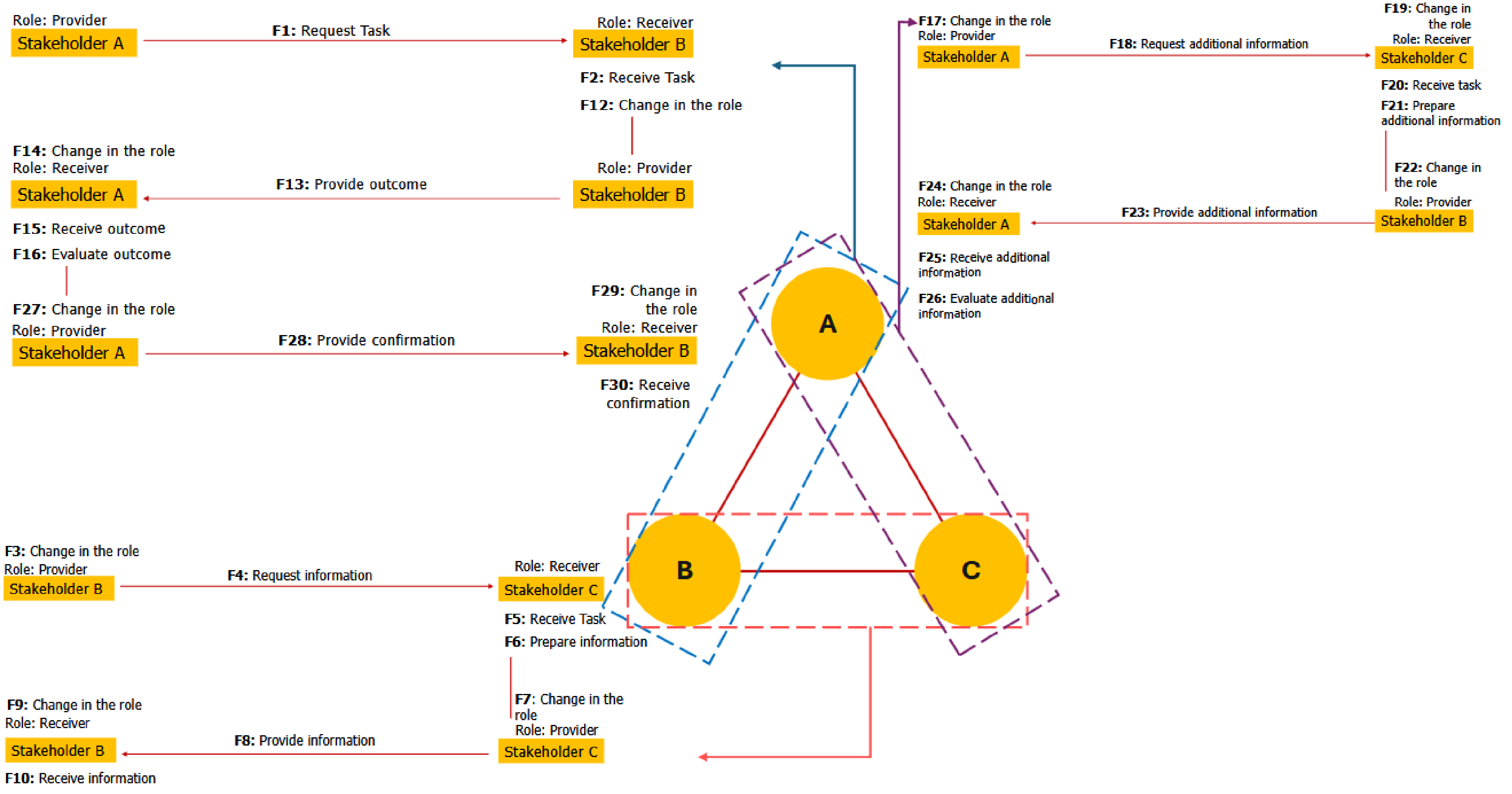

Figure 1: An example of the exchange network with the types of functions

Note: This diagram is also included in Yaldiz and Chakrabarti (2024b), which is currently under review.

To generalise the function identification process across different case studies, a standard approach has been developed, as shown in Figures 2, 3, and 4. Functions are identified based on the available information in each case study. The functions are described as "Request Task" or "Request Information." The corresponding function is then described as "Receive the Task" or "Receive Information."

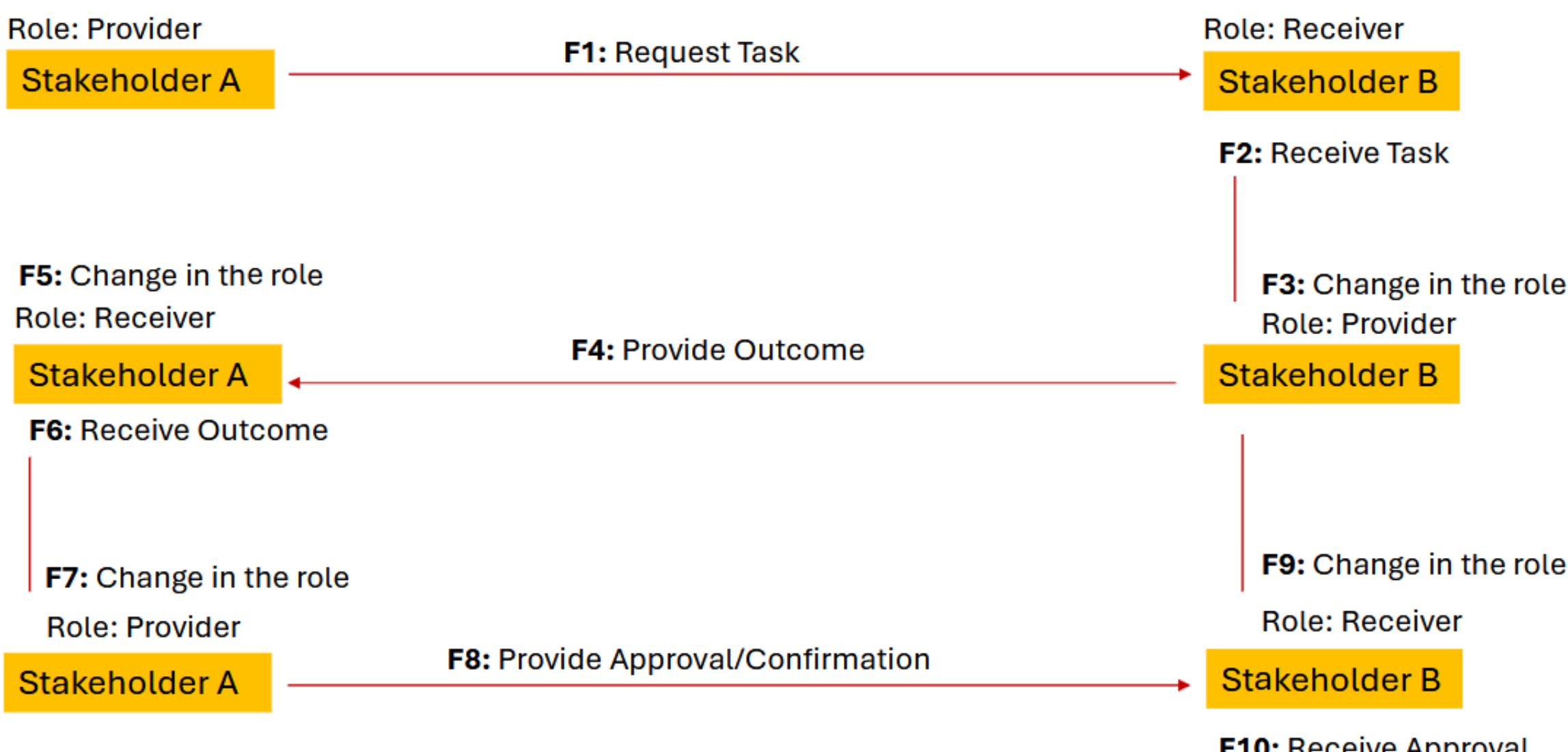

Figure 2: Generalisation of function analysis -1

After these steps (shown in Figure 2), the next functions should focus on evaluating the received information or making a decision about attempting another task, to allow for further detailing. As demonstrated in Figure 3, additional information can be requested, followed by the provision of that information. With this, a maximum of two levels of detailing in function analysis can be supported. Additionally, as shown in Figure 4, function identification can be followed by requesting data or information separately from the initial task request. The data or information request can also be followed by providing the information. The roles of stakeholders should be updated by a following function, such as “Change in the Role from Receiver to Provider" or "Change in the Role from Provider to Receiver," depending on the roles of the stakeholder group. For example, if Stakeholder Group A requests a task from Stakeholder Group B, Stakeholder A acts as the provider, and Stakeholder B acts as the receiver. When Stakeholder B provides the outcome, their role should change from receiver to provider.

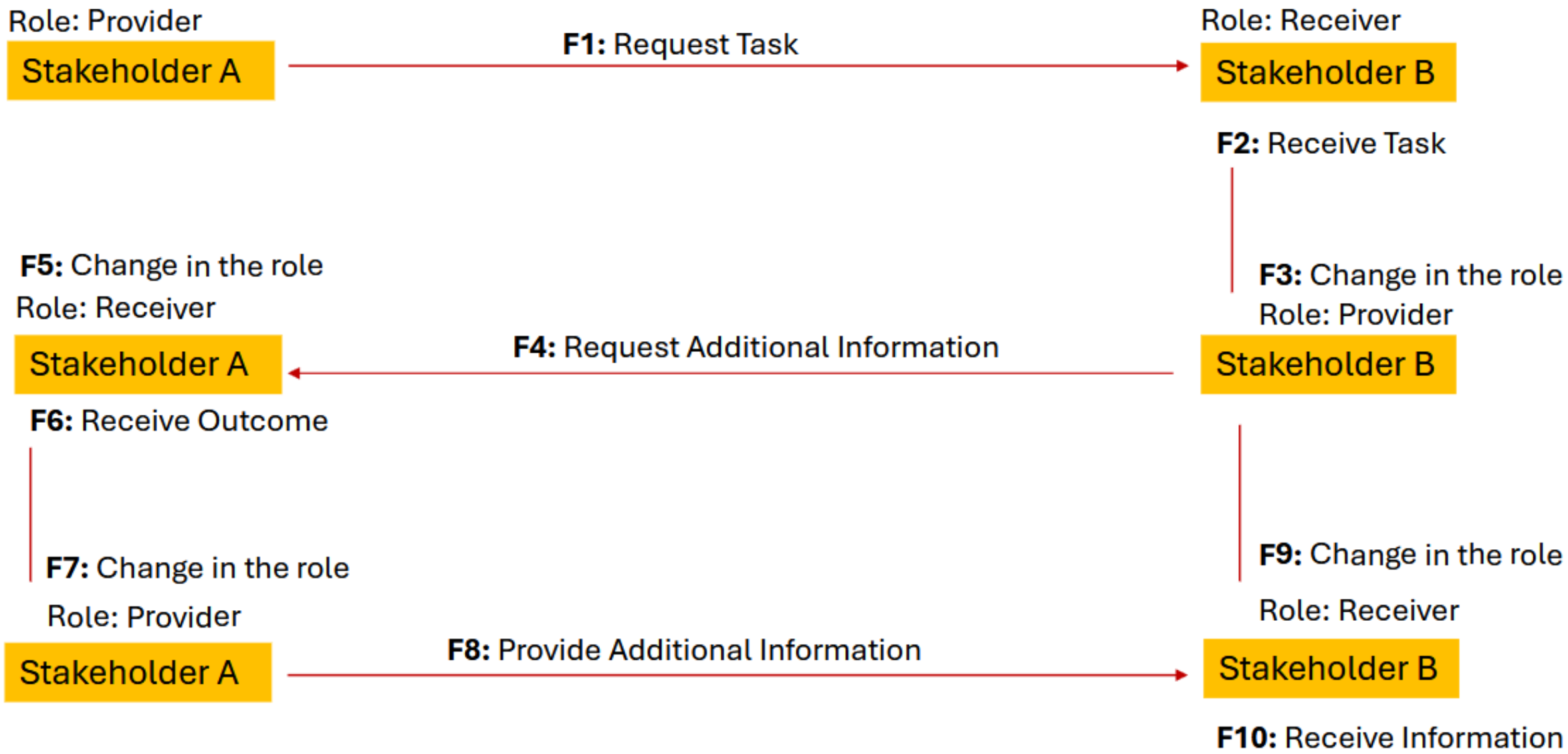

Figure 3: Generalisation of function analysis -2

This change in the roles of stakeholders is visualised in the diagrams related to function generalisation in Figures 1, 2, 3, 4 and 5. In function pairs, the roles of stakeholders should be considered carefully for each function. Providers ensure that resources or opportunities are available for receivers to benefit from, while receivers access these opportunities based on their resources, needs, and capabilities. There can also be functions involving only one stakeholder group (individual functions), and in such cases, a role change function is not required.

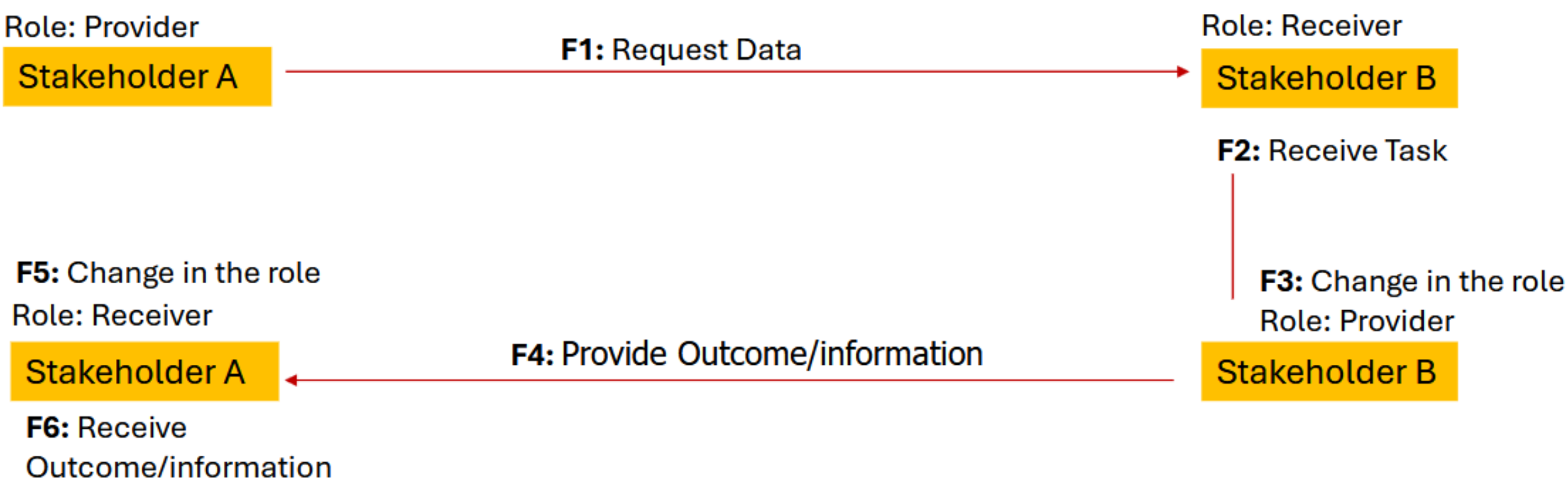

Figure 4: Generalisation of function analysis -3

For example, based on the following content, the function analysis is presented in Table 1:
Content: Based on the feedback from users after using the first prototype and on the observations, the second prototype was developed. "...Trials were conducted on the second prototype after fabrication. The results of trials show that the machine performance has increased and it is more user-friendly than the other two." (Tak et al., 2019a).

Table 1: Function analysis

| | | | | |
|---|---|---|---|---|
| F286&F287 | F286 | Provide training (information) about the second modified machine | Stakeholder A | Provider |
| | F287 | Receive information | Stakeholder B | Receiver |
| F288&F289 | F288 | Request (task) to use the modified machine | Stakeholder A | Provider |
| | F289 | Receive the request | Stakeholder B | Receiver |
| X | F290 | Use the machine | Stakeholder B | - |
| X | F291 | Observe the users | Stakeholder A | - |
| F292&F293 | F292 | Request to provide feedback about the modified machine | Stakeholder A | Provider |
| | F293 | Receive the task | Stakeholder B | Receiver |
| X | F294 | Change in the role (Provider → Receiver) | Stakeholder A | Receiver |
| X | F295 | Change in the role (Receiver → Provider) | Stakeholder B | Provider |

In order to identify the group and number of needs, the needs and resources of stakeholders for each function must be considered. For example, as presented in Table 1, the need of Stakeholder A to attempt Function 1, "Provide training about the second modified machine," is related to responsibilities, duties, and work, which can be grouped under "Participation and Having" based on the Matrix of Needs and Satisfiers (Max-Neef, 1992). The resources of Stakeholder A to attempt Function 1 are position, experience, and capability. However, in this case, position is prioritised, as it enables Stakeholder A to be involved in providing training. This prioritisation reflects a hierarchy of resources, where position comes first. Due to Stakeholder A's position, Stakeholder B was involved in receiving training about the modified machine.
Another example can be observed in Figure 5. The need for Stakeholder A to be involved in Function F7 is also related to their duty and can be grouped under "Participation and Having." The main resource for attempting Function F7 is the position, even though experience and capability may also contribute. The impact of F7 on Stakeholder B leads to Function F8, where Stakeholder B needs to fulfil the requirement for using a modified version of the machine. This need is also connected to "Participation and Having," as the machine supports income generation for users. In this context, the responsibility associated with operating the machine can be considered under the same need group. The resource of Stakeholder B for F8 is the experience of using both the previous and modified versions of the machine.

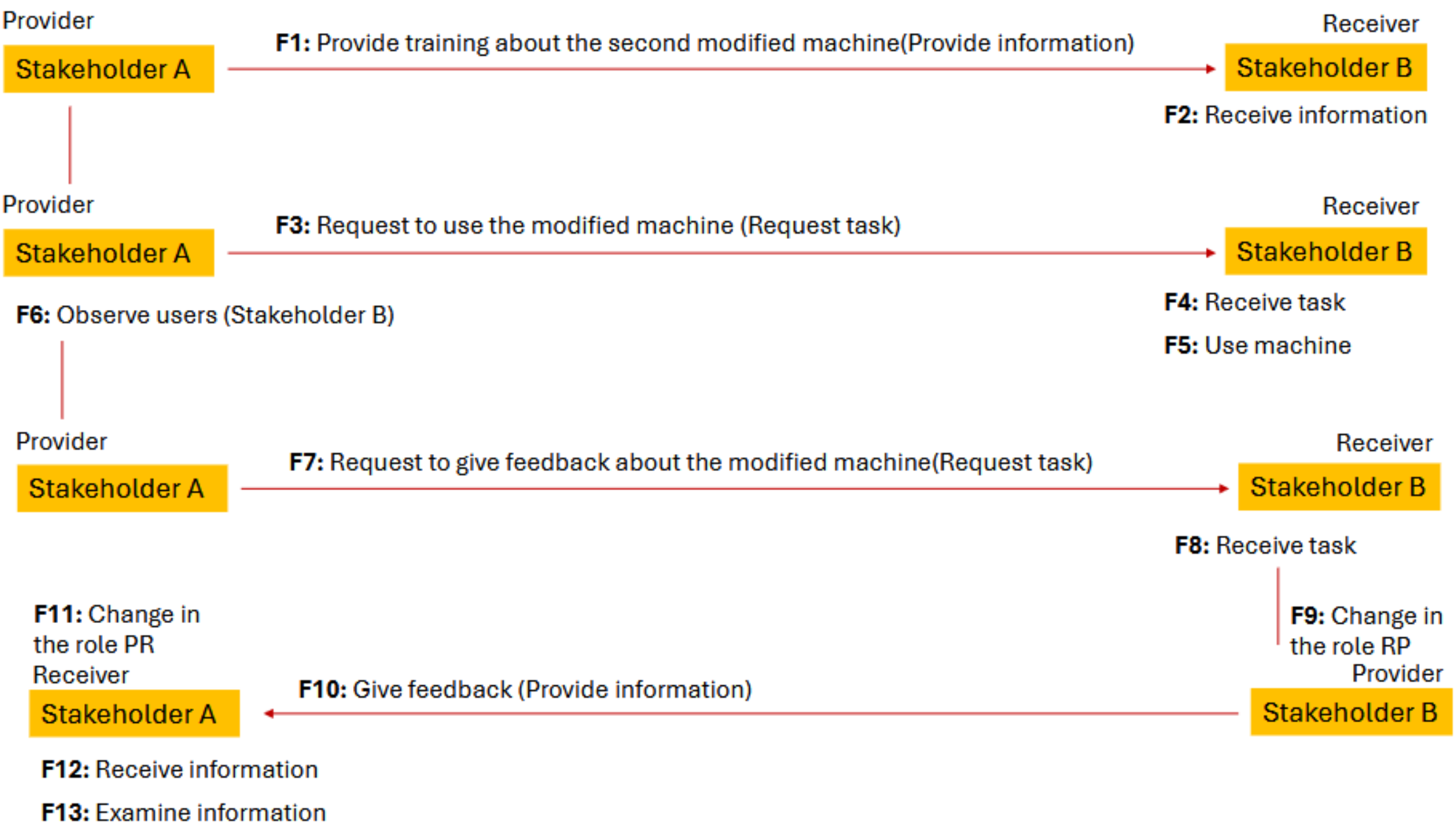

Figure 5: Interaction diagram for function analysis

Note: Some content in this supplementary file also appears in the supplementary material of Yaldiz and Chakrabarti (2024b, under review), as the two studies are built on the same conceptual background.

**Supplementary File 4**

Ten real-life case studies were selected to identify the functions (see main document Section 4 and Supplementary File 3). A brief summary of these case studies is provided below. The summary highlights the purpose of each project, the involved stakeholders, the total number of identified functions, and the evaluated lifecycle phases.

- CS1

Case Study 1 (Ogawa and Piller, 2006; Fuchs and Schreier, 2011) aims to increase user engagement in the design and decision-making stages of new product development through an online platform. In this way, the risks associated with the product lifecycle can be reduced. In this business model, users are included in the process as co-creators, evaluators, and resources. This case study highlights an approach to enhancing stakeholder involvement in the co-creation process.
Based on the information in Ogawa and Piller (2006) and Fuchs and Schreier (2011), functions are identified for planning, development and production phases in the lifecycle process of the product. The total number of functions is 179 for 11 stakeholders.

- CS2

Case Study 2 (Saini et al., 2019) focuses on improving traditional rural technology through the development of the Modified Bageshwari Wool Charkha. The aim is to enhance the efficiency and usability of the spinning process for local people. In this model, stakeholders are accepted as local users, researchers as designers and developers, workshop workers, NGOs, and others. This case study highlights an approach to increasing stakeholder empowerment by supporting regularity in their income through the development of the machine based on the local people's needs and desires. In this way, the study improves technology adoption in rural areas by promoting the inclusivity of local people to support their empowerment.
Based on the information provided by Saini, Singal, Ali, and Joshi (2019), functions are identified for the planning, development, production, and utilisation phases of the product lifecycle. A total of 385 functions are identified for 11 stakeholders.

- CS3

Case Study 3 (Singh and Gosain, 2019) focuses on the development of a low-cost groundwater-level measuring device aimed at supporting water management in rural areas. The purpose is to provide an affordable and accessible solution for local communities to monitor groundwater resources more effectively. In this model, stakeholders are accepted as local users, researchers as designers and developers, workshop workers, NGOs, and others. This case study highlights an approach to enhancing stakeholder empowerment by addressing local needs through an inclusive lifecycle process.
Based on the information provided by Singh and Gosain (2019), functions are identified for the planning, development, production, and utilisation phases of the product lifecycle. A total of 205 functions are identified for 8 stakeholders.

- CS4

Case Study 4 (Mukherji, 2022) focuses on the development of customised energy solutions to meet the specific needs of families who could not afford standard products for lighting. The case study is based on an inclusive business model, which aims to transform lives and create livelihoods by addressing both functional needs and financial constraints. In this study, stakeholders such as local people, technicians as designers and developers, and financial supporters such as local banks, and others are included. This case study highlights an approach to enhancing the inclusivity of local people who face financial constraints. It empowers them to improve their livelihoods in rural areas by addressing their needs through an inclusive lifecycle process.

Based on the information provided by Mukherji (2022), functions are identified for the planning, development, and production phases of the product lifecycle. A total of 300 functions are identified for 8 stakeholders.

- CS5

Case Study 5 (Kotwal et al., 2019) focuses on the development of an improved street food cart design to enhance food safety and hygiene for street vendors. The case study is based on the development of the food cart as part of a student class project, considering the needs and budget of the vendors. In this study, stakeholders such as local street food vendors, students as designers and developers, customers, and others are included. This case study highlights an approach to enhancing the inclusivity of local vendors by improving their working conditions and aiming to improve food safety through an inclusive lifecycle process.
Based on the information provided by Kotwal et al. (2019), functions are identified for the planning and development phases of the product lifecycle. A total of 168 functions are identified for 7 stakeholders.

- CS6

Case Study 6 (Bhat et al., 2019) focuses on the development of a low-cost full-face mask to improve the safety of stone carvers. The case study is based on the design of a face mask intended to reduce health risks caused by dust exposure during stone carving activities. In this study, stakeholders such as stone carvers, employees, researchers as designers and developers, and manufacturers are included. This case study highlights an approach to enhancing the inclusivity of local workers by providing a solution to improve their health and safety conditions through an inclusive lifecycle process.
Based on the information provided by Bhat et al. (2019), functions are identified for the planning, development, and production phases of the product lifecycle. A total of 177 functions are identified for 4 stakeholders.

- CS7

Case Study 7 (Tak et al., 2019b) focuses on the development of a cow lift designed to support downer cows that are unable to stand due to injury or illness. The case study is based on the design and implementation of rural technologies to improve animal welfare and support farmers in rural areas. In this study, stakeholders such as local farmers, researchers as designers and developers, veterinarians, workshop workers, and others are included. This case study highlights an approach to enhancing the inclusivity of local people (farmers) by addressing problems that impact their well-being and animal welfare. It aims to improve the situation by supporting an effective solution that can be utilised by all local people who face similar challenges. The solution is developed through their inclusion in an empowering lifecycle process.
Based on the information provided by Tak, Bhandakkar, and Khanolkar (2019), functions are identified for the planning, development, production, and utilisation phases of the product lifecycle. A total of 309 functions are identified for 8 stakeholders.

- CS8

Case Study 8 (Tak et al., 2019a; Haque, Tak, and Rao, 2016) focuses on improving the livelihoods of tribal communities by enhancing the supply chain and developing a Hirda decortication machine. The case study is based on supply chain improvements designed to increase production potential, reduce manual labour, and create better market access for tribal people. In this study, stakeholders such as local people, researchers as designers and developers, workshop workers, vendors, NGOs, and others are included. This case study highlights an approach to enhancing the inclusivity of local communities by addressing livelihood challenges through product development and supply chain improvements within an inclusive lifecycle process. The approach aims to empower stakeholders by involving them directly in the development and improvement processes.

Based on the information provided by Tak et al. (2019) and Haque, Tak, and Rao (2016), functions are identified for the planning, development, production, and utilisation phases of the product lifecycle. A total of 304 functions are identified for 11 stakeholders.

- CS9

Case Study 9 (Tak et al., 2019c; Khanolkar, Rao, and Ghosh, 2018) focuses on the development of floating fish cages for inland waters to create livelihood opportunities for tribal communities. The case study is based on the design and implementation of floating fish cage technology to enhance fish farming practices and generate additional income for local people. In this study, stakeholders such as local tribal communities, researchers as designers and developers, workshop workers, NGOs, and others are included. This case study highlights an approach to enhancing the inclusivity of rural communities by addressing livelihood challenges through product development and resource management within an inclusive lifecycle process. The approach aims to empower stakeholders by involving them directly in the development and improvement processes to support the regularity of their income generation.
Based on the information provided by Tak et al. (2019) and Khanolkar, Rao, and Ghosh (2018), functions are identified for the planning, development, production, and utilisation phases of the product lifecycle. A total of 316 functions are identified for 11 stakeholders.

- CS10

Case Study 10 (Ogawa and Piller, 2006; Fuchs and Schreier, 2011; Nishikawa, Schreier, and Ogawa, 2013) focuses on the development of customer-driven products to reduce the risks associated with new product development by empowering users through their inclusion in the lifecycle process. The case study is based on an inclusive business model, where customers are actively involved in generating product ideas, evaluating them, and choosing final products for production. In this study, stakeholders such as customers (users, co-creators, evaluators, and resources), designers, managers, and others are included. This case study highlights an approach to enhancing the inclusivity of customers by involving them directly in the co-creation and decision-making processes within an inclusive lifecycle process. The approach aims to empower stakeholders by providing them with opportunities to contribute to idea generation and decision-making.
Based on the information provided by Ogawa and Piller (2006), Fuchs and Schreier (2011), and Nishikawa, Schreier, and Ogawa (2013), functions are identified for the planning, development, and production phases of the product lifecycle. A total of 347 functions are identified for 15 stakeholders.

As summarised through the purposes of each case study, the case studies adopt two approaches to empower people. The first approach, as seen in Case Study 1 and Case Study 10, is the inclusion of people in different roles to contribute to idea generation and decision-making processes, which also supports companies by reducing the risks associated with customer engagement. The second approach, followed in the remaining case studies, is to include people in the lifecycle processes to develop a product based on their needs, feedback, and budget, in order to support their income generation. This highlights the connection between the lifecycle processes of different products. It starts with developing a product (the lifecycle of the machine), then using the machine (the utilisation phase of the machine), and producing a different product by using the machine (combined with the lifecycle of the produced product). After that, the utilisation phase follows, eventually leading to the start of the recirculation phases for all products. Therefore, the inclusivity of people is supported through empowering lifecycle processes by improving their incomes and increasing their contributions to decision-making. In addition, as explained in the brief summaries of the case studies, information about the recirculation phases of products could not be found. However, the planning phase should ideally also include information or strategies for the recirculation phase, since it is an essential part of the lifecycle process.

Table 1: Summary of Case Study Characteristics: Functions, Stakeholders, and Lifecycle Phases

| **Case Study (CS)** | **#Function** | **#Stakeholders** | **LC Phases** |
|---|---|---|---|
| 1 | 179 | 11 | Planning - Development - Production |
| 2 | 385 | 9 | Planning - Development - Production - Utilisation |
| 3 | 205 | 8 | Planning - Development - Production - Utilisation |
| 4 | 300 | 8 | Planning - Development - Production - Utilisation |
| 5 | 168 | 7 | Planning - Development |
| 6 | 177 | 4 | Planning - Development - Production |
| 7 | 309 | 8 | Planning - Development - Production - Utilisation |
| 8 | 304 | 11 | Planning - Development - Production - Utilisation |
| 9 | 316 | 11 | Planning - Development - Production - Utilisation |
| 10 | 347 | 15 | Planning - Development - Production |

Table 2: Distribution of Functions Across Lifecycle Phases in Case Studies

| **Lifecycle Phases** | **#Functions** | | | | | | | | | |
|---|---|---|---|---|---|---|---|---|---|---|
| | **CS1** | **CS2** | **CS3** | **CS4** | **CS5** | **CS6** | **CS7** | **CS8** | **CS9** | **CS10** |
| Planning | 89 | 82 | 78 | 175 | 89 | 76 | 105 | 112 | 165 | 162 |
| Development | 69 | 204 | 30 | 58 | 79 | 72 | 59 | 58 | 40 | 175 |
| Production | 21 | 69 | 48 | 51 | 0 | 29 | 89 | 88 | 63 | 10 |
| Utilisation | 0 | 30 | 49 | 16 | 0 | 0 | 56 | 46 | 48 | 0 |
| Recirculation | 0 | 0 | 0 | 0 | 0 | 0 | 0 | 0 | 0 | 0 |

## References

Bhat, S., Doshi, N., Bharadwaj, C.D., Singh, S.N., Patel, Y. and Saha, S.K. (2019) 'Design of a low-cost full-face mask for stone carvers', in *Rural Technology Development and Delivery: RuTAG and Its Synergy with Other Initiatives*. Singapore: Springer, pp. 279–285.

Fuchs, C. and Schreier, M. (2011) 'Customer empowerment in new product development', *Journal of Product Innovation Management*, 28(1), pp. 17–32.

Haque, T., Tak, P. and Rao, A.B. (2016) 'Better livelihood opportunities for tribals through supply chain interventions of Hirda', in *2016 IEEE Region 10 Humanitarian Technology Conference (R10-HTC)*, pp. 1–5.

Khanolkar, R.S., Rao, A.B. and Ghosh, S. (2018) 'RuTAG IIT Bombay floating fish cages for livelihood opportunities for tribals in Dimbhe area', in *Techno-Societal 2016: Proceedings of the International Conference on Advanced Technologies for Societal Applications*. Cham: Springer, pp. 27–35.

Kotwal, V., Satya, S., Naik, S.N., Dahiya, A. and Kumar, J. (2019) 'Street food cart design: A critical component of food safety', in *Rural Technology Development and Delivery: RuTAG and Its Synergy with Other Initiatives*. Singapore: Springer, pp. 263–278.

Mukherji, S. (2022) 'Empowering communities', in *Inclusive Business Models: Transforming Lives and Creating Livelihoods*. Cambridge: Cambridge University Press, pp. 128–159.

Nishikawa, H., Schreier, M. and Ogawa, S. (2013) 'User-generated versus designer-generated products: A performance assessment at Muji', *International Journal of Research in Marketing*, 30(2), pp. 160–167.

Ogawa, S. and Piller, F.T. (2006) 'Reducing the risks of new product development', *MIT Sloan Management Review*, 47(2).

Saini, R.P., Singal, S.K., Ali, I. and Joshi, R.C. (2019) 'Development of modified Bageshwari wool Charkha', in *Rural Technology Development and Delivery: RuTAG and Its Synergy with Other Initiatives*. Singapore: Springer, pp. 347–357.

Singh, D.P. and Gosain, A.K. (2019) 'Development of a low-cost groundwater-level measuring device', in *Rural Technology Development and Delivery: RuTAG and Its Synergy with Other Initiatives*. Singapore: Springer, pp. 225–235.

Tak, P.P. et al. (2019a) 'Study of supply chain, production potential of Hirda and design of Hirda decortication machine for livelihood generation for tribal people', in *Rural Technology Development and Delivery: RuTAG and Its Synergy with Other Initiatives*. Singapore: Springer, pp. 249–261.

Tak, P.P., Bhandakkar, T.K. and Khanolkar, R.S. (2019b) 'Designing a cow lift for downer cow: Experience of working on a rural technology', in *Rural Technology Development and Delivery: RuTAG and Its Synergy with Other Initiatives*. Singapore: Springer, pp. 323–333.

Tak, P.P. et al. (2019c) 'Evolution of "floating fish cages for inland waters" developed by RuTAG IIT Bombay', in *Rural Technology Development and Delivery: RuTAG and Its Synergy with Other Initiatives*. Singapore: Springer, pp. 237–247.

**Supplementary File 4**

**Framework Application Results Across 10 Case Studies**

**CS1**

| Stakeholders | E1 | E2 | E3 | E4 | I1 | I2 | I3 | I4 |
|---|---|---|---|---|---|---|---|---|
| A | 107 | 16.5 | 695.5 | 695.5 | 34 | 24 | 102 | 42 |
| B | 7 | 1.5 | 3.5 | 3.5 | 15 | 45 | 45 | 9 |
| C | 5 | 1 | 2.5 | 2.5 | 6 | 12 | 12 | 4 |
| D | 14 | 1.5 | 10 | 10 | 8 | 16 | 16 | 4 |
| E | 15 | 2 | 20.5 | 20.5 | 6 | 8 | 12 | 6 |
| F | 8 | 1.5 | 6.5 | 6.5 | 6 | 12 | 12 | 4 |
| G | 2 | 0.5 | 1 | 1 | 5 | 5 | 5 | 1 |
| H | 61 | 10 | 293 | 293 | 12 | 36 | 36 | 9 |
| I | 5 | 1 | 5 | 5 | 4 | 2 | 4 | 2 |
| J | 26 | 3.5 | 47 | 47 | 6 | 8 | 12 | 6 |
| K | 26 | 3.5 | 91 | 91 | 6 | 2 | 6 | 3 |

**CS2**

| Stakeholders | E1 | E2 | E3 | E4 | I1 | I2 | I3 | I4 |
|---|---|---|---|---|---|---|---|---|
| A | 98 | 11.5 | 335.5 | 335.5 | 21 | 48 | 84 | 28 |
| B | 46 | 7 | 84 | 84 | 24 | 64 | 96 | 24 |
| C | 17 | 2.5 | 42.5 | 42.5 | 5 | 5 | 5 | 1 |
| D | 89 | 14 | 401.5 | 401.5 | 30 | 48 | 120 | 40 |
| E | 140 | 22.25 | 849.5 | 849.5 | 60 | 64 | 240 | 60 |
| F | 9 | 1 | 9 | 9 | 5 | 5 | 5 | 1 |
| G | 75 | 11 | 416 | 416 | 6 | 6 | 8 | 4 |
| H | 73 | 12.5 | 547 | 547 | 6 | 6 | 8 | 4 |
| I | 35 | 6 | 130.5 | 130.5 | 2 | 3 | 3 | 3 |

**CS3**

| Stakeholders | E1 | E2 | E3 | E4 | I1 | I2 | I3 | I4 |
|---|---|---|---|---|---|---|---|---|
| A | 14 | 1.5 | 21 | 21 | 8 | 4 | 8 | 4 |
| B | 41 | 9 | 190 | 190 | 24 | 24 | 48 | 16 |
| C | 84 | 13 | 400.5 | 400.5 | 24 | 32 | 64 | 16 |
| D | 118 | 21 | 726 | 726 | 28 | 64 | 112 | 28 |
| E | 8 | 1.5 | 12 | 12 | 1 | 1 | 1 | 1 |
| F | 19 | 3.5 | 66.5 | 66.5 | 4 | 2 | 4 | 2 |
| G | 15 | 2.5 | 37.5 | 37.5 | 5 | 5 | 5 | 1 |
| H | 33 | 5 | 165 | 165 | 4 | 2 | 4 | 2 |

**CS4**

| Stakeholders | E1 | E2 | E3 | E4 | I1 | I2 | I3 | I4 |
|---|---|---|---|---|---|---|---|---|
| A | 42 | 5 | 210 | 210 | 20 | 5 | 20 | 4 |
| B | 40 | 7.5 | 109.5 | 109.5 | 40 | 160 | 160 | 32 |
| C | 59 | 8.5 | 501.5 | 501.5 | 8 | 2 | 8 | 4 |
| D | 201 | 30.5 | 2363.5 | 2363.5 | 22 | 32 | 88 | 44 |
| E | 57 | 8 | 456 | 456 | 10 | 2 | 10 | 5 |
| F | 33 | 6 | 198 | 198 | 6 | 2 | 6 | 3 |
| G | 15 | 3 | 45 | 45 | 1 | 1 | 1 | 1 |
| H | 69 | 11.5 | 328.5 | 328.5 | 28 | 64 | 112 | 28 |

**CS5**

| Stakeholders | E1 | E2 | E3 | E4 | I1 | I2 | I3 | I4 |
|---|---|---|---|---|---|---|---|---|
| A | 99 | 14 | 840 | 840 | 25 | 20 | 50 | 10 |
| B | 14 | 2.5 | 35 | 35 | 4 | 4 | 4 | 2 |
| C | 14 | 2.5 | 35 | 35 | 3 | 3 | 3 | 1 |
| D | 14 | 2.5 | 35 | 35 | 3 | 3 | 3 | 1 |
| E | 14 | 2.5 | 35 | 35 | 4 | 4 | 4 | 2 |
| F | 62 | 9.5 | 589 | 589 | 6 | 3 | 6 | 2 |
| G | 32 | 5.5 | 176 | 176 | 4 | 4 | 4 | 2 |

**CS6**

| Stakeholders | E1 | E2 | E3 | E4 | I1 | I2 | I3 | I4 |
|---|---|---|---|---|---|---|---|---|
| A | 137 | 19.5 | 1032.5 | 1032.5 | 30 | 45 | 90 | 18 |
| B | 15 | 2.5 | 37.5 | 37.5 | 3 | 3 | 3 | 1 |
| C | 75 | 11.5 | 331.5 | 331.5 | 6 | 18 | 18 | 9 |
| D | 30 | 5.5 | 88.5 | 88.5 | 18 | 36 | 36 | 12 |

**CS7**

| Stakeholders | E1 | E2 | E3 | E4 | I1 | I2 | I3 | I4 |
|---|---|---|---|---|---|---|---|---|
| A | 204 | 28 | 1709.5 | 1709.5 | 36 | 64 | 144 | 36 |
| B | 74 | 11 | 344 | 344 | 15 | 27 | 45 | 15 |
| C | 73 | 10 | 335.5 | 335.5 | 15 | 27 | 45 | 15 |
| D | 15 | 2.5 | 37.5 | 37.5 | 5 | 5 | 5 | 1 |
| E | 26 | 4 | 104 | 104 | 4 | 4 | 4 | 1 |
| F | 16 | 3 | 48 | 48 | 1 | 1 | 1 | 1 |
| G | 38 | 7 | 266 | 266 | 2 | 1 | 2 | 2 |
| H | 61 | 8.5 | 518.5 | 518.5 | 4 | 2 | 4 | 2 |

**CS8**

| Stakeholders | E1 | E2 | E3 | E4 | I1 | I2 | I3 | I4 |
|---|---|---|---|---|---|---|---|---|
| A | 176 | 26 | 1269 | 1269 | 28 | 64 | 112 | 28 |
| B | 97 | 16 | 849.5 | 849.5 | 36 | 36 | 108 | 27 |
| C | 8 | 1.5 | 12 | 12 | 9 | 9 | 9 | 3 |
| D | 8 | 1.5 | 12 | 12 | 9 | 9 | 9 | 3 |
| E | 8 | 1.5 | 12 | 12 | 5 | 5 | 5 | 1 |
| F | 8 | 1.5 | 12 | 12 | 5 | 5 | 5 | 1 |
| G | 15 | 2.5 | 37.5 | 37.5 | 5 | 5 | 5 | 1 |
| H | 49 | 7.5 | 220.5 | 220.5 | 36 | 36 | 72 | 24 |
| I | 16 | 3 | 48 | 48 | 1 | 1 | 1 | 1 |
| J | 38 | 7 | 266 | 266 | 4 | 2 | 4 | 2 |
| K | 60 | 8.5 | 510 | 510 | 4 | 2 | 4 | 2 |

**CS9**

| Stakeholders | E1 | E2 | E3 | E4 | I1 | I2 | I3 | I4 |
|---|---|---|---|---|---|---|---|---|
| A | 111 | 14 | 601 | 601 | 44 | 64 | 176 | 44 |
| B | 116 | 15.5 | 529.5 | 529.5 | 33 | 48 | 132 | 44 |
| C | 68 | 6 | 231 | 232 | 14 | 24 | 56 | 28 |
| D | 13 | 2 | 26 | 26 | 4 | 2 | 4 | 2 |
| E | 62 | 8.5 | 349 | 349 | 20 | 20 | 40 | 8 |
| F | 157 | 21.5 | 1086 | 1086 | 52 | 64 | 208 | 52 |
| G | 6 | 1 | 6 | 6 | 2 | 2 | 2 | 1 |
| H | 15 | 2.5 | 37.5 | 37.5 | 5 | 5 | 5 | 1 |
| I | 5 | 1 | 5 | 5 | 1 | 1 | 1 | 1 |
| J | 14 | 2.5 | 35 | 35 | 4 | 2 | 4 | 2 |
| K | 16 | 1.5 | 24 | 24 | 2 | 2 | 2 | 1 |

**CS10**

| Stakeholders | E1 | E2 | E3 | E4 | I1 | I2 | I3 | I4 |
|---|---|---|---|---|---|---|---|---|
| A | 78 | 13.75 | 509.75 | 509.75 | 33 | 27 | 99 | 33 |
| B | 2 | 0.5 | 1 | 1 | 5 | 5 | 5 | 1 |
| C | 2 | 0.5 | 1 | 1 | 3 | 3 | 3 | 1 |
| D | 6 | 1 | 6 | 6 | 4 | 4 | 4 | 1 |
| E | 22 | 1.5 | 33 | 33 | 4 | 2 | 4 | 2 |
| F | 38 | 7.5 | 246.5 | 246.5 | 14 | 8 | 28 | 14 |
| G | 15 | 1.5 | 22.5 | 22.5 | 6 | 3 | 6 | 2 |
| H | 142 | 23.5 | 1662.5 | 1662.5 | 16 | 18 | 48 | 24 |
| I | 120 | 19.5 | 1274.5 | 1274.5 | 18 | 12 | 36 | 12 |
| J | 22 | 4 | 47 | 47 | 4 | 8 | 8 | 4 |
| K | 12 | 1.75 | 21 | 21 | 8 | 4 | 8 | 2 |
| L | 44 | 7 | 166 | 166 | 12 | 8 | 24 | 12 |
| M | 13 | 2 | 13 | 13 | 6 | 12 | 12 | 4 |
| N | 38 | 5.5 | 153.5 | 153.5 | 6 | 8 | 12 | 6 |
| O | 13 | 2 | 26 | 26 | 5 | 5 | 5 | 1 |

**Supplementary File 5**

**Application of the Framework in Case Study 3 (CS3)**

| Function Pair | Function No | Stakeholder | Role | Lifecycle Phase | #Incl. Stk. | #Task | EI | Means | Freq. | Ext. | Groups of the needs | # need groups | Groups of the sources | # resource groups | DL | Power |
|---|---|---|---|---|---|---|---|---|---|---|---|---|---|---|---|---|
| F1&F2 | F1 | D | Provider | Planning | 2 | 1 | 2 | 1 | 29 | 29 | Participation and Doing | 2 | Position | 1 | 0.5 | 14.5 |
| | F2 | C | Receiver | Planning | 2 | 1 | 2 | 1 | 22 | 22 | Understanding and Doing | 2 | Position | 1 | 0.5 | 11 |
| F3&F4 &F5 | F3 | D | Provider | Planning | 3 | 1 | 3 | 1 | 29 | 29 | Participation and Doing | 2 | Position | 1 | 0.5 | 14.5 |
| | F4 | C | Receiver | Planning | 3 | 1 | 3 | 1 | 22 | 22 | Participation and Doing | 2 | Position | 1 | 0.5 | 11 |
| | F5 | A | Receiver | Planning | 3 | 1 | 3 | 1 | 8 | 8 | Creation and Interacting | 2 | Capability | 1 | 0.5 | 4 |
| X | F6 | D | Receiver | Planning | 1 | 1 | 1 | 1 | 29 | 29 | Participation and Having | 2 | Position | 1 | 0 | 0 |
| X | F7 | C | Provider | Planning | 1 | 1 | 1 | 1 | 22 | 22 | Participation and Having | 2 | Position | 1 | 0 | 0 |
| x | F8 | A | Provider | Planning | 1 | 1 | 1 | 1 | 8 | 8 | Creation and Interacting | 2 | Capability | 1 | 0 | 0 |
| F9&F10 | F9 | C | Provider | Planning | 2 | 1 | 2 | 1 | 22 | 22 | Participation and Doing | 2 | Position, Capability | 2 | 0.5 | 11 |
| | F10 | B | Receiver | Planning | 2 | 1 | 2 | 1 | 9 | 9 | Creation and Interacting | 2 | Capability | 1 | 0.5 | 4.5 |
| X | F11 | C | Receiver | Planning | 1 | 1 | 1 | 1 | 22 | 22 | Participation and Having | 2 | Position | 1 | 0 | 1 |
| X | F12 | B | Provider | Planning | 1 | 1 | 1 | 1 | 9 | 9 | Creation and Interacting | 2 | Capability | 1 | 0 | 0 |
| F13&F14 | F13 | B | Provider | Planning | 2 | 1 | 2 | 1 | 9 | 9 | Creation and Interacting | 2 | Capability | 1 | 0.5 | 4.5 |
| | F14 | C | Receiver | Planning | 2 | 1 | 2 | 1 | 22 | 22 | Participation and Doing | 2 | Position | 1 | 0.5 | 11 |
| X | F15 | C | Provider | Planning | 1 | 1 | 1 | 1 | 22 | 22 | Participation and Having | 2 | Position | 1 | 0 | 0 |
| F16 | F16 | C | Provider | Planning | 3 | 1 | 3 | 1 | 22 | 22 | Participation and Doing | 2 | Position, Capability | 2 | 0.5 | 11 |
| &F17 &F18 | F17 | A | Provider | Planning | 3 | 1 | 3 | 1 | 8 | 8 | Creation and Interacting | 2 | Capability | 1 | 0.5 | 4 |
| | F18 | D | Receiver | Planning | 3 | 1 | 3 | 1 | 29 | 29 | Participation and Doing | 2 | Position | 1 | 0.5 | 14.5 |
| X | F19 | D | - | Planning | 1 | 1 | 1 | 1 | 29 | 29 | Participation and Having | 2 | Position | 1 | 0 | 0 |
| X | F20 | D | Provider | Planning | 1 | 1 | 1 | 1 | 29 | 29 | Participation and Having | 2 | Position | 1 | 0 | 0 |
| X | F21 | A | Receiver | Planning | 1 | 1 | 1 | 1 | 8 | 8 | Creation and Interacting | 2 | Capability | 1 | 0 | 0 |
| F22 &F23 | F22 | D | Provider | Planning | 2 | 1 | 2 | 1 | 29 | 29 | Participation and Doing | 2 | Position | 1 | 0.5 | 14.5 |
| | F23 | A | Receiver | Planning | 2 | 1 | 2 | 1 | 8 | 8 | Creation and Interacting | 2 | Capability | 1 | 0.5 | 4 |
| X | F24 | A | - | Planning | 1 | 1 | 1 | 1 | 8 | 8 | Creation and Interacting | 2 | Capability | 1 | 0 | 0 |
| X | F25 | A | - | Planning | 1 | 1 | 1 | 1 | 8 | 8 | Creation and Interacting | 2 | Capability | 1 | 0 | 0 |
| X | F26 | D | - | Planning | 1 | 1 | 1 | 1 | 29 | 29 | Participation and Having | 2 | Position | 1 | 0 | 0 |
| F27 &F28 | F27 | A | Provider | Planning | 2 | 1 | 2 | 1 | 8 | 8 | Participation and Having | 2 | Position | 1 | 0 | 0 |
| | F28 | D | Receiver | Planning | 2 | 1 | 2 | 1 | 29 | 29 | Participation and Doing | 2 | Position | 1 | 0.5 | 14.5 |
| X | F29 | D | - | Planning | 1 | 1 | 1 | 1 | 29 | 29 | Participation and Having | 2 | Position | 1 | 0 | 0 |
| F30 &F31 | F30 | C | Provider | Planning | 2 | 1 | 2 | 1 | 22 | 22 | Participation and Doing | 2 | Position, Capability | 2 | 0.5 | 11 |
| | F31 | D | Receiver | Planning | 2 | 1 | 2 | 1 | 29 | 29 | Participation and Doing | 2 | Position | 1 | 0.5 | 14.5 |
| X | F32 | C | Receiver | Planning | 1 | 1 | 1 | 1 | 22 | 22 | Participation and Having | 2 | Position | 1 | 0 | 0 |
| X | F33 | D | Provider | Planning | 1 | 1 | 1 | 1 | 29 | 29 | Participation and Having | 2 | Position | 1 | 0 | 0 |
| F34 &F35 | F34 | D | Provider | Planning | 2 | 1 | 2 | 1 | 29 | 29 | Participation and Doing | 2 | Position | 1 | 0.5 | 14.5 |
| | F35 | C | Receiver | Planning | 2 | 1 | 2 | 1 | 22 | 22 | Participation and Doing | 2 | Position, Capability | 2 | 0.5 | 11 |

| X | F36 | D | Receiver | Planning | 1 | 1 | 1 | 1 | 29 | 29 | Participation and Having | 2 | Position | 1 | 0 | 0 |
|---|---|---|---|---|---|---|---|---|---|---|---|---|---|---|---|---|
| X | F37 | C | Provider | Planning | 1 | 1 | 1 | 1 | 22 | 22 | Participation and Having | 2 | Position | 1 | 0 | 0 |
| X | F38 | B | Receiver | Planning | 1 | 1 | 1 | 1 | 9 | 9 | Creation and Interacting | 2 | Capability | 1 | 0 | 0 |
| F39 &F40 | F39 | C | Provider | Planning | 2 | 1 | 2 | 1 | 22 | 22 | Participation and Doing | 2 | Position, Capability | 2 | 0.5 | 11 |
| | F40 | B | Receiver | Planning | 2 | 1 | 2 | 1 | 9 | 9 | Creation and Interacting | 2 | Capability | 1 | 0.5 | 4.5 |
| X | F41 | B | Provider | Planning | 1 | 1 | 1 | 1 | 9 | 9 | Creation and Interacting | 2 | Capability | 1 | 0 | 0 |
| X | F42 | C | Receiver | Planning | 1 | 1 | 1 | 1 | 22 | 22 | Participation and Having | 2 | Position | 1 | 0 | 0 |
| F43 &F44 | F43 | B | Provider | Planning | 2 | 1 | 2 | 1 | 9 | 9 | Creation and Interacting | 2 | Capability | 1 | 0.5 | 4.5 |
| | F44 | C | Receiver | Planning | 2 | 1 | 2 | 1 | 22 | 22 | Participation and Doing | 2 | Position, Capability | 2 | 0.5 | 11 |
| X | F45 | C | - | Planning | 1 | 1 | 1 | 1 | 22 | 22 | Participation and Having | 2 | Position | 1 | 0 | 0 |
| X | F46 | C | - | Planning | 1 | 1 | 1 | 1 | 22 | 22 | Participation and Having | 2 | Position | 1 | 0 | 0 |
| X | F47 | C | - | Planning | 1 | 1 | 1 | 1 | 22 | 22 | Participation and Having | 2 | Position | 1 | 0 | 0 |
| X | F48 | C | Provider | Planning | 1 | 1 | 1 | 1 | 22 | 22 | Participation and Having | 2 | Position | 1 | 0 | 0 |
| F49 &F50 | F49 | C | Provider | Planning | 2 | 1 | 2 | 1 | 22 | 22 | Participation and Doing | 2 | Position, Capability | 2 | 0.5 | 11 |
| | F50 | D | Receiver | Planning | 2 | 1 | 2 | 1 | 29 | 29 | Participation and Doing | 2 | Position | 1 | 0.5 | 14.5 |
| X | F51 | D | - | Planning | 1 | 1 | 1 | 1 | 29 | 29 | Understanding and Doing | 2 | Position | 1 | 0 | 0 |
| X | F52 | D | - | Planning | 1 | 1 | 1 | 1 | 29 | 29 | Understanding and Doing | 2 | Position | 1 | 0 | 0 |
| X | F53 | D | Provider | Planning | 1 | 1 | 1 | 1 | 29 | | Participation and Doing | 2 | Position | 1 | 0 | 0 |
| F54 | F54 | D | Provider | Planning | 2 | 1 | 2 | 1 | 29 | 29 | Participation and Having | 2 | Position | 1 | 0.5 | 14.5 |

| &F55 | F55 | G | Receiver | Planning | 2 | 1 | 2 | 1 | 10 | 10 | Participation and Having | 2 | Position | 1 | 0.5 | 5 |
|---|---|---|---|---|---|---|---|---|---|---|---|---|---|---|---|---|
| F56 &F57 | F56 | D | Provider | Planning | 2 | 1 | 2 | 1 | 29 | 29 | Participation and Having | 2 | Position | 1 | 0.5 | 14.5 |
| | F57 | G | Receiver | Planning | 2 | 1 | 2 | 1 | 10 | 10 | Participation and Having | 2 | Position | 1 | 0.5 | 5 |
| X | F58 | G | - | Planning | 1 | 1 | 1 | 1 | 10 | 10 | Understanding and Doing | 2 | Position | 1 | 0 | 0 |
| X | F59 | G | Provider | Planning | 1 | 1 | 1 | 1 | 10 | 10 | Participation and Having | 2 | Position | 1 | 0 | 0 |
| X | F60 | D | Receiver | Planning | 1 | 1 | 1 | 1 | 29 | 29 | Participation and Having | 2 | Position | 1 | 0 | 0 |
| F61 &F62 | F61 | G | Provider | Planning | 2 | 1 | 2 | 1 | 10 | 10 | Participation and Having | 2 | Position | 1 | 0.5 | 5 |
| | F62 | D | Receiver | Planning | 2 | 1 | 2 | 1 | 29 | 29 | Participation and Having | 2 | Position | 1 | 0.5 | 14.5 |
| X | F63 | D | - | Planning | 1 | 1 | 1 | 1 | 29 | 29 | Understanding and Doing | 2 | Position | 1 | 0 | 0 |
| X | F64 | D | Provider | Planning | 1 | 1 | 1 | 1 | 29 | 29 | Participation and Having | 2 | Position | 1 | 0 | 0 |
| X | F65 | G | Receiver | Planning | 1 | 1 | 1 | 1 | 10 | 10 | Participation and Having | 2 | Position | 1 | 0 | 0 |
| F66 &F67 | F66 | D | Provider | Planning | 2 | 1 | 2 | 1 | 29 | 29 | Participation and Having | 2 | Position | 1 | 0.5 | 14.5 |
| | F67 | G | Receiver | Planning | 2 | 1 | 2 | 1 | 10 | 10 | Participation and Having | 2 | Position | 1 | 0.5 | 5 |
| X | F68 | G | - | Planning | 1 | 1 | 1 | 1 | 10 | 10 | Understanding and Doing | 2 | Position | 1 | 0 | 0 |
| X | F69 | G | Provider | Planning | 1 | 1 | 1 | 1 | 10 | 10 | Participation and Having | 2 | Position | 1 | 0 | 0 |
| X | F70 | D | Receiver | Planning | 1 | 1 | 1 | 1 | 29 | 29 | Participation and Having | 2 | Position | 1 | 0 | 0 |
| F71&F72 | F71 | G | Provider | Planning | 2 | 1 | 2 | 1 | 10 | 10 | Participation and Having | 2 | Position | 1 | 0.5 | 5 |
| | F72 | D | Receiver | Planning | 2 | 1 | 2 | 1 | 29 | 29 | Participation and Having | 2 | Position | 1 | 0.5 | 14.5 |
| X | F73 | D | Provider | Planning | 1 | 1 | 1 | 1 | 29 | 29 | Participation and Having | 2 | Position | 1 | 0 | 0 |
| X | F74 | C | Receiver | Planning | 1 | 1 | 1 | 1 | 22 | 22 | Participation and Having | 2 | Position | 1 | 0 | 0 |
| X | F75 | B | Receiver | Planning | 1 | 1 | 1 | 1 | 9 | 9 | Creation and Interacting | 2 | Capability | 1 | 0 | 0 |
| F76 &F77 &F78 | F76 | D | Provider | Planning | 3 | 1 | 3 | 1 | 29 | 29 | Participation and Having | 2 | Position | 1 | 0.5 | 14.5 |
| | F77 | C | Receiver | Planning | 3 | 1 | 3 | 1 | 22 | 22 | Participation and Having | 2 | Position | 1 | 0.5 | 11 |
| | F78 | B | Receiver | Planning | 3 | 1 | 3 | 1 | 9 | 9 | Creation and Interacting | 2 | Capability | 1 | 0.5 | 4.5 |
| X | F79 | D | - | Development | 1 | 1 | 1 | 1 | 20 | 20 | Creation and Doing | 2 | sition, Capabi | 2 | 0 | 0 |
| X | F80 | D | - | Development | 1 | 1 | 1 | 1 | 20 | 20 | Creation and Doing | 2 | sition, Capabi | 2 | 0 | 0 |
| X | F81 | D | - | Development | 1 | 1 | 1 | 1 | 20 | 20 | Creation and Doing | 2 | sition, Capabi | 2 | 0 | 0 |
| X | F82 | D | - | Development | 1 | 1 | 1 | 1 | 20 | 20 | Creation and Doing | 2 | sition, Capabi | 2 | 0 | 0 |
| X | F83 | D | - | Development | 1 | 1 | 1 | 1 | 20 | 20 | Creation and Doing | 2 | sition, Capabi | 2 | 0 | 0 |
| X | F84 | D | - | Development | 1 | 1 | 1 | 1 | 20 | 20 | Creation and Doing | 2 | sition, Capabi | 2 | 0 | 0 |
| F85 &F86 | F85 | D | Provider | Development | 2 | 1 | 2 | 1 | 20 | 20 | Participation and Doing | 2 | Position | 1 | 0.5 | 10 |
| | F86 | C | Receiver | Development | 2 | 1 | 2 | 1 | 10 | 10 | Participation and Doing | 2 | Position | 1 | 0.5 | 5 |
| X | F87 | C | - | Development | 1 | 1 | 1 | 1 | 10 | 10 | Understanding and Doing | 2 | Position | 1 | 0 | 0 |
| F88 &F89 | F88 | D | Provider | Development | 2 | 1 | 2 | 1 | 20 | 20 | Participation and Doing | 2 | Position | 1 | 0.5 | 10 |
| | F89 | C | Receiver | Development | 2 | 1 | 2 | 1 | 10 | 10 | Participation and Doing | 2 | Position | 1 | 0.5 | 5 |
| X | F90 | C | Provider | Development | 1 | 1 | 1 | 1 | 10 | 10 | Participation and Doing | 2 | Position | 1 | 0 | 0 |
| X | F91 | D | Receiver | Development | 1 | 1 | 1 | 1 | 20 | 20 | Participation and Doing | 2 | Position | 1 | 0 | 0 |
| F92 | F92 | C | Provider | Development | 2 | 1 | 2 | 1 | 10 | 10 | Participation and Doing | 2 | Position | 1 | 0.5 | 5 |
| &F93 | F93 | D | Receiver | Development | 2 | 1 | 2 | 1 | 20 | 20 | Participation and Doing | 2 | Position | 1 | 0.5 | 10 |
| X | F94 | D | Provider | Development | 1 | 1 | 1 | 1 | 20 | 20 | Participation and Doing | 2 | Position | 1 | 0 | 0 |
| X | F95 | C | Receiver | Development | 1 | 1 | 1 | 1 | 10 | 10 | Participation and Doing | 2 | Position | 1 | 0 | 0 |
| F96 &F97 | F96 | D | Provider | Development | 2 | 1 | 2 | 1 | 20 | 20 | Participation and Doing | 2 | Position | 1 | 0.5 | 10 |
| | F97 | C | Receiver | Development | 2 | 1 | 2 | 1 | 10 | 10 | Participation and Doing | 2 | Position | 1 | 0.5 | 5 |
| X | F98 | C | - | Development | 1 | 1 | 1 | 1 | 10 | 10 | Understanding and Doing | 2 | Position | 1 | 0 | 0 |
| X | F99 | C | Provider | Development | 1 | 1 | 1 | 1 | 10 | 10 | Participation and Doing | 2 | Position | 1 | 0 | 0 |
| X | F100 | D | Receiver | Development | 1 | 1 | 1 | 1 | 20 | 20 | Participation and Doing | 2 | Position | 1 | 0 | 0 |
| F101 &F102 | F101 | C | Provider | Development | 2 | 1 | 2 | 1 | 10 | 10 | Participation and Doing | 2 | Position | 1 | 0.5 | 5 |
| | F102 | D | Receiver | Development | 2 | 1 | 2 | 1 | 20 | 20 | Participation and Doing | 2 | Position | 1 | 0.5 | 10 |
| X | F103 | D | - | Development | 1 | 1 | 1 | 1 | 20 | 20 | Understanding and Doing | 2 | Position | 1 | 0 | 0 |
| X | F104 | D | - | Development | 1 | 1 | 1 | 1 | 20 | 20 | Creation and Doing | 2 | sition, Capabi | 2 | 0 | 0 |
| X | F105 | D | - | Development | 1 | 1 | 1 | 1 | 20 | 20 | Creation and Doing | 2 | sition, Capabi | 2 | 0 | 0 |
| X | F106 | D | - | Development | 1 | 1 | 1 | 1 | 20 | 20 | Creation and Doing | 2 | sition, Capabi | 2 | 0 | 0 |
| X | F107 | D | - | Development | 1 | 1 | 1 | 1 | 20 | 20 | Creation and Doing | 2 | sition, Capabi | 2 | 0 | 0 |
| X | F108 | D | Provider | Development | 1 | 1 | 1 | 1 | 20 | 20 | Participation and Doing | 2 | Position | 1 | 0 | 0 |
| F109 &F110 | F109 | D | Provider | Production | 2 | 1 | 2 | 1 | 10 | 10 | Participation and Having | 2 | Position | 1 | 0.5 | 5 |
| | F110 | H | Receiver | Production | 2 | 1 | 2 | 1 | 21 | 21 | Participation and Having | 2 | Position | 1 | 0.5 | 10.5 |
| X | F111 | H | - | Production | 1 | 1 | 1 | 1 | 21 | 21 | erstanding and D | 2 | Position | 1 | 0 | 0 |
| F112 | F112 | D | Provider | Production | 2 | 1 | 2 | 1 | 10 | 10 | Participation and Having | 2 | Position | 1 | 0.5 | 5 |

| &F113 | F113 | H | Receiver | Production | 2 | 1 | 2 | 1 | 21 | 21 | Participation and Having | 2 | Position | 1 | 0.5 | 10.5 |
|---|---|---|---|---|---|---|---|---|---|---|---|---|---|---|---|---|
| X | F114 | H | Provider | Production | 1 | 1 | 1 | 1 | 21 | 21 | Participation and Having | 2 | Position | 1 | 0 | 0 |
| X | F115 | D | Receiver | Production | 1 | 1 | 1 | 1 | 10 | 10 | Participation and Having | 2 | Position | 1 | 0 | 0 |
| X | F116 | H | - | Production | 1 | 1 | 1 | 1 | 21 | 21 | Participation and Having | 2 | Position | 1 | 0 | 0 |
| F117 &F118 | F117 | H | Provider | Production | 2 | 1 | 2 | 1 | 21 | 21 | Participation and Having | 2 | Position | 1 | 0.5 | 10.5 |
| | F118 | D | Receiver | Production | 2 | 1 | 2 | 1 | 10 | 10 | Participation and Having | 2 | Position | 1 | 0.5 | 5 |
| F119 &F120 | F119 | H | Provider | Production | 2 | 1 | 2 | 1 | 21 | 21 | Participation and Having | 2 | Position | 1 | 0.5 | 10.5 |
| | F120 | D | Receiver | Production | 2 | 1 | 2 | 1 | 10 | 10 | Participation and Having | 2 | Position | 1 | 0.5 | 5 |
| X | F121 | D | - | Production | 1 | 1 | 1 | 1 | 10 | 10 | Understanding and Doing | 2 | Position | 1 | 0 | 0 |
| X | F122 | D | Provider | Production | 1 | 1 | 1 | 1 | 10 | 10 | Participation and Having | 2 | Position | 1 | 0 | 0 |
| X | F123 | H | Receiver | Production | 1 | 1 | 1 | 1 | 21 | 21 | Participation and Having | 2 | Position | 1 | 0 | 0 |
| F124 &F125 | F124 | D | Provider | Production | 2 | 1 | 2 | 1 | 10 | 10 | Participation and Having | 2 | Position | 1 | 0.5 | 5 |
| | F125 | H | Receiver | Production | 2 | 1 | 2 | 1 | 21 | 21 | Participation and Having | 2 | Position | 1 | 0.5 | 10.5 |
| X | F126 | H | Provider | Production | 1 | 1 | 1 | 1 | 21 | 21 | Participation and Having | 2 | Position | 1 | 0 | 0 |
| F127 &F128 | F127 | H | Provider | Production | 2 | 1 | 2 | 1 | 21 | 21 | Participation and Having | 2 | Position | 1 | 0.5 | 10.5 |
| | F128 | F | Receiver | Production | 2 | 1 | 2 | 1 | 12 | 12 | Participation and Having | 2 | Position | 1 | 0.5 | 6 |
| F129 &F130 | F129 | H | Provider | Production | 2 | 1 | 2 | 1 | 21 | 21 | Participation and Having | 2 | Position | 1 | 0.5 | 10.5 |
| | F130 | F | Receiver | Production | 2 | 1 | 2 | 1 | 12 | 12 | Participation and Having | 2 | Position | 1 | 0.5 | 6 |
| X | F131 | F | Provider | Production | 1 | 1 | 1 | 1 | 12 | 12 | Participation and Having | 2 | Position | 1 | 0 | 0 |

<table>
<tr><td>X</td><td>F132</td><td>H</td><td>Receiver</td><td>Production</td><td>1</td><td>1</td><td>1</td><td>1</td><td>21</td><td>21</td><td>Participation and Having</td><td>2</td><td>Position</td><td>1</td><td>0</td><td>0</td></tr>
<tr><td rowspan="2">F133 &F134</td><td>F133</td><td>F</td><td>Provider</td><td>Production</td><td>2</td><td>1</td><td>2</td><td>1</td><td>12</td><td>12</td><td>Participation and Having</td><td>2</td><td>Position</td><td>1</td><td>0.5</td><td>6</td></tr>
<tr><td>F134</td><td>E</td><td>Receiver</td><td>Production</td><td>2</td><td>1</td><td>2</td><td>1</td><td>5</td><td>5</td><td>Participation and Having</td><td>2</td><td>Position</td><td>1</td><td>0.5</td><td>2.5</td></tr>
<tr><td rowspan="2">F135 &F136</td><td>F135</td><td>F</td><td>Provider</td><td>Production</td><td>2</td><td>1</td><td>2</td><td>1</td><td>12</td><td>12</td><td>Participation and Having</td><td>2</td><td>Position</td><td>1</td><td>0.5</td><td>6</td></tr>
<tr><td>F136</td><td>E</td><td>Receiver</td><td>Production</td><td>2</td><td>1</td><td>2</td><td>1</td><td>5</td><td>5</td><td>Participation and Having</td><td>2</td><td>Position</td><td>1</td><td>0.5</td><td>2.5</td></tr>
<tr><td>X</td><td>F137</td><td>E</td><td>-</td><td>Production</td><td>1</td><td>1</td><td>1</td><td>1</td><td>5</td><td>5</td><td>Participation and Having</td><td>2</td><td>Position</td><td>1</td><td>0</td><td>0</td></tr>
<tr><td>X</td><td>F138</td><td>E</td><td>Provider</td><td>Production</td><td>1</td><td>1</td><td>1</td><td>1</td><td>5</td><td>5</td><td>Participation and Having</td><td>2</td><td>Position</td><td>1</td><td>0</td><td>0</td></tr>
<tr><td>X</td><td>F139</td><td>F</td><td>Receiver</td><td>Production</td><td>1</td><td>1</td><td>1</td><td>1</td><td>12</td><td>12</td><td>Participation and Having</td><td>2</td><td>Position</td><td>1</td><td>0</td><td>0</td></tr>
<tr><td rowspan="2">F140 &F141</td><td>F140</td><td>E</td><td>Provider</td><td>Production</td><td>2</td><td>1</td><td>2</td><td>1</td><td>5</td><td>5</td><td>Participation and Having</td><td>2</td><td>Position</td><td>1</td><td>0.5</td><td>2.5</td></tr>
<tr><td>F141</td><td>F</td><td>Receiver</td><td>Production</td><td>2</td><td>1</td><td>2</td><td>1</td><td>12</td><td>12</td><td>Participation and Having</td><td>2</td><td>Position</td><td>1</td><td>0.5</td><td>6</td></tr>
<tr><td>X</td><td>F142</td><td>F</td><td>-</td><td>Production</td><td>1</td><td>1</td><td>1</td><td>1</td><td>12</td><td>12</td><td>Participation and Having</td><td>2</td><td>Position</td><td>1</td><td>0</td><td>0</td></tr>
<tr><td>X</td><td>F143</td><td>F</td><td>Provider</td><td>Production</td><td>1</td><td>1</td><td>1</td><td>1</td><td>12</td><td>12</td><td>Participation and Having</td><td>2</td><td>Position</td><td>1</td><td>0</td><td>0</td></tr>
<tr><td rowspan="2">F144 &F145</td><td>F144</td><td>F</td><td>Provider</td><td>Production</td><td>2</td><td>1</td><td>2</td><td>1</td><td>12</td><td>12</td><td>Participation and Having</td><td>2</td><td>Position</td><td>1</td><td>0.5</td><td>6</td></tr>
<tr><td>F145</td><td>H</td><td>Receiver</td><td>Production</td><td>2</td><td>1</td><td>2</td><td>1</td><td>21</td><td>21</td><td>Participation and Having</td><td>2</td><td>Position</td><td>1</td><td>0.5</td><td>10.5</td></tr>
<tr><td>X</td><td>F146</td><td>H</td><td>-</td><td>Production</td><td>1</td><td>1</td><td>1</td><td>1</td><td>21</td><td>21</td><td>Participation and Having</td><td>2</td><td>Position</td><td>1</td><td>0</td><td>0</td></tr>
<tr><td>X</td><td>F147</td><td>H</td><td>Provider</td><td>Production</td><td>1</td><td>1</td><td>1</td><td>1</td><td>21</td><td>21</td><td>Participation and Having</td><td>2</td><td>Position</td><td>1</td><td>0</td><td>0</td></tr>
<tr><td>X</td><td>F148</td><td>F</td><td>Receiver</td><td>Production</td><td>1</td><td>1</td><td>1</td><td>1</td><td>12</td><td>12</td><td>Participation and Having</td><td>2</td><td>Position</td><td>1</td><td>0</td><td>0</td></tr>
<tr><td rowspan="2">F149 &F150</td><td>F149</td><td>H</td><td>Provider</td><td>Production</td><td>2</td><td>1</td><td>2</td><td>1</td><td>21</td><td>21</td><td>Participation and Having</td><td>2</td><td>Position</td><td>1</td><td>0.5</td><td>10.5</td></tr>
<tr><td>F150</td><td>F</td><td>Receiver</td><td>Production</td><td>2</td><td>1</td><td>2</td><td>1</td><td>12</td><td>12</td><td>Participation and Having</td><td>2</td><td>Position</td><td>1</td><td>0.5</td><td>6</td></tr>
</table>

| | | | | | | | | | | | | | | | | |
|---|---|---|---|---|---|---|---|---|---|---|---|---|---|---|---|---|
| X | F151 | H | - | Production | 1 | 2 | 2 | 1 | 21 | 21 | Creation and Doing | 2 | Position | 1 | 0 | 0 |
| X | F152 | H | - | Production | 1 | 2 | 2 | 1 | 21 | 21 | Creation and Doing | 2 | Position | 1 | 0 | 0 |
| X | F153 | H | - | Production | 1 | 1 | 1 | 1 | 21 | 21 | Creation and Doing | 2 | Position | 1 | 0 | 0 |
| X | F154 | D | Receiver | Production | 1 | 1 | 1 | 1 | 10 | 10 | Participation and Having | 2 | Position | 1 | 0 | 0 |
| F155 &F156 | F155 | H | Provider | Production | 2 | 1 | 2 | 1 | 21 | 21 | Participation and Having | 2 | Position | 1 | 0.5 | 10.5 |
| | F156 | D | Receiver | Production | 2 | 1 | 2 | 1 | 10 | 10 | Participation and Having | 2 | Position | 1 | 0.5 | 5 |
| X | F157 | D | Provider | Utilisation | 1 | 1 | 1 | 1 | 18 | 18 | Participation and Having | 2 | Position | 1 | 0 | 0 |
| X | F158 | C | Receiver | Utilisation | 1 | 1 | 1 | 1 | 18 | 18 | Participation and Having | 2 | Position | 1 | 0 | 0 |
| F159 &F160 | F159 | D | Provider | Utilisation | 2 | 1 | 2 | 1 | 18 | 18 | Participation and Having | 2 | Position | 1 | 0.5 | 9 |
| | F160 | C | Receiver | Utilisation | 2 | 1 | 2 | 1 | 18 | 18 | Participation and Doing | 2 | Position, Capability | 2 | 0.5 | 9 |
| F161 &F162 | F161 | D | Provider | Utilisation | 2 | 1 | 2 | 1 | 18 | 18 | Participation and Having | 2 | Position | 1 | 0.5 | 9 |
| | F162 | C | Receiver | Utilisation | 2 | 1 | 2 | 1 | 18 | 18 | Participation and Doing | 2 | Position, Capability | 2 | 0.5 | 9 |
| x | F163 | C | - | Utilisation | 1 | 1 | 1 | 1 | 18 | 18 | Participation and Having | 2 | Position | 1 | 0 | 0 |
| X | F164 | C | Provider | Utilisation | 1 | 1 | 1 | 1 | 18 | 18 | Participation and Having | 2 | Position | 1 | 0 | 0 |
| X | F165 | B | Receiver | Utilisation | 1 | 1 | 1 | 1 | 13 | 13 | Creation and Interacting | 2 | Capability | 1 | 0 | 0 |
| F166&F16 | F166 | C | Provider | Utilisation | 2 | 1 | 2 | 1 | 18 | 18 | Participation and Having | 2 | Position | 1 | 0.5 | 9 |
| | F167 | B | Receiver | Utilisation | 2 | 1 | 2 | 1 | 13 | 13 | Creation and Interacting | 2 | Capability | 1 | 0.5 | 6.5 |
| F168 | F168 | C | Provider | Utilisation | 2 | 1 | 2 | 1 | 18 | 18 | Participation and Having | 2 | Position | 1 | 0.5 | 9 |

| &F169 | F169 | B | Receiver | Utilisation | 2 | 1 | 2 | 1 | 13 | 13 | Creation and Interacting | 2 | Capability | 1 | 0.5 | 6.5 |
|---|---|---|---|---|---|---|---|---|---|---|---|---|---|---|---|---|
| X | F170 | D | Receiver | Utilisation | 1 | 1 | 1 | 1 | 18 | 18 | Participation and Having | 2 | Position | 1 | 0 | 0 |
| F171 &F172 | F171 | C | Provider | Utilisation | 2 | 1 | 2 | 1 | 18 | 18 | Participation and Doing | 2 | Position, Capability | 2 | 0.5 | 9 |
| | F172 | D | Receiver | Utilisation | 2 | 1 | 2 | 1 | 18 | 18 | Participation and Having | 2 | Position | 1 | 0.5 | 9 |
| X | F173 | D | Provider | Utilisation | 1 | 1 | 1 | 1 | 18 | 18 | Participation and Having | 2 | Position | 1 | 0 | 0 |
| X | F174 | C | Receiver | Utilisation | 1 | 1 | 1 | 1 | 18 | 18 | Participation and Having | 2 | Position | 1 | 0 | 0 |
| X | F175 | D | - | Utilisation | 1 | 1 | 1 | 1 | 18 | 18 | Participation and Having | 2 | Position | 1 | 0 | 0 |
| X | F176 | D | - | Utilisation | 1 | 1 | 1 | 1 | 18 | 18 | Participation and Having | 2 | Position | 1 | 0 | 0 |
| F177 &F178 &F179 | F177 | D | Provider | Utilisation | 3 | 2 | 6 | 1 | 18 | 18 | Participation and Having | 2 | Position | 1 | 0.5 | 9 |
| | F178 | C | Receiver | Utilisation | 3 | 2 | 6 | 1 | 18 | 18 | Participation and Doing | 2 | Position, Capability | 2 | 0.5 | 9 |
| | F179 | B | Receiver | Utilisation | 3 | 2 | 6 | 1 | 13 | 13 | Creation and Interacting | 2 | Capability | 1 | 0.5 | 6.5 |
| X | F180 | B | Provider | Utilisation | 1 | 1 | 1 | 1 | 13 | 13 | Creation and Interacting | 2 | Capability | 1 | 0 | 0 |
| X | F181 | D | Receiver | Utilisation | 1 | 1 | 1 | 1 | 18 | 18 | Participation and Having | 2 | Position | 1 | 0 | 0 |
| F182 &F183 | F182 | B | Provider | Utilisation | 2 | 1 | 2 | 1 | 13 | 13 | Creation and Interacting | 2 | Capability | 1 | 0.5 | 6.5 |
| | F183 | D | Receiver | Utilisation | 2 | 1 | 2 | 1 | 18 | 18 | Participation and Having | 2 | Position | 1 | 0.5 | 9 |
| X | F184 | D | Provider | Utilisation | 1 | 1 | 1 | 1 | 18 | 18 | Participation and Having | 2 | Position | 1 | 0 | 0 |
| X | F185 | B | Receiver | Utilisation | 1 | 1 | 1 | 1 | 13 | 13 | Creation and Interacting | 2 | Capability | 1 | 0 | 0 |
| F186 &F187 | F186 | D | Provider | Utilisation | 2 | 1 | 2 | 1 | 18 | 18 | Participation and Having | 2 | Position | 1 | 0.5 | 9 |
| | F187 | B | Receiver | Utilisation | 2 | 1 | 2 | 1 | 13 | 13 | Creation and Interacting | 2 | Capability | 1 | 0.5 | 6.5 |
| F188&& F189 &F190 | F188 | D | Provider | Utilisation | 3 | 1 | 3 | 1 | 18 | 18 | Participation and Having | 2 | Position | 1 | 0.5 | 9 |
| | F189 | C | Receiver | Utilisation | 3 | 1 | 3 | 1 | 18 | 18 | Participation and Doing | 2 | Position, Capability | 2 | 0.5 | 9 |
| | F190 | B | Receiver | Utilisation | 3 | 1 | 3 | 1 | 13 | 13 | Creation and Interacting | 2 | Capability | 1 | 0.5 | 6.5 |
| X | F191 | C | - | Utilisation | 1 | 1 | 1 | 1 | 18 | 18 | Participation and Having | 2 | Position | 1 | 0 | 0 |
| X | F192 | B | - | Utilisation | 1 | 1 | 1 | 1 | 13 | 13 | Creation and Interacting | 2 | Capability | 1 | 0 | 0 |
| X | F193 | C | Provider | Utilisation | 1 | 1 | 1 | 1 | 18 | 18 | Participation and Having | 2 | Position | 1 | 0 | 0 |
| X | F194 | D | Receiver | Utilisation | 1 | 1 | 1 | 1 | 18 | 18 | Participation and Having | 2 | Position | 1 | 0 | 0 |
| F195 &F196 | F195 | C | Provider | Utilisation | 2 | 1 | 2 | 1 | 18 | 18 | Participation and Doing | 2 | Position, Capability | 2 | 0.5 | 9 |
| | F196 | B | Receiver | Utilisation | 2 | 1 | 2 | 1 | 13 | 13 | Creation and Interacting | 2 | Capability | 1 | 0.5 | 6.5 |
| X | F197 | B | Provider | Utilisation | 1 | 1 | 1 | 1 | 13 | 13 | Creation and Interacting | 2 | Capability | 1 | 0 | 0 |
| X | F198 | C | Receiver | Utilisation | 1 | 1 | 1 | 1 | 18 | 18 | Participation and Having | 2 | Position | 1 | 0 | 0 |
| F199 &F200 | F199 | B | Provider | Utilisation | 2 | 1 | 2 | 1 | 13 | 13 | Creation and Interacting | 2 | Capability | 1 | 0.5 | 6.5 |
| | F200 | C | Receiver | Utilisation | 2 | 1 | 2 | 1 | 18 | 18 | Participation and Doing | 2 | Position, Capability | 2 | 0.5 | 9 |
| X | F201 | C | Provider | Utilisation | 1 | 1 | 1 | 1 | 18 | 18 | Participation and Having | 2 | Position | 1 | 0 | 0 |
| F202 | F202 | C | Provider | Utilisation | 2 | 1 | 2 | 1 | 18 | 18 | Participation and Doing | 2 | Position, Capability | 2 | 0.5 | 9 |

| &F203 | F203 | D | Receiver | Utilisation | 2 | 1 | 2 | 1 | 18 | 18 | Participation and Having | 2 | Position | 1 | 0.5 | 9 |
|---|---|---|---|---|---|---|---|---|---|---|---|---|---|---|---|---|
| X | F204 | D | - | Utilisation | 1 | 1 | 1 | 1 | 18 | 18 | Participation and Having | 2 | Position | 1 | 0 | 0 |
| X | F205 | D | - | Utilisation | 1 | 1 | 1 | 1 | 18 | 18 | Participation and Doing | 2 | Position | 1 | 0 | 0 |

**Supplementary File 5**

**Application of the Framework in Case Study 3 (CS3) - Function Identification**

| Function No | Function Explanation |
|---|---|
| F1 | Provide information (during workshop) |
| F2 | Receive information |
| F3 | Request to share the problems with existing conditions and products |
| F4 | Receive the task |
| F5 | |
| F6 | Change in the role PR |
| F7 | Change in the role RP |
| F8 | Change in the role RP |
| F9 | Request to provide problems and complaints about conditions and existing product |
| F10 | Receive the task |
| F11 | Change in the role PR |
| F12 | Change in the role RP |
| F13 | Sharing problems about the existing product |
| F14 | Receiving the problems |
| F15 | Change in the role RP |
| F16 | Provide information about the problems with existing groundwater level measuring device |

| F17 | Provide information about the problems with existing groundwater level measuring device |
| --- | --- |
| F18 | Receive the information |
| F19 | Examine information |
| F20 | Change in the role RP |
| F21 | Change in the role PR |
| F22 | Request to provide additional information about the usage of the existing product |
| F23 | Receive the task |
| F24 | Have a discussion with local people about the product problems |
| F25 | Change in the role RP |
| F26 | Change in the role PR |
| F27 | Provide additional information |
| F28 | Receive additional information for modification |
| F29 | Examine information |
| F30 | Request to provide solution - improvement in product quality |
| F31 | Receive the task |
| F32 | Change in the role PR |
| F33 | Change in the role RP |
| F34 | Request to provide additional informaion |
| F35 | Receive task |
| F36 | Change in the role PR |
| F37 | Change in the role RP |
| F38 | Change in the role PR |
| F39 | Request to provide additional information about the problems |
| F40 | Receive the task |
| F41 | Change in the role RP |
| F42 | Change in the role PR |
| F43 | Provide additional information |
| F44 | Receive additional information |
| F45 | Examine information |
| F46 | Include more details about the existing product |
| F47 | Combine the whole data |
| F48 | Change in the role RP |
| F49 | Provide additional information - encountered problems with the device developed by NGO |
| F50 | Receive additional information |
| F51 | Examine the information |
| F52 | Prepare a plan to modify the GLM device |
| F53 | Change in the role RP |
| F54 | Provide information |
| F55 | Receive information |
| F56 | Request the fund for upgrading GLM device |
| F57 | Receive the task |
| F58 | Examine the information |
| F59 | Change in the role RP |
| F60 | Change in the role PR |
| F61 | Request additional information |
| F62 | Receive the request |
| F63 | Prepare additional information |
| F64 | Change in the role RP |
| F65 | Change in the role PR |
| F66 | Provide additional information |
| F67 | Receive additional information |
| F68 | Examine the additional information |
| F69 | Change in the role RP |
| F70 | Change in the role PR |
| F71 | Provide aproval |
| F72 | Receive approval |
| F73 | Change in the role RP |

| F74 | Change in the role PR |
|---|---|
| F75 | Change in the role PR |
| F76 | Provide information about the acceptance of the fund |
| F77 | Receive the information |
| F78 | |
| F79 | Examine additional information about the problems of the device |
| F80 | Analsying requirements based on the condtions |
| F81 | Addressing solutions for the defined problems |
| F82 | Defining modifications to solve the problems |
| F83 | Conceptualisation based on the determined modifications |
| F84 | Preparing the CAD model |
| F85 | Provide information about the determined modification |
| F86 | Receive the information |
| F87 | Examine the information |
| F88 | Request to provide suggestions |
| F89 | Receive the task |
| F90 | Change in the role RP |
| F91 | Change in the role PR |
| F92 | Request additional information about the working principle |

| F93 | Receive the task |
|---|---|
| F94 | Change in the role RP |
| F95 | Change in the role PR |
| F96 | Provide additional information |
| F97 | Receive additional information |
| F98 | Examine additional information |
| F99 | Change in the role RP |
| F100 | Change in the role PR |
| F101 | Provide information/suggestion |
| F102 | Receive the information |
| F103 | Examine the information |
| F104 | Update the modified design based on the suggestion |
| F105 | Detailing product attributes |
| F106 | Metarial selection |
| F107 | Preparing detailed drawings for prototyping |
| F108 | Change in the role RP |
| F109 | Provide drawings |
| F110 | Receive drawings |
| F111 | Examine drawings |
| F112 | Request to produce the first prototype for trial |

| F113 | Receive the request |
|---|---|
| F114 | Change in the role RP |
| F115 | Change in the role PR |
| F116 | Prepare a list of materials for production |
| F117 | Provıde the list |
| F118 | Receive the list |
| F119 | Request confirmation |
| F120 | Receive the task |
| F121 | Examine the list |
| F122 | Change in the role RP |
| F123 | Change in the role PR |
| F124 | Confirm the list |
| F125 | Receive the confirmation |
| F126 | Change in the role RP |
| F127 | Provide the list |
| F128 | Receive the list |
| F129 | Request to supply the materials |
| F130 | Receive the task |
| F131 | Change in the role RP |

| F132 | Change in the role PR |
| --- | --- |
| F133 | Provide the list |
| F134 | Receive the list |
| F135 | Request to supply the materials |
| F136 | Receive the task |
| F137 | Collect materials |
| F138 | Change in the role RP |
| F139 | Change in the role PR |
| F140 | Provide ordered materials |
| F141 | Receive the order |
| F142 | Control the received materials |
| F143 | Change in the role RP |
| F144 | Provide ordered materials |
| F145 | Receive the order |
| F146 | Control the received materials |
| F147 | Change in the role RP |
| F148 | Change in the role PR |
| F149 | Provide confirmation |
| F150 | Receive confirmation |

| F151 | Preparation of manufacturing and assembly |
|---|---|
| F152 | Manufacturing and assembly of components |
| F153 | Completing the first prototype for trial |
| F154 | Change in the role PR |
| F155 | Handling the first prototype |
| F156 | Receive the prototype |
| F157 | Change in the role RP |
| F158 | Change in the role PR |
| F159 | Provide information about the upgraded model |
| F160 | Receive information |
| F161 | Request help from NGO to test the upgraded device |
| F162 | Receive the task |
| F163 | Prepare facilities for testing the upgraded device |
| F164 | Change in the role RP |
| F165 | Change in the role PR |
| F166 | Inform about device testing |
| F167 | Receive the information |
| F168 | Request to participate during testing the device |

<table>
<tr><td>F169</td><td>Receive the request</td></tr>
<tr><td>F170</td><td>Change in the role PR</td></tr>
<tr><td>F171</td><td>Provide facilities and support to test the device</td></tr>
<tr><td>F172</td><td>Receive the outcome</td></tr>
<tr><td>F173</td><td>Change in the role RP</td></tr>
<tr><td>F174</td><td>Change in the role PR</td></tr>
<tr><td>F175</td><td>Installation of the device</td></tr>
<tr><td>F176</td><td>Test the device</td></tr>
<tr><td>F177</td><td>Provide the device and information about the usage</td></tr>
<tr><td>F178</td><td rowspan="2">Receive information and device</td></tr>
<tr><td>F179</td></tr>
<tr><td>F180</td><td>Change in the role RP</td></tr>
<tr><td>F181</td><td>Change in the role PR</td></tr>
<tr><td>F182</td><td>Request to provide additional information about the device</td></tr>
<tr><td>F183</td><td>Receive the task</td></tr>
<tr><td>F184</td><td>Change in the role RP</td></tr>
<tr><td>F185</td><td>Change in the role PR</td></tr>
</table>

| F186 | Provide additional informational |
|---|---|
| F187 | Receive additional information |
| F188 | Request to provide feedback |
| F189 | Receive the request |
| F190 | |
| F191 | Use the device (for two years) |
| F192 | Use the device (for two years) |
| F193 | Change in the role RP |
| F194 | Change in the role PR |
| F195 | Request to provide feedback |
| F196 | Receive the task |
| F197 | Change in the role RP |
| F198 | Change in the role PR |
| F199 | Provide information/feedback |
| F200 | Receive information |
| F201 | Change in the role RP |
| F202 | Provide feedback |

| F203 | Receive the feedback |
| --- | --- |
| F204 | Examining feedback |
| F205 | Identifying vendors and suppliers for mass production |

**Supplementary File 5**

**Analysis of Inclusivity and Empowerment Metrics Across Lifecycle Phases in CS3**

| **Planning** | | | | | | | | |
|---|---|---|---|---|---|---|---|---|
| | A | B | C | D | E | F | G | H |
| Diversity | 2 | 2 | 1 | 1 | 0 | 0 | 1 | 0 |
| Hierarchy | 2 | 3 | 4 | 4 | 0 | 0 | 5 | 0 |
| LC | 1 | 2 | 3 | 4 | 0 | 0 | 1 | 0 |
| # people | 2 | 2 | 3 | 3 | 0 | 0 | 1 | 0 |
| | | | | | | | | |
| I1 | 8 | 12 | 12 | 12 | 0 | 0 | 5 | 0 |
| I2 | 4 | 12 | 12 | 16 | 0 | 0 | 5 | 0 |
| I3 | 8 | 24 | 36 | 48 | 0 | 0 | 5 | 0 |
| I4 | 4 | 8 | 9 | 12 | 0 | 0 | 1 | 0 |
| #I | 24 | 56 | 69 | 88 | 0 | 0 | 16 | 0 |
| | 253 | | | | | | | |
| Means | 1 | 1 | 1 | 1 | 0 | 0 | 1 | 0 |
| Impact | 14 | 15 | 36 | 46 | 0 | 0 | 15 | 0 |
| DL | 1.5 | 4 | 5.5 | 9 | 0 | 0 | 2.5 | 0 |
| | | | | | | | | |
| E1 | 14 | 15 | 36 | 46 | 0 | 0 | 15 | 0 |
| E2 | 1.5 | 4 | 5.5 | 9 | 0 | 0 | 2.5 | 0 |
| E3 | 21 | 60 | 198 | 414 | 0 | 0 | 37.5 | 0 |
| E4 | 21 | 60 | 198 | 414 | 0 | 0 | 37.5 | 0 |
| #E | 57.5 | 139 | 437.5 | 883 | 0 | 0 | 92.5 | 0 |
| | 1609.5 | | | | | | | |

| Development | | | | | | | | |
|---|---|---|---|---|---|---|---|---|
| | A | B | C | D | E | F | G | H |
| Diversity | 0 | 0 | 1 | 1 | 0 | 0 | 0 | 0 |
| Hierarchy | 0 | 0 | 4 | 4 | 0 | 0 | 0 | 0 |
| LC | 0 | 0 | 3 | 4 | 0 | 0 | 0 | 0 |
| # people | 0 | 0 | 1 | 1 | 0 | 0 | 0 | 0 |
| | | | | | | | | |
| I1 | 0 | 0 | 4 | 4 | 0 | 0 | 0 | 0 |
| I2 | 0 | 0 | 12 | 16 | 0 | 0 | 0 | 0 |
| I3 | 0 | 0 | 12 | 16 | 0 | 0 | 0 | 0 |
| I4 | 0 | 0 | 3 | 4 | 0 | 0 | 0 | 0 |
| #I | 0 | 0 | 31 | 40 | 0 | 0 | 0 | 0 |
| | 71 | | | | | | | |
| Means | 0 | 0 | 1 | 1 | 0 | 0 | 0 | 0 |
| Impact | 0 | 0 | 15 | 25 | 0 | 0 | 0 | 0 |
| DL | 0 | 0 | 2.5 | 2.5 | 0 | 0 | 0 | 0 |
| | | | | | | | | |
| E1 | 0 | 0 | 15 | 25 | 0 | 0 | 0 | 0 |
| E2 | 0 | 0 | 2.5 | 2.5 | 0 | 0 | 0 | 0 |
| E3 | 0 | 0 | 37.5 | 62.5 | 0 | 0 | 0 | 0 |
| E4 | 0 | 0 | 37.5 | 62.5 | 0 | 0 | 0 | 0 |
| #E | 0 | 0 | 92.5 | 152.5 | 0 | 0 | 0 | 0 |
| | 245 | | | | | | | |

| Production | | | | | | | | |
|---|---|---|---|---|---|---|---|---|
| | A | B | C | D | E | F | G | H |
| Diversity | 0 | 0 | 0 | 1 | 1 | 1 | 0 | 1 |
| Hierarchy | 0 | 0 | 0 | 4 | 1 | 2 | 0 | 2 |
| LC | 0 | 0 | 0 | 4 | 1 | 1 | 0 | 1 |
| # people | 0 | 0 | 0 | 1 | 1 | 2 | 0 | 2 |
| | | | | | | | | |
| I1 | 0 | 0 | 0 | 4 | 1 | 4 | 0 | 4 |
| I2 | 0 | 0 | 0 | 16 | 1 | 2 | 0 | 2 |
| I3 | 0 | 0 | 0 | 16 | 1 | 4 | 0 | 4 |
| I4 | 0 | 0 | 0 | 4 | 1 | 2 | 0 | 2 |
| #I | 0 | 0 | 0 | 40 | 4 | 12 | 0 | 12 |
| | 68 | | | | | | | |
| Means | 0 | 0 | 0 | 1 | 1 | 1 | 0 | 1 |
| Impact | 0 | 0 | 0 | 16 | 8 | 19 | 0 | 33 |
| DL | 0 | 0 | 0 | 3 | 1.5 | 3.5 | 0 | 5 |
| | | | | | | | | |
| E1 | 0 | 0 | 0 | 16 | 8 | 19 | 0 | 33 |
| E2 | 0 | 0 | 0 | 3 | 1.5 | 3.5 | 0 | 5 |
| E3 | 0 | 0 | 0 | 48 | 12 | 66.5 | 0 | 165 |
| E4 | 0 | 0 | 0 | 48 | 12 | 66.5 | 0 | 165 |
| #E | 0 | 0 | 0 | 115 | 33.5 | 155.5 | 0 | 368 |
| | 672 | | | | | | | |

| **Utilisation** | | | | | | | | |
|---|---|---|---|---|---|---|---|---|
| | A | B | C | D | E | F | G | H |
| Diversity | 0 | 2 | 1 | 1 | 0 | 0 | 0 | 0 |
| Hierarchy | 0 | 3 | 4 | 4 | 0 | 0 | 0 | 0 |
| LC | 0 | 2 | 2 | 4 | 0 | 0 | 0 | 0 |
| # people | 0 | 2 | 2 | 2 | 0 | 0 | 0 | 0 |
| | | | | | | | | |
| I1 | 0 | 12 | 8 | 8 | 0 | 0 | 0 | 0 |
| I2 | 0 | 12 | 8 | 16 | 0 | 0 | 0 | 0 |
| I3 | 0 | 24 | 16 | 32 | 0 | 0 | 0 | 0 |
| I4 | 0 | 8 | 4 | 8 | 0 | 0 | 0 | 0 |
| #I | 0 | 56 | 36 | 64 | 0 | 0 | 0 | 0 |
| | 156 | | | | | | | |
| Means | 0 | 1 | 1 | 1 | 0 | 0 | 0 | 0 |
| Impact | 0 | 26 | 33 | 31 | 0 | 0 | 0 | 0 |
| DL | 0 | 5 | 5 | 6.5 | 0 | 0 | 0 | 0 |
| | | | | | | | | |
| E1 | 0 | 26 | 33 | 31 | 0 | 0 | 0 | 0 |
| E2 | 0 | 5 | 5 | 6.5 | 0 | 0 | 0 | 0 |
| E3 | 0 | 130 | 165 | 201.5 | 0 | 0 | 0 | 0 |
| E4 | 0 | 130 | 165 | 201.5 | 0 | 0 | 0 | 0 |
| #E | 0 | 291 | 368 | 440.5 | 0 | 0 | 0 | 0 |
| | 1099.5 | | | | | | | |

**The framework application for all case studies can be accessed via the following link:**

**CS1-10.rar**

**Supplementary File 6**

**Summary of Linear Models for E4–I3 Relationships Across Case Studies (Details of Figure 6, see main content)**

| Dataset | Slope | Intercept | R_Squared | Formula |
|---|---|---|---|---|
| CS1 | 0.12 | 10.83 | 0.8 | I3 = 0.12 × E4 + 10.83 |
| CS2 | 0.19 | 2.65 | 0.44 | I3 = 0.19 × E4 + 2.65 |
| CS3 | 0.16 | -1.07 | 0.92 | I3 = 0.16 × E4 - 1.07 |
| CS4 | 0.02 | 42.39 | 0.04 | I3 = 0.02 × E4 + 42.39 |
| CS5 | 0.04 | -0.22 | 0.67 | I3 = 0.04 × E4 - 0.22 |
| CS6 | 0.08 | 8.71 | 0.83 | I3 = 0.08 × E4 + 8.71 |
| CS7 | 0.08 | -3.87 | 0.86 | I3 = 0.08 × E4 - 3.87 |
| CS8 | 0.09 | 4.34 | 0.7 | I3 = 0.09 × E4 + 4.34 |
| CS9 | 0.21 | 0.22 | 0.93 | I3 = 0.21 × E4 + 0.22 |
| CS10 | 0.03 | 11.94 | 0.34 | I3 = 0.03 × E4 + 11.94 |

**Comparison of Slopes and $R^2$ Values to Evaluate the Hypothesis**

| Dataset | Slope | Intercept | R_Squared | Level of Support |
|---|---|---|---|---|
| CS1 | 0.12 | 10.83 | 0.8 | Strong support |
| CS2 | 0.19 | 2.65 | 0.44 | Partial support |
| CS3 | 0.16 | -1.07 | 0.92 | Strong support |
| CS4 | 0.02 | 42.39 | 0.04 | Weak / no support |
| CS5 | 0.04 | -0.22 | 0.67 | Weak / no support |
| CS6 | 0.08 | 8.71 | 0.83 | Strong support |
| CS7 | 0.08 | -3.87 | 0.86 | Strong support |
| CS8 | 0.09 | 4.34 | 0.7 | Strong support |
| CS9 | 0.21 | 0.22 | 0.93 | Strong support |
| CS10 | 0.03 | 11.94 | 0.34 | Weak / no support |

The regression results show varying levels of support for the positive relationship between inclusivity and empowerment across the case studies. Strong support is found in CS1, CS3, CS6, CS7, CS8, and CS9, where both the slope and $R^2$ values are high, indicating that increases in inclusivity lead to consistent and significant increases in empowerment. Partial support is seen in CS2, where the slope is steep but the model fit is moderate, suggesting that the positive influence of inclusivity depends on certain conditions. In contrast, CS4, CS5, and CS10 provide weak or no support, as either the slope is nearly flat or the $R^2$ is too low, meaning that other barriers may need to be addressed before inclusivity can have a reliable impact on empowerment. In cases like CS6, CS7, and CS8, where the slope is low (around 0.08–0.09) but the $R^2$ is high ($\geq 0.70$), the findings imply a slow but steady positive effect. Here, inclusivity explains most of the variation in empowerment, although the per-unit effect is modest. These cases still confirm the hypothesis, but they suggest that continuous or higher levels of inclusivity are required to achieve meaningful empowerment gains.